%% file: main_arxiv.tex
\documentclass[10pt,journal,twocolumn]{IEEEtran}
\newif\ifCLASSOPTIONromanappendices \CLASSOPTIONromanappendicestrue
 \usepackage{bm}
\usepackage{stfloats}
\usepackage{relsize}
\usepackage{amsthm}

\usepackage[utf8x]{inputenc}
\usepackage[T1]{fontenc}
\usepackage{url}
\usepackage{xfrac}
\usepackage{multicol}
\usepackage{xcolor}
\usepackage{ifthen}
\usepackage{amsfonts}
\usepackage{amssymb}
\usepackage{amsmath,hyperref}
\usepackage{sansmath}
\usepackage{siunitx}
\usepackage{mathtools}
\usepackage{nccmath}
\usepackage{cite}
\usepackage{systeme}
\usepackage{spalign}
\usepackage{caption}
\usepackage{algpseudocode}
\usepackage{algorithm}

\DeclareMathOperator*{\argmax}{argmax}
\algnewcommand\algorithmicforeach{\textbf{for each}}
\algdef{S}[FOR]{ForEach}[1]{\algorithmicforeach\ #1\ \algorithmicdo}

\usepackage{cancel,xcolor}

\usepackage{eqparbox}
\newdimen{\algindent}
\algnewcommand\LeftComment[2]{%
\hspace{#1\algindent}$\triangleright$ \eqparbox{COMMENT}{#2} \hfill %
}
\algnewcommand\LeftCommentNoTriangle[2]{%
\hspace{#1\algindent} \eqparbox{COMMENT}{#2} \hfill %
}
\algnewcommand\LeftCommentNoIntent[1]{%
$\triangleright$ \eqparbox{COMMENT}{#1} \hfill %
}
\usepackage{verbatim}
\usepackage{tikz}
\usetikzlibrary{positioning,quotes,calc,fit,shapes,shadows,arrows}
\tikzset{block/.style={draw,very thick,text width=2cm,minimum height=4cm,align=center},
         line/.style={-latex}}
\tikzset{blockV/.style={draw,very thick,text width=2cm,minimum height=2cm, minimum width=4cm,align=center},
         line/.style={-latex}}
\tikzset{blockExt/.style={draw,very thick,minimum height=1cm, minimum width=1cm,align=center},
         line/.style={-latex}}
\usetikzlibrary{patterns}  
\usepgflibrary{arrows}
\usetikzlibrary{bayesnet}

\definecolor{light-gray}{HTML}{E0E0E0}

\makeatletter
\newcommand\notsotiny{\@setfontsize\notsotiny{6.82}{7.5}}
\makeatother
\newcommand{\B}[1]{\boldsymbol{#1}}

\newcommand\norm[1]{\left\lVert#1\right\rVert}

\makeatletter
\newcommand{\labeltarget}[1]{\Hy@raisedlink{\hypertarget{#1}{}}}
\makeatother

\begin{document}
\title{Nonlocal Reconfigurable Intelligent Surfaces for Wireless Communication: \\Modeling and Physical Layer Aspects}

\author{Amine Mezghani, Faouzi Bellili,
and Ekram Hossain
 \vspace{0.3cm}
\\\small University of Manitoba,  Winnipeg, MB, R3T 5V6, Canada.  
  \vspace{0.1cm}
  \\\small Emails: 
  \{amine.mezghani,faouzi.bellili,ekram.hossain\}@umanitoba.ca
  \vspace{-0.5cm}
}
\maketitle
\begin{abstract}

Conventional  Reconfigurable intelligent surfaces (RIS) for wireless communications have a local position-dependent (phase-gradient) scattering response on the surface. We consider more general RIS structures, called nonlocal (or redirective) RIS, that are capable of selectively manipulate the impinging waves depending on the incident angle. Redirective RIS have nonlocal wavefront-selective scattering behavior and can be implemented using multilayer arrays such as metalenses. 
We demonstrate that this more sophisticated type of surfaces has several advantages such as: lower overhead through coodebook-based reconfigurability, decoupled wave manipulations, and higher efficiency in multiuser scenarios via multifunctional operation.
Additionally, redirective RIS architectures greatly  benefit form the directional nature of wave propagation at high frequencies and can support integrated fronthaul and access (IFA) networks most efficiently. We also discuss the scalability and compactness issues and propose efficient nonlocal RIS architectures such as fractionated lens-based RIS and mirror-backed phase-masks structures that do not require additional control complexity and overhead while still offering better performance than conventional local RIS.

\end{abstract}



\section{Introduction}
The network capacity and coverage of previous cellular systems have been expanded mainly by deploying more access points and adding more frequency bands. Despite the rapid technological progress, a cost-effective way to deliver ubiquitous multi-Gpbs mobile data transmissions anywhere/anytime
remains however still
a challenging problem, since all current cellular
technologies become very capital intensive
beyond a certain average rate per user threshold \cite{OUGHTON201950}.

Higher network capacity and peak data rates are
theoretically conceivable at the mmWave and THz spectrum, due to the availability of larger bandwidths and the realizability of more directional transmissions in 3D. The main challenge for these spectrum parts is, however, the reduced propagation range and exacerbated blockage issues
in  addition to the radio frequency (RF) power limitations, 
especially in the uplink. As a consequence, mmWave/TeraHz mobile access intrinsically requires small-cell deployment with a line-of-sight (LOS) path or strong mirror-like reflective paths and is currently targeting ultra-high-density areas. Another issue is the traffic variation and the strongly asymmetric performance between the uplink and the downlink, thereby limiting the deployment scenarios. 

It becomes clear then that more innovations in network design and communication techniques are needed to enable wider deployment of mmWave/TeraHz technologies. In fact, a simple expansion by adding radio nodes with more backhaul infrastructure significantly increases cost and complexity while potentially leading to an underutilized system \cite{OUGHTON201950} due to the stronger impact of time-variant user distribution and traffic fluctuations on dense networks. It is also worth mentioning that user-centric generalized coverage by multiple base stations is also critical even in lower frequencies. In addition, controlled cell range extension for network resilience (to node failures for instance) or mobile broadband coverage in rural areas is also an open problem even at sub-6GHz frequencies necessitating concepts for low-cost flexible network edge components. This calls for innovation in network design that could enable an access point to multiplex across different areas, optimizing both spatial coverage and temporal utilization of the  network.

To address these issues, the 
use of reconfigurable intelligent surfaces (RIS) \cite{Renzo2020,Dajer_2022}, also known as smart passive and active (network-controlled) repeaters, is being widely investigated in academia and industry as a potentially low-cost, low-power stopgap solution for mitigating the coverage issue of future networks. While densification is unavoidable for widespread 
service availability, 
another equally important  goal is to reduce the number of required sophisticated and energy-intensive nodes/cells in the network and distribute them sparsely and flexibly (closer to existing fiber points of presence) so as to make mobile mmWave/TeraHz communication economically 
viable. To be considered as a cheaper, competitive, and economically scalable alternative to small cells for network expansion,  RIS-based edge nodes scattered around a central baseband node
have to meet the following criteria \cite{Dajer_2022}
\begin{itemize}
    \item Low-power/quasi-passive  (e.g., solar-powered), low-cost, low-maintenance, lightweight, and easy-to-deploy and operate nodes for instance on an aerial platform 
    \item  Full-duplex operation, i.e., on-channel manipulation (as opposed to standard relays) and with very low latency (data forwarding with no processing delay) 
  \item Restricted RF and physical layer capabilities (transparent layer-0 nodes with only control channel receiver and no transmitter chain)
  \item In-band or out-of-band IoT-dedicated remote control link (e.g., via a small control bandwidth partition), which 
  can be decoded without demodulating the entire waveform).
\end{itemize}


Furthermore, the form factor and aesthetic aspects of future wireless equipment will also be critical to not further impact the urban landscape and to keep wireless radiations and deployments at levels that are acceptable by the general public while meeting the levels already imposed by regulations.

\begin{figure}
    \centering
    \input{fig8}
\caption{Closing the coverage gap in rural areas with RIS-enabled multiple data pipes that exploit the extra capacity of lightly loaded base stations.} 
\label{RIS_rural}
\end{figure}
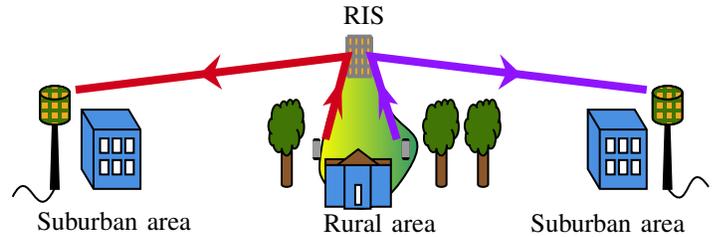

In this context, new antenna designs and deployment strategies are being considered to enable multipurpose reconfigurable intelligent active/passive surfaces/scatterers (RIS) that can favorably alter wave propagation.  These antennas are ultrathin surfaces with subwavelength (i.e., microscopic) structures, also known as metasurfaces with 2D applied/induced magnetic and electric currents along the surface \cite{Renzo2020}.  In the context of passive metasurfaces, the subwavelength structure offers a flexible and efficient wavefront manipulation such as beam-steering of the incident wave by optimizing the surface impedance distribution.
A popular use case for RIS is when there are obstructions between the transmitter and the receiver that lead to a communication failure. By bending/deflecting RF beams in 3D, one can easily navigate around the obstructions
along appropriate trajectories (c.f. Fig.~\ref{RIS_street}). For sub-6 GHz frequencies, RIS can be used to eliminate the coverage gap in rural areas by closing multiple links to far-away base stations on the edge of the network that are generally underutilized as shown in Fig.~\ref{RIS_rural}. 

\begin{figure}
    \centering
    \input{fig5}
\caption{Streetlight-mounted RIS nodes assisting an ultra-high capacity cell site.}
\label{RIS_street}
\end{figure}
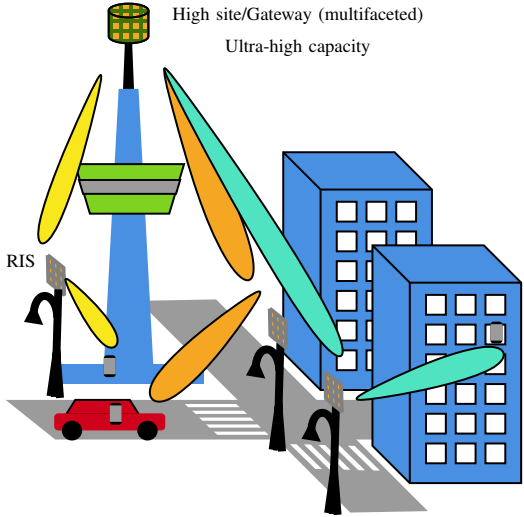

While the majority of prior work has focused on phased reflective\footnote{For simplicity, transmissive RIS configurations are also categorized as being reflective throughout this paper.} RIS architecture with local wave manipulation \cite{Renzo2020},  recent research papers have considered nonlocal designs such as directive reconfigurable surfaces  \cite{Abri_2017,Tasci_2022}, layered RIS with multi-plane wave conversion \cite{stacked_liu_2022,stacked_an_2023}, as well RIS with non-diagonal phase-shift matrices \cite{Li_2022}.  However, the existing RIS-based directive communication designs rely,  for instance, on the use of separate reception and re-transmission arrays that are connected in a back-to-back manner, thereby resulting in a very spatially selective wave manipulation. In contrast to reflective configurations, redirective systems are very common in practice and are known as \textit{bent-pipe} (through) relays in satellite communications and as (network-controlled) \textit{smart repeaters} in the 5G industry \cite{Flamini2022,Ho_2019}. From the metamaterial perspective, redirective antenna systems can be regarded as nonlocal metasurfaces due to their angle-dependent processing \cite{Kwon_2018}. Nonlocal RIS design offers additional degrees of freedom to design wave manipulations that are not possible with conventional local designs. However, designs based on layered RIS \cite{stacked_liu_2022,stacked_an_2023} suffer from higher complexity, control overhead, and losses as compared with the conventional local design. 

Local metasurfaces only provide pointwise scattering responses and have restricted functionalities due to the local power conservation requirement. By contrast, nonlocal metasurfaces allow for more general responses that are angular (or wave vector) dependent \cite{Kwon_2018,Chen_2021,Popov_2021}. Throughout the paper, RIS design that is implemented using local metasurfaces is also described as reflective, whereas nonlocal metasurface-based design is also termed redirective RIS. 

The question of which type of RIS scattering, configuration, and controllability to use for cost-effective mmWave//TeraHz networks with low control and training overhead remains, however, an open research question. This paper is an attempt to provide a systematic comparison of several aspects of these two different types of RIS deployment. While some questions remain open (regarding compactness and scalability), our conclusion is quite in favor of network configurations with redirective RIS-based quasi-passive nodes, which provide better scaling in terms of aperture, bandwidth, control, and estimation overhead. We also show that surprisingly simple salable nonlocal RIS designs that do not require additional complexity or overhead as compared to local designs are potentially possible while offering better performance.    

\noindent
\textbf{Notation}:
Vectors and matrices are denoted by lower- and upper-case italic boldface letters.  The operators $(\bullet)^\mathsf{T}$, $(\bullet)^\mathsf{H}$, $\textrm{tr}(\bullet)$, and $(\bullet)^*$ stand for transpose, Hermitian (i.e., conjugate transpose), trace, and complex conjugate, respectively.  The identity matrix is denoted as ${\bf I}$ and the size will be understood from the context. All vectors are in column-wise orientation by default and $x_i$ is the $i$-th element of $\B{x}$.    The $i$-th column of a  matrix $\B{X}$ is denoted as $\boldsymbol{x}_i$ while  $\left[\B{X}\right]_{i,j}$ stands for its ($i$th, $j$th) entry.  We represent the Kronecker and Hadamard product of vectors and matrices by the operators  "$\otimes$" and "$\odot$", respectively. Additionally, both $\textrm{Diag}(\boldsymbol{X})$ and $\textrm{Diag}(\boldsymbol{x})$ return a diagonal matrix by, respectively, keeping only the diagonal elements in  $\boldsymbol{X}$, or placing the vector $\boldsymbol{x}$ on its main diagonal. Finally, ${\rm j}$ is the imaginary unit (i.e., $ {\rm j}^2= -1$), and the notation $\triangleq$  is used for definitions.

\section{Physically-Consistent Modeling of RIS}

\begin{figure}
    \centering
    \input{fig1}
\caption{Antenna scattering with $m$ accessible ports.}
\label{antenna_scattering}
\end{figure}
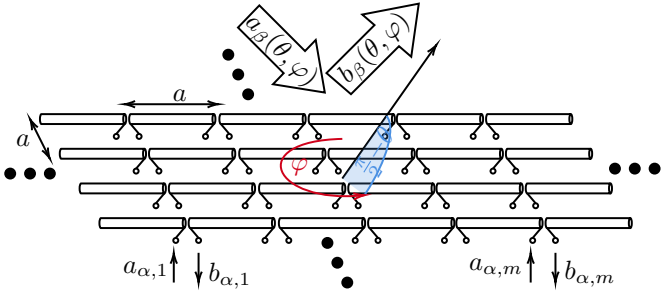

The modeling of antenna arrays or reconfigurable surfaces is becoming critical for the design and optimization of future wireless systems \cite{Renzo2020, Dajer_2022}. In this section, we provide a general modeling methodology for antenna arrays as well as reconfigurable surfaces based on the linear scattering theory. Contrary to prior work \cite{Gradoni_2021}, we adapt the linear scattering description rather than the impedance description since it is easier to measure and gives a more informative and intuitive interpretation of the model. In particular, we show that the conventional model that ignores the mutual coupling effects can be regarded as a first-order approximation of the physically consistent model. 
\subsection{Linear Scattering Model}

As shown in Fig~\ref{antenna_scattering}, the electromagnetic properties of an antenna array or a scatterer are described by its radiating/receiving patterns, the space-side scattering pattern, as well as the electrical multi-port behavior of its terminals \cite{Kerns1960TheoryOD} (assuming lumped elements connected to the structure). For simplicity, we treat the electromagnetic field, which is generally a vectorial complex quantity, as a complex-valued scalar quantity (e.g., considering a single polarization). 
At each $m$th accessible port, the forward and backward traveling (voltage) wave phasors along the antenna feed line are denoted by $a_{\alpha,m}$ and $b_{\alpha,m}$, respectively.
At a distant sphere, the angular transmission characteristic of a certain polarization when the antenna port $m$ is connected to a wave-source amplitude $a_{\alpha,m}$ (in $ [\sqrt{\rm W}]$) at port $m$ is defined as (also known as complex pattern)
\begin{equation}
    s_m(\theta,\varphi)=\lim\limits_{r \rightarrow \infty} -{\rm j} r {\rm e}^{{\rm j}k r} \frac{E^{(m)}(\theta,\varphi,r)}{a_{\alpha,m}} \sqrt{1/\eta_0},
\end{equation}
where $\eta_0$ is the characteristic impedance of free space and $E^{(m)}(\theta,\varphi,r)$ is the corresponding generated electrical far-field. These element-embedded complex patterns, $s_m(\theta,\varphi)$, are stacked together to form the receiving/transmitting characteristic vector $\B{s}_{\alpha\beta}(\theta,\varphi)$.
Assuming single polarization hemispherical radiation, the general  antenna array can now be characterized by the linear scattering description for the terminal side
\begin{equation}
 \B{b}_\alpha = \B{S}_{\alpha\alpha} \B{a}_\alpha + \int_{0}^{\frac{\pi}{2}} \int_{-\pi}^{\pi} \!\!  \B{s}_{\alpha\beta}(\theta,\varphi)  a_\beta(\theta,\varphi)    
{\rm sin}\theta {\rm d} \varphi {\rm d} \theta,
\end{equation}
where $ \B{a}_\alpha\triangleq [a_{\alpha,1},\ldots, a_{\alpha,M}]^T$ and $\B{S}_{\alpha\alpha}$ is the scattering matrix . For the space-side, the angular spectra of the outgoing propagating wave phasors $b_\beta(\theta,\varphi)$ are expressed as a linear functional of the incoming one $a_\beta(\theta,\varphi)$ as well as the port incident phasor  
\begin{equation}
\begin{aligned}
 b_\beta(\theta,\varphi) \,=~&   \B{s}_{\alpha\beta}(\theta,\varphi)^{\mathsf{T}}  \B{a}_\alpha  + \\
 &\int_{0}^{\frac{\pi}{2}} \!\! \!\!  \int_{-\pi}^{\pi} \!\!\! s_{\beta\beta}(\theta,\varphi,\theta',\varphi')  a_\beta(\theta',\varphi')    
{\rm sin}\theta' {\rm d} \varphi' {\rm d} \theta',
\end{aligned}
\end{equation}
where $s_{\beta\beta}(\theta,\varphi,\theta',\varphi')$ is the wave back-scattering characteristic which is in general hard to determine experimentally. Ideally, $s_{\beta\beta}(\theta,\varphi,\theta',\varphi')$ is effectively negligible or is just considered as part of the fixed (direct) communication channel. In the context of RIS deployment, the antenna ports are terminated by loads characterized by the load scattering matrix $\B{S}_L$, i.e., $\B{b}_\alpha=\B{S}_L\B{a}_\alpha$. Consequently, we get the total scattering  
\begin{equation}
\begin{aligned}
 b_\beta(\theta,\varphi) \,=~&   \B{s}_{\alpha\beta}(\theta,\varphi)^{\mathsf{T}}  (\B{S}_L^{-1}-\B{S}_{\alpha\alpha})^{-1} \times \\
& \int_{0}^{\frac{\pi}{2}} \!\! \!\!  \int_{-\pi}^{\pi} \!\!\! \B{s}_{\alpha\beta}(\theta',\varphi')  a_\beta(\theta',\varphi')    
{\rm sin}\theta' {\rm d} \varphi' {\rm d}\theta' \,+ \\
 &\underbrace{\int_{0}^{\frac{\pi}{2}} \!\! \!\!  \int_{-\pi}^{\pi} \!\!\! s_{\beta\beta}(\theta,\varphi,\theta',\varphi')  a_\beta(\theta',\varphi')   
{\rm sin}\theta' {\rm d} \varphi' {\rm d} \theta'}_{\textrm{Residual scattering}}.
\end{aligned}
\label{exact_scattering_model}
\end{equation}

For passive RIS, the spectral radius of this scattering operator ($ a_\beta(\theta,\varphi)\longrightarrow b_\beta(\theta,\varphi)$) is less or equal to one. That is, the scattered power is less than or equal to the impinging power regardless of the impinging wave 
\begin{equation}
\begin{aligned}
\int_{0}^{\frac{\pi}{2}} \!\! \!\!  \int_{-\pi}^{\pi} \!\!\!  |b_\beta(\theta,\varphi)|^2
{\rm sin}\theta {\rm d} \varphi {\rm d}\theta \leq \int_{0}^{\frac{\pi}{2}} \!\! \!\!  \int_{-\pi}^{\pi} \!\!\!  |a_\beta(\theta,\varphi)|^2
{\rm sin}\theta {\rm d} \varphi {\rm d}\theta.
\end{aligned}
\end{equation}
If the antenna is lossless then we have equality and the operator is unitary with all eigenvalues having unit absolute values. A simplistic approach that is commonly used in prior work is to use the Neumann series approximation
\begin{equation}
\begin{aligned}(\B{S}_L^{-1}-\B{S}_{\alpha\alpha})^{-1}\!=\!\!\sum_{k=0}^\infty (\B{S}_L \B{S}_{\alpha\alpha})^k \B{S}_L\!=\!\B{S}_L \!+\! \B{S}_L \B{S}_{\alpha\alpha} \B{S}_L\!+\ldots, 
\end{aligned}
\label{multiple_reflections}
\end{equation}
\begin{equation}
\begin{aligned}
\textrm{thus, }  b_\beta(\theta,\varphi)_\textrm{main-refl.} \,&\approx~   \B{s}_{\alpha\beta}(\theta,\varphi)^{\mathsf{T}}  \B{S}_L \times \\
& \int_{0}^{\frac{\pi}{2}} \!\! \!\!  \int_{-\pi}^{\pi} \!\!\! \B{s}_{\alpha\beta}(\theta',\varphi')  a_\beta(\theta',\varphi')    
{\rm sin}\theta' {\rm d} \varphi' {\rm d}\theta',
\end{aligned}
\label{naive_approx}
\end{equation}
which neglects the residual backscattering and the high-order multiple reflections between the antenna and load given by (\ref{multiple_reflections}). This applies for $\B{S}_{\alpha\alpha}=\B{0}$ (or $\rho (\B{S}_L \B{S}_{\alpha\alpha}) \ll 1$), i.e., perfectly matched array (or high losses). In general, the exact expression might give better results since $(\B{S}_L^{-1}-\B{S}_{\alpha\alpha})^{-1}$ might have singular values larger than 1 as opposed to $\B{S}_L$.
Nevertheless, the main reflection approximation in (\ref{naive_approx}) has a more tractable mathematical structure for optimizing the reactive load  $\B{S}_L$, the reason for which it is widely used in the literature.

\subsection{Determination of the Scattering Parameters in the Far-field}

Consider a quasi-continuous infinite unidirectional antenna surface with element-wise far-field effective area $A_{e}(\theta, \varphi)$   for the elevation (aspect) and azimuth angles $\theta$ and $\varphi$ for all elements while neglecting the edge effects (i.e., $A_{e}(\theta, \varphi)$ is the embedded pattern of inner elements). The embedded far-field array response of an $\sqrt{M} \times \sqrt{M}$ uniform planar array is $\B{s}_{\alpha\beta}(\theta,\varphi)=\sqrt{A_{e}(\theta, \varphi)} \boldsymbol{a}\left( \theta, \varphi \right)$ 
with steering vector
\begin{equation}
\begin{aligned}    
   \!\!\!\!& \boldsymbol{a}\left( \theta, \varphi \right)   \,=\,\begin{bmatrix}  1 \\ e^{-2\pi {\rm j} k_y } \\ \vdots  \\   e^{-2\pi {\rm j} k_y (\sqrt{M}-1) }  \end{bmatrix}  \otimes  \begin{bmatrix}  1 \\ e^{-2\pi {\rm j}  k_x } \\ \vdots  \\   e^{-2\pi {\rm j} k_x (\sqrt{M}-1) }  \end{bmatrix}, 
    \end{aligned}  \!\!\!\!
    \label{array_response}
\end{equation}
with spatial frequencies $k_y=\frac{a}{\lambda}  \sin \theta \sin \varphi$, $k_x=\frac{a}{\lambda}  \sin \theta \cos \varphi$, square inter-element spacing $a$ and wavelength $\lambda=c_0/f$. Assuming lossless antennas with the property 
\begin{equation}
\begin{aligned}
\B{S}_{\alpha\alpha}\B{S}_{\alpha\alpha}^{\mathsf{H}}+ \underbrace{\int_{0}^{\frac{\pi}{2}} \!\! \!\!  \int_{-\pi}^{\pi} \!\!\! \B{s}_{\alpha\beta}(\theta,\varphi)   \B{s}_{\alpha\beta}(\theta,\varphi)^{\mathsf{H}}
{\rm sin}\theta {\rm d} \varphi {\rm d}\theta}_{\substack{\\\mathlarger{\triangleq}\\\B{B}}} = {\bf I},
\end{aligned}
\label{matrix_B}
\end{equation}
then a relationship between the embedded pattern coupling matrix $\B{B}$ and the S-matrix $\B{S}_{\alpha\alpha}$ is given by  
\begin{equation}
    \B{B}={\bf I}-\B{S}_{\alpha\alpha}\B{S}_{\alpha\alpha}^{\mathsf{H}}=\B{U}\B{\Lambda}\B{U}^{\mathsf{H}}.
\end{equation}
Due to reciprocity, i.e.,  $\B{S}_{\alpha\alpha}=\B{S}_{\alpha\alpha}^{\mathsf{T}}$, we can obtain when all eigenvalues of $\B{B}$ are distinct 
\begin{equation}
    \B{S}_{\alpha\alpha}=\B{U}{\bf Diag}({\rm e}^{{\rm j} \alpha_1},\ldots,{\rm e}^{{\rm j} \alpha_M})\sqrt{{\bf I}-\B{\Lambda}}\B{U}^{\mathsf{T}},
\label{s_matrix}
\end{equation}
with arbitrary phases $\alpha_1\cdots \alpha_M$.

A possible admissible effective area that is compatible with the passivity condition  $\B{B}\preccurlyeq {\bf I}$ is the cosine-shaped pattern, also known as the normal gain with  uniformly illuminated aperture:
\begin{equation}
 A_{e}(\theta, \varphi)=a^2 \cos \theta.
\end{equation}
The cosine accounts for the projected effective area in the desired direction. This effective area of an embedded antenna element is considered as an upper limit and applies only for antenna arrays that can illuminate one half-space separately (e.g., antenna array backed by an infinite perfect reflector) and with sufficiently oversampled (quasi-continuous) aperture \cite{Kildal2016,Craeye2011}.

A practical implementation using connected dipoles that approximately achieves this behavior can be found in \cite{Neto2006} and is illustrated in Fig.~\ref{antenna_scattering}. Connected dipoles, however, do not achieve wide-angle scanning performance according to the cosine pattern. This requires generally a more sophisticated design based on magneto-electric dipoles with magneto-electric coupling effect \cite{Asadchy_2016,Henry_2016}. 
Explicitly, for the case with $ A_e(\theta,\varphi) =a^2 \cos \theta $, the  matrix $\B{B}$ is  obtained from the evaluation of the spherical  integral in (\ref{matrix_B}) as follows \cite{williams2019communication,Mezghani_2020}:
\begin{equation}
\begin{aligned}
&[\B{B}]_{k+\sqrt{M}(\ell-1), m+\sqrt{M}(n-1) }  = \\
 & \begin{cases}
    \frac{\pi a^2}{\lambda^2}, & \text{for } k=m \text{ and } \ell=n \\
    \frac{a}{\lambda} \frac{ J_1(\frac{2\pi a}{\lambda}\sqrt{(k-m)^2+(\ell-n)^2})}{\sqrt{(k-m)^2+(\ell-n)^2}}, & \text{otherwise, }  
  \end{cases}
\end{aligned}
\end{equation}
where $J_1(.)$ is the  Bessel function of first kind and first order. The scattering matrix $\B{S}_{\alpha\alpha}$ can be then obtained from (\ref{s_matrix}). 

In the large system limit, the matrix $\B{S}_{\alpha\alpha}$ is 
 partially isometric, i.e.,  $\B{S}_{\alpha\alpha}\B{S}_{\alpha\alpha}^{\mathsf{H}}$ is idempotent with eigenvalues that are either 0 or 1 \cite{williams2019communication}. Since the product of partial isometries is a contraction that is not necessarily a power-conserving transformation \cite{Kuo_1989}, we have
\begin{equation}
\begin{aligned}
\int_{0}^{\frac{\pi}{2}} \!\! \!\!  \int_{-\pi}^{\pi} \!\!\!  |b_\beta(\theta,\varphi)_\textrm{main-refl.}|^2
{\rm sin}\theta {\rm d} \varphi {\rm d}\theta \leq \!\! \int_{0}^{\frac{\pi}{2}} \!\! \!\!  \int_{-\pi}^{\pi} \!\!\!  |b_\beta(\theta,\varphi)|^2
{\rm sin}\theta {\rm d} \varphi {\rm d}\theta.
\end{aligned}
\end{equation}
In other words, the approximation in (\ref{naive_approx}) that does not take into account mutual couplings might violate the power conservation law even for this ideal RIS model with cosine-shaped pattern. 

\subsection{Numerical examples}
 In this subsection, we provide some numerical results to illustrate the impact of mutual coupling on the achievable rate performance.  Consider a base station with 32 antennas communicating to $K$ users via a reflective RIS node of size $8\lambda\times8 \lambda$ where no direct link is available. The carrier frequency is 30~GHz, the bandwidth is 300~MHz, and the BS transmit power is 10~W. We assume non-line-of-sight propagation over a 50-meter distance between the BS and the RIS nodes with 10 multipath components and $2.5$ pathloss exponent, whereas the users' terminals are in a line-of-sight condition to the RIS node with 5 meters distance. The end-to-end channel is given by (ignoring the direct link and the residual scattering)
 \begin{equation}
 \B{H}_{\rm 2-hop}=\B{H}_{\rm UE-RIS}(\B{S}_L^{-1}-\B{S}_{\alpha\alpha})^{-1}\B{H}_{\rm RIS-BS},
 \end{equation}
 where $\B{H}_{\rm UE-RIS}$ and $\B{H}_{\rm RIS-BS}$ are the far-field channel matrices between the users and RIS node, and between the RIS node and the base stations, respectively. These two channel matrices are defined when the RIS ports are terminated (and sensed) by loads equal to the reference impedance.


In the case of reflective RIS configuration, alternating optimization can be used to optimize the digital precoder $\B{P}$ at the BS, the RIS phase shifts $\B{S}_{\alpha\alpha}$, and the common receiver gain $\alpha$ in terms of minimum sum mean square error \cite{Haseeb_2021}. The phase shifters optimization is performed using the gradient projection procedure. That is
 \begin{equation}
\min\limits_{\B{S}_L} \norm{\alpha\B{H}_{\rm UE-RIS}{(\B{S}_L^{-1}-\B{S}_{\alpha\alpha})^{-1}}\B{H}_{\rm RIS-BS}\B{P}-{\bf I}_K}_{\rm F}^2, 
\label{MSE}
\end{equation} 
leading to the gradient descent step  
\begin{equation}
\begin{aligned}
\B{S}_L^{\ell+1}=\B{S}_L^{\ell}-\mu \cdot  {\rm Diag}(\B{\Delta}^\ell),
\end{aligned}
\end{equation}
with 
\begin{equation}
\begin{aligned}
&\B{\Delta}=\\
&\B{S}_L^{-\rm H}(\B{S}_L^{-1}-\B{S}_{\alpha\alpha})^{-\rm H}     \alpha\B{H}_{\rm UE-RIS}^{\rm H}    \Big(\alpha\B{H}_{\rm UE-RIS}{(\B{S}_L^{-1}-\B{S}_{\alpha\alpha})^{-1}}\\
&\B{H}_{\rm RIS-BS}\B{P}-{\bf I}_K\Big) \B{P}^{\rm H} \B{H}_{\rm RIS-BS}^{\rm H}      (\B{S}_L^{-1}-\B{S}_{\alpha\alpha})^{-\rm H}\B{S}_L^{-\rm H},
\end{aligned}
\end{equation}
 followed by a projection of the diagonal elements of $\B{S}_L^{\ell+1}$ on the unit circle.

\begin{figure}
    \centering
    \includegraphics[width=0.9\linewidth]{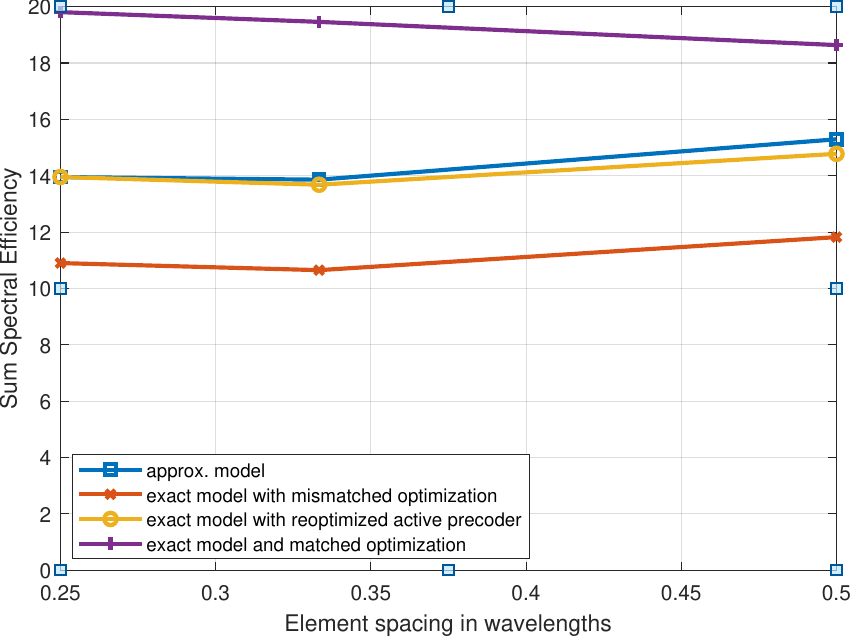}
\caption{Performance of RIS-aided 4-users downlink with ideal phase shifters. Mutual coupling has a significant impact.}
\label{4users_no_losses}
\end{figure}

\begin{figure}
    \centering
    \includegraphics[width=0.9\linewidth]{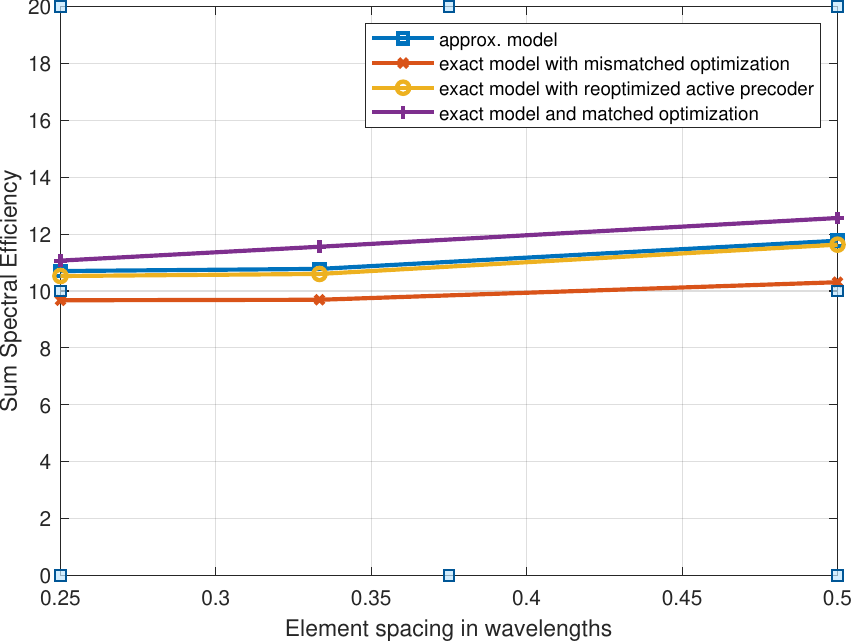}
\caption{Performance of RIS-aided 4-users downlink with 3dB lossy phase shifters. Mutual coupling has less impact.}
\label{4users_lossy}
\end{figure}

\begin{figure}
    \centering
    \includegraphics[width=0.9\linewidth]{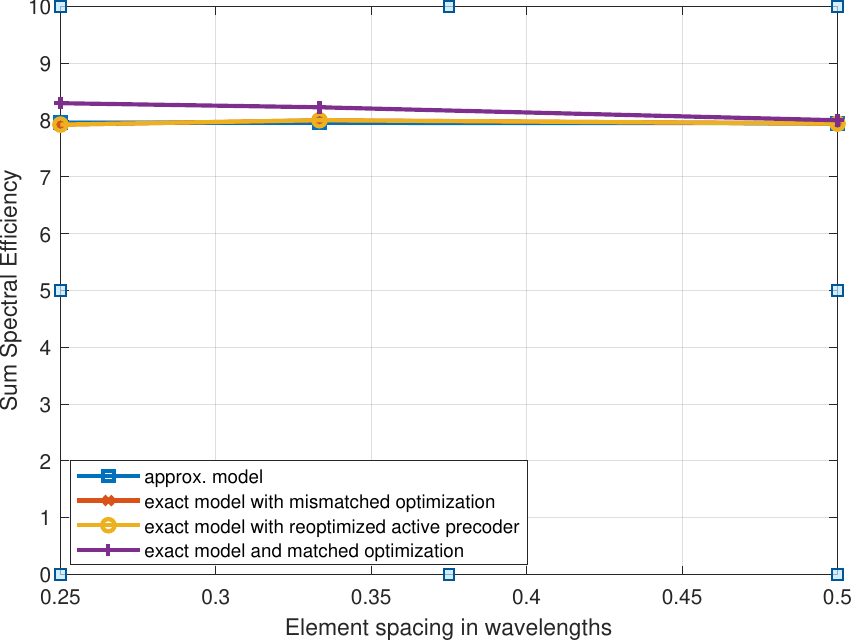}
\caption{Performance of RIS-aided single-user downlink with ideal phase shifters. Mutual coupling has a marginal impact.}
\label{1user_no_losses}
\end{figure}

 In Fig.~\ref{4users_no_losses}, we consider a 4-user scenario and a reflective RIS with perfect (lossless) phase shifters.
 When the mutual coupling is ignored in the optimization ($\B{S}_{\alpha\alpha}={\bf 0}$), we observe a significant degradation from the "idealistic" (i.e., approximate) performance in \cite{Haseeb_2021}  where the approximate model (\ref{naive_approx}) has been assumed.  This is mainly due to multi-user interference caused by the multiple reflections. If the mutual coupling is taken into account only in the active BS precoder, then the original performance is almost fully restored. Surprisingly, if the mutual coupling is taken into account in both BS and RIS nodes (matched optimization), we obtain even better results than the "idealistic" case, which can be attributed to the additional signal power offered by the multiple reflections. This means that mutual coupling is beneficial if properly taken into account. Nevertheless, if phase shifters are lossy, then these performance discrepancies are less pronounced as shown in Fig.~\ref{4users_lossy}, where the phase shifters have 3dB loss, i.e., $\B{S}_L^{\rm H}\B{S}_L \preccurlyeq \frac{1}{2}{\bf I}$. This results from the exponential power decay of the multiple reflections. Likewise, the impact of mutual coupling is marginal in the single-user case even without losses,  as shown in Fig.~\ref{1user_no_losses}.    

\subsection{RIS Decoupling With an Analog DFT Transformation}
As mentioned before, the
coupling matrix $\B{B}$ and equivalently the scattering matrix $\B{S}_{\alpha\alpha}$  have a nearly partial isometric structure with most of the eigenvalues having zero or unit amplitudes \cite{williams2019communication}. Due to their 2D Toeplitz structure, they can be nearly diagonalized by a 2D DFT. In other words, being coupled due to spatial oversampling,  the antenna elements can be decoupled by applying an analog DFT (e.g., a multi-metasurface-based flat lens \cite{Chisum_2020}) and taking only the non-zero eigenmodes, i.e., the visible region of the spatial frequencies with $k_x^2+k_y^2\leq \frac{a^2}{\lambda^2}$. This leads to a reduced beam coupling and less number of uncoupled ports for the same aperture since oversampling is avoided. Therefore, the simple model in (\ref{naive_approx}) becomes nearly exact. We call this RIS architecture with DFT processing a redirective RIS (RedRIS), which can be implemented using a lens \cite{Ho_2019} as discussed in the next section. The lens acts as a beamforming as well as decoupling network for the ports of the reconfigurable lumped components.

\section{Reflective RIS vs. Redirective RIS}

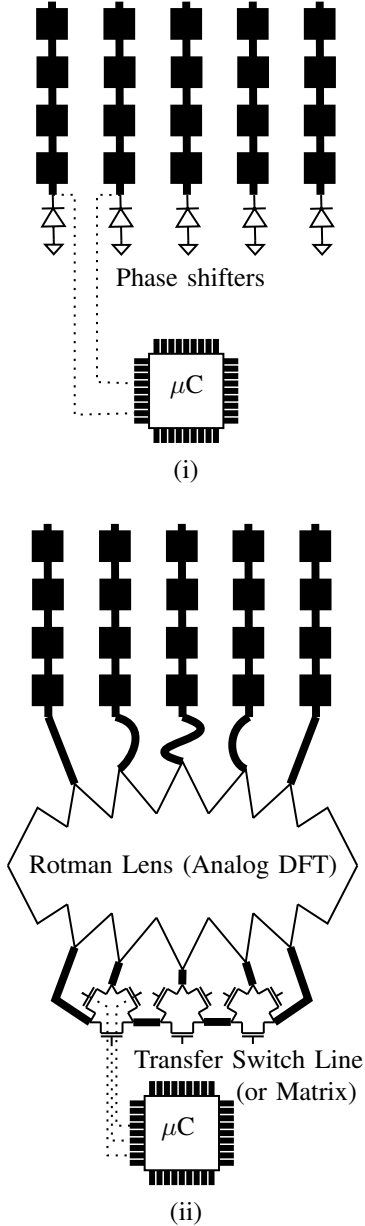
\begin{figure}
    \centering
    \input{fig4}\\
    (i) \\
\vspace{0.5cm}
    \input{fig3}\\
    (ii)
\caption{RIS configurations: (i) Reflective/local phased RIS (ii) Redirective/nonlocal lens-based switched RIS with predesigned beams using an RF post-processing Rotman lens and back-to-back switches. Notice the longer RF interconnections needed for the redirective lens-based RIS architecture.}
\label{lens_based}
\end{figure}

Based on more general concepts from transformation electromagnetics, we consider two types of reconfigurable RF termination scattering matrices for $\B{S}_L$ as shown in Fig.~\ref{lens_based}. These are $i)$ a phased control with a diagonal unitary $\B{S}_L$ and $ii)$ a switched DFT-based angular-selective  termination, i.e.,
\begin{equation}
    \B{S}_L=\B{F} \B{S}_L'\B{F},
\end{equation}
with $\B{F}$ being a DFT matrix and $\B{S}_L'$ a partial symmetric permutation matrix representing the back-to-back connections between the beam ports.

\subsection{RIS Size Requirement and the Bandwidth Dilemma for Phase Control}

Consider the case where a RIS-based node is used to mitigate a blockage situation between the transmitter and the receiver and has distances $d_{\rm Tx}$ and  $d_{\rm Rx}$ to them, respectively. The size of a passive RIS needs to be in the range of the primary Fresnel zone to restore the
line-of-sight-like (quasi-line-of-sight) condition and resolve the blockage. That is, the size must be in the order of $\sqrt{\lambda\frac{d_{\rm Rx}d_{\rm Tx}}{d_{\rm Tx}+d_{\rm Rx}}}$ since the active propagation medium is confined within the first Fresnel ellipsoid. The required size is largest when the RIS is placed halfway between the transmitter and the receiver ($L_{\rm max}=\frac{1}{2}\sqrt{\lambda (d_{\rm Rx}+d_{\rm Tx})}$). This rough estimate is in agreement with recent system-level simulations found in \cite{Sihlbom_2021}.

Unfortunately, the performance of electrically large phased RIS in terms of bandwidth is well-known to be mediocre since temporal processing is limited \cite{Hum_2014}. Therefore,  deployment of configurable thin but large surfaces in wireless communications is challenging and techniques for wideband adaptive operation with true-delay elements tend to be very sophisticated and lead to larger physical thickness \cite{Hum_2014} or require analog baseband processing with switched capacitor circuits \cite{Fikes_2021}.
More precisely, assuming a phased RIS implementation without using true-delay elements,  the symbol duration must be greater than the propagation delay differences across the length of the array \cite{Fathnan:18}, leading to the following limitation on the fractional bandwidth  
\begin{equation}
    \frac{B}{f_c} \leq \frac{\lambda}{L |\sin \theta_{\rm I}-\sin \theta_{\rm R}|}\approx \frac{2}{|\sin \theta_{\rm I}-\sin \theta_{\rm R}|}\sqrt{\frac{\lambda}{d_{\rm Rx}+d_{\rm Tx}}},
\end{equation}
where $\theta_I$ and $\theta_R$  are the incident and reflected angles, respectively.  

If densification of active nodes shall be avoided, this result would predict a very limited fractional bandwidth at higher frequencies which contradicts the purpose of using them. As a remedy, multi-hop RIS-aided transmission can be used to $i)$ reduce the inter-node distance with passive distributed nodes, $ii)$ lower the element count, and $iii)$ improve the bandwidth.  In general, to keep a reasonable fractional bandwidth with this approach, the density of active/passive nodes (base stations or RIS) still needs to increase with $1/\lambda^2$. 

The size-bandwidth trade-off can be relaxed by allowing for thicker RIS structures such as lens-based array structures while maintaining a reasonably dense node deployment. The switched beam structures can be used to trade off bit resolution for bandwidth and robustness. 
Nevertheless, as we will show later in Section~\ref{control_overhead},   higher control (as well as channel estimation) overhead is needed for a larger element count when multi-user mobile broadband access is intended. If we assume the same transmission slot size, then the optimal element count can only scale with the frequency of operation up to some limit. Densification of high-frequency nodes is therefore still needed regardless of the propagation environment and cannot be fully traded by a larger RIS aperture. 

\subsection{Redirective RIS as a Key Enabler for  Active Wave Amplification With Low Noise and Interference Enhancement}
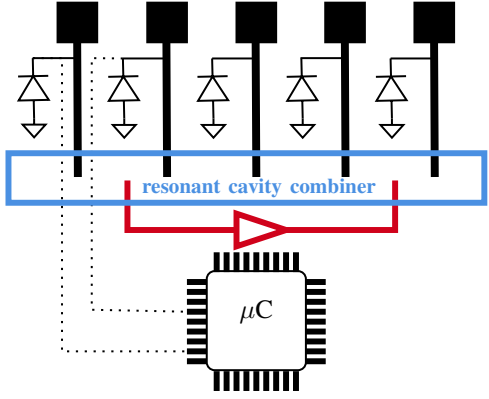
\begin{figure}
    \centering
    \input{fig7}
\caption{Nonlocal Gain-assisted RIS node via coupling.}
\label{RIS_amp}
\end{figure}

Recently, there has been significant interest in integrating
weakly active amplifiers/gain devices (or bi-directional amplifiers \cite{Pang_2020,Tasci_2022}) within the RIS system as shown in Fig.~\ref{RIS_amp}. This provides several advantages such as increasing the overall
gain of the RIS node, compensating for losses (such as conductive and switching losses), reducing the size of the aperture \cite{Hum_2014}, and enabling wave splitting for signal routing as discussed later. In general, amplification might lead to significant degradation of the SNR, but since the SNR is very high at the RIS node (much higher than the receiver), this degradation is generally not critical.

However, stability is a more serious issue due to power mismatch and possible coupling between the beam ports. To avoid potential oscillatory behavior, the following stability constraint in terms of spectral radius must be fulfilled  (c.f. \ref{multiple_reflections})
\begin{equation}\label{stability_condition}
    \rho(\B{S}_L \B{S}_{\alpha\alpha}) <1.
\end{equation}
In other words, all eigenvalues of $\B{S}_L \B{S}_{\alpha\alpha}$  must be smaller than one. Since, the reflection coefficient $[\B{S}_{\alpha\alpha}]_{n,n}$ is typically much larger than the mutual coupling transmission coefficient $[\B{S}_{\alpha\alpha}]_{n,k}$, zero-diagonal active loading matrix $\B{S}_L$ is preferred, which can be implemented using two-port amplifiers \cite{Hum_2014}. This is an important advantage of the redirective RIS architectures with high port-to-port isolation ($>20$dB) as compared to reflective RIS architectures which would require extremely challenging reflection-mode amplifiers \cite{Hum_2014}. In addition, two separate active redirective RIS antennas could be used for the uplink and downlink. In general, to fulfill the stability condition in (\ref{stability_condition}), only gains up to 40dB are feasible and the re-radiated power is in the range of tens of milliwatts or less.  Such a propagation environment with a high density of weakly active nodes can be regarded as a lumped gain media. It is worth mentioning that the uplink coverage is much more problematic at higher frequencies than the downlink (due to the TDMA/TDD access technique with limited time allocated to the user’s transmission). Therefore, electronic amplification
is more required for the uplink and less for the downlink, thereby leading to nonreciprocal design and allowing better ways for isolating both
communications directions from each other.    

Based on the linear RIS scattering model in (\ref{exact_scattering_model}), the radiated thermal noise radiation density, $N(\theta,\varphi)$, in the far-field due to amplification can be evaluated as (with $N_0$ being the noise power density and  $N_F^{\infty}$ being the asymptotic noise figure under infinite amplification):
\begin{equation}
\begin{aligned}
 &N(\theta,\varphi)=N_F^{\infty}N_0\cdot \max\Big(\B{s}_{\alpha\beta}(\theta,\varphi)^{\mathsf{T}}(\B{S}_L^{-1}-\B{S}_{\alpha\alpha})^{-1}  \\
 &~~~~~~~~~~~~~~~~ \B{B}(\B{S}_L^{-1}-\B{S}_{\alpha\alpha})^{-\mathsf{H}} \B{s}_{\alpha\beta}(\theta,\varphi)^*-\frac{A}{\lambda^2}\cos \theta,0\Big),
\end{aligned}
\end{equation}
where we subtracted the re-radiated non-amplified thermal noise from the RIS area $A$. Evidently, for passive RIS, we have $N(\theta,\varphi)=0$. 

Likewise, if we consider an environment with isotropic interference having the angular power density $N_I$, then by neglecting the residual scattering  the re-radiated  amplified interference is obtained as: 

\begin{equation}
\begin{aligned}
 &\!\!\!\!\!I(\theta,\varphi)\,=\,N_I \cdot  \B{s}_{\alpha\beta}(\theta,\varphi)^{\mathsf{T}}(\B{S}_L^{-1}-\B{S}_{\alpha\alpha})^{-1}\times  \\
 &~~~~~~~~~~~~~~~~~~~~~~~~~~~~\B{B}(\B{S}_L^{-1}-\B{S}_{\alpha\alpha})^{-\mathsf{H}} \B{s}_{\alpha\beta}(\theta,\varphi)^*.
\end{aligned}
\end{equation}
With a full rank matrix $\B{S}_L$ as in the case of reflective surfaces, the interference is enhanced linearly with the RIS area and can travel further as argued in  \cite{Torres2022}. 
Unlike reflective RIS which reflects all impinging waves including undesired signals,
RedRIS 
redirects the target wave only (spatial channelizer), thereby reducing the impact on other radios.

\subsection{Redirective RIS as an Enabler for Low Control Overhead}
\label{control_overhead}
In RedRIS architectures, the termination scattering matrix $\B{S}_L$
is low rank and only a small number of beam connections are selected from a predefined codebook. This makes the control overhead scales only logarithmically with the RedRIS size. Furthermore, the overhead can be substantially reduced if the beam configurations of the BS-RIS channel are static or are updated at a slow refreshment rate while still having the UEs-RIS channels switched at each time slot.
This semi-static configuration can enable a much higher gain for the fronthaul channel than the access channel (see discussion later). With fixed fronthaul configuration, the control overhead is then dominated by the access link setting information in each transmission slot.  As the carrier frequency and bandwidth increase, the slot/dwell time duration 
becomes shorter while the  number of slot-wise transmitted symbols $N_s$
does not change significantly.  
We note that due to analog beamforming TDMA is the widely
adopted scheme for mmWave multiple access. Thus, super-fast access beam switching/hopping is still required even when users' channels are not dynamic.

Aside from the control overhead savings, another important benefit of the semi-static configuration is the lower channel estimation overhead and the quasi-unrestricted performance improvement stemming from the increased
 directionality 
in the BS-RIS link. In addition, the semi-static configuration with wide beams for the access link and narrow beams for the base station link can provide coverage over
wider areas with enhanced resilience to mobility. In addition, a very high beamforming gain might not be meaningful at the access side of active RIS nodes due to the stricter EIRP limitations usually imposed by regulations (as compared to fronthaul links).

Consider now an in-band control channel for the access-side configuration  with maximum beamforming gain $M_A$ utilizing the following fraction $B_c$ of the total bandwidth $B_w$
\begin{equation}
    \frac{B_c}{B_w}\,=\,\frac{b_A \log_2 M_A }{\eta_B N_s}.
\end{equation}
Here $b_A \log_2 M_A$ is the number of control bits per slot and RedRIS node while $\eta_B$ is the spectral efficiency of the control channel. The logarithmic scaling of the control overhead with the access antenna gain $M_A$ is due to the sparsity of the desired directional scattering. 
A directional control channel can be  used to control multiple $K$ RedRIS nodes based on space division multiplexing (SDMA with pencil beams)  \cite{Gao_2015} leading to the following sum-rate

\begin{equation}
  R^{\rm redirective}=  K\left(1-\frac{b_A \log_2 M_A }{\eta_B N_s} \right) \log_2 \! \left(1\!+\!\frac{G_c P_T M_B M_A }{B_wN_0}\!\!\right)\!\!,
\end{equation}
where $G_c$ is the isotropic channel gain, $P_T$ is the transmit power,  and $M_B$ is the RedRIS beamforming gain for the fronthaul link.
Using the high SNR approximation to optimize the spectral efficiency with respect to the access antenna gain $M_A$, we get the following approximation for the optimal gain of the access link:

\begin{equation}
    M_{A,\rm opt}^{\rm redirective} \approx \frac{2^{\frac{\eta_B N_s}{2b_A}}}{\sqrt{\frac{G_c P_T M_B }{B_wN_0}}},
\end{equation}
revealing an exponential scaling in terms of slot size $N_s$. 

Consider now the case of reflective RIS configuration with the same number of antennas for fronthaul and access, i.e., $M_A=M_B$. Assuming that the control overhead scales linearly with the number of antennas \cite{Zappone_2020}\footnote{Codebook-based control with logarithmic overhead scaling is generally not trivial in the case that the reflective RIS is shared by multiple users unless the RIS is subdivided into subarrays with substantial gain losses.}, it follows that the sum rate is given by
\begin{equation}
  R^{\rm reflective}=  K\left(1-\frac{b_A  M_A }{\eta_B N_s} \right) \log_2 \left(1+\frac{G_c P_T M_A^2 }{B_wN_0}\right),
\end{equation}
with the following asymptotic optimum access beamforming gain at high SNR  
\begin{equation}
    M_{A,\rm opt}^{\rm reflective} \approx \frac{\frac{\eta_B N_s}{b_A}}{W\left(\frac{\eta_B N_s}{b_A}\sqrt{\frac{G_c P_T}{B_wN_0}  }\right)} ,
\end{equation}
in which $W(\cdot)$ is the Lambert function. 
Clearly, the reflective RIS 
yields a poor sublinear scaling in terms of slot length. 

\subsection{Further control overhead reduction using pre-scheduled intelligent  beam-hopping techniques with statistical channel state information}

In practice, real-time reconfigurability in the range of milliseconds might be still difficult to achieve as it requires stringent timing requirements for the control channel. Alternatively, beam-hopping techniques that are popular in satellite communications \cite{Rohde_2019} can be considered. Beam-hopping consists of serving sequentially users spots in turn according to a predetermined schedule. The periodic beam hopping time plan can be determined and updated based on the varying traffic demand and the RIS scattering pattern can be optimized based on long-term statistical channel information \cite{Cao_2022} which also reduces the training overhead (c.f. Section~\ref{estimation}). Therefore, the reconfiguration needs to be done only occasionally with long cycle times and the requirements on the control channel are significantly relaxed. To allow for initial access, all potential beam directions are sequentially illuminated and scanned (beam sweeping) during multiple synchronization signal blocks (SSB). This results in substantial initial access latency and a long beam-hopping period. Therefore, the RIS node is designed to support a medium number of wide initial access wide beams or, alternatively, a permanent directive link is dedicated between the access point and the RIS node. While the control overhead is reduced, synchronous operation (for instance via GPS) between the RIS nodes and the donor nodes is still required. A notable advantage of the redirective RIS system is the simultaneous beam hopping of multiple beams at full aperture gain, particularly when the RIS node is shared among several donor sites (e.g. Fig~\ref{RIS_rural}) as explained in the next subsection.

\subsection{Redirective RIS as a Key Enabler for Multifunctional Wave Manipulation}

\begin{figure}
    \centering
    \input{fig6}
\caption{RedRIS-aided X-channel at the cell edge
enabling interference mitigation, traffic balancing 
and improved coverage.}
\label{RIS_multilink}
\end{figure}
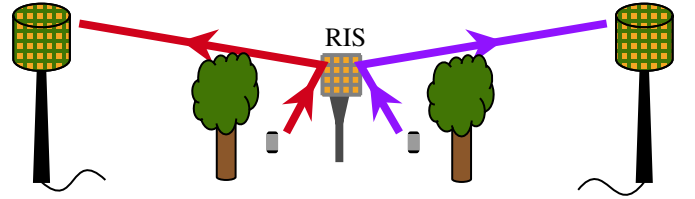

In this section, we highlight the multi-beam data forwarding advantage of 
RedRIS. More precisely, we show that 
RedRIS-aided systems can concurrently provide multiple beam routes with a shared full aperture gain thereby enabling a 
wirelessly routed network (c.f. Fig.~\ref{RIS_multilink}). 
In fact, a major advantage of redirective RIS deployment is the engineering and reshaping of the ambient propagation 
environment by a minimalistic approach (i.e., via low-rank scattering matrices) without affecting the entire angular domain as in the reflective RIS case. The rank of the scattering matrix is therefore not just a key indicator of how much the environment is affected by the presence of a RedRIS node in terms of noise and interference but also a key indicator for scalability in terms of parallel processing of multiple waves.

Consider for instance a lens-based RedRIS serving multiple (say $K$) users, each of which is served by a different base station, possibly belonging to different operators.
Assume that each user has a certain desired surface configuration  $\B{S}_{L,k}=\B{F}\B{S}_k'\B{F}$. The following least square error optimization can be considered to enable these $K$ routes simultaneously:   
\begin{equation}
\begin{aligned}
   & \min\limits_{\B{S}_L'} \quad \sum_{k=1}^{K} \|\B{F}\B{S}_L'\B{F}-\B{F}\B{S}_k'\B{F}\|_{\rm F}^2  \\
    &\quad \textrm{s.t. } \B{S}_k' \textrm{ is a partial symmetric perumtation.} 
\end{aligned}    
\end{equation}
Due to the sparseness sparsity and directionality of the channel, we can assume $\B{S}_k'$ to be a partial permutation over non-overlapping sets, thereby leading to the following solution:
\begin{equation}
\begin{aligned}
\B{S}_L'&= \sum_{k=1}^{K}\B{S}_k'.
\end{aligned}    
\end{equation}
leading to zero residual error. When the switch is in the off-state, the unused beam ports are terminated by a matched load (i.e., absorptive switch), generating near-zero scattering in all uninvolved directions.

The optimization of the matrices $\B{S}_k'$ is however combinatorial.
We provide instead a greedy approach to perform this task for the case that the order of the switching matrix $\B{S}_L'$ is equal to the number of users. Consider for instance a $K$-users interference channel \cite{jiang2022interference} where single-antenna transmitters and receivers are linked via the RIS channels $\B{H}_{\rm RIS-TX}$ and $\B{H}_{\rm RX-RIS}$. The transmit and noise power of each transmitter/receiver pair is $P$ and $\sigma_n^2$.  
With the initialization $\B{S}_L'^{(0)}=\B{0}$, the switching matrix is optimized successively by minimizing the signal-to-interference and noise ratio in a given order, as follows
\begin{equation}
\begin{aligned}
&\B{S}_k'=\argmax\limits_{\B{S};~ \B{S}\B{S}_L'^{(k-1)}=\B{0}} \frac{|[\B{H}^{(k)}]_{k,k}|^2}{\frac{\sigma_n^2}{P}+\sum_{\ell\neq k}|[\B{H}^{(k)}]_{k,\ell}|^2+|[\B{H}^{(k)}]_{\ell,k}|^2}\\
&\B{S}_L'^{(k)}=\B{S}_L'^{(k-1)}+\B{S}_k',
\end{aligned}
\end{equation}
where $\B{H}^{(k)}=\B{H}_{\rm RX-RIS}\B{F}(\B{S}_L'^{(k-1)}+\B{S} )\B{F}\B{H}_{\rm RIS-TX}$, and the matrix $\B{S}$ to be optimized is a symmetric matrix with at most two nonzero elements equal to one. Therefore each maximization requires in the order of $M^2/2$ evaluations. The final matrix $\B{S}_L'$ is obtained after $K$ successive maximizations. 

For the case of reflective surfaces without the analog DFT, however, the optimization becomes  
\begin{equation}
\begin{aligned}
   & \min\limits_{\B{S}_L} \quad \sum_{k=1}^{K} \|\B{S}_L-\B{S}_k\|_{\rm F}^2  \\
    &\quad \textrm{s.t. } \B{S}_L \textrm{ is a diagonal unitary matrix.} 
\end{aligned}    
\end{equation}
The solution to this optimization is given by 
\begin{equation}
\begin{aligned}
\B{S}_L&=  \left(\sum_{k=1}^{K}\B{S}_k\sum_{k=1}^{K}\B{S}_k^{\mathsf{H}}\right)^{\!\!-\frac{1}{2}}\sum_{k=1}^{K}\B{S}_k  
\approx \frac{1}{\sqrt{K}}\sum_{k=1}^{K}\B{S}_k.
\end{aligned}    
\end{equation}
A more optimal approach is to use projected gradient descent to minimize the sum mean square error (cf. (\ref{MSE})) with respect to $\B{S}_L$ directly. 
The substantial gain loss of factor $K$ also means that most of the impinging power is scattered in unwanted radiations. Reflective surfaces are thus not suitable for wirelessly routing multiple plane waves simultaneously. In addition, interference nulling requires that $M$ grows quadratically with the number of users $K$ \cite{jiang2022interference}. The multi-user advantage of the nonlocal RedRIS as compared to the local RIS configuration is highlighted in Fig.~\ref{rate_vs_users} for the case $P/\sigma_n^2=20$ and a square surface of $M=256$ half-wavelength-spaced elements. For this plot, we have line-of-sight channels with unit pathloss between users and RIS antennas together with the cosine-shaped pattern. In addition, users are uniformly distributed on a spherical cap centered by the RIS with apex angle of $120^{\circ}$.   

\subsection{Redirective RIS: Scalability as a Major Issue for Switched Lens-Based RIS}
After highlighting several advantages of the directive RIS architecture, we shall discuss its disadvantages as compared to the reflective RIS configuration. In addition, to the need for a (metasurface) lens for analog DFT processing, the major issue is the need for longer RF interconnections (see Fig.~\ref{lens_based}) and a multistage-switching network for conductive RF routing which is in general quite challenging at high frequencies. Switching matrices are used in several applications such as satellite communications \cite{Nardo_2013}. As the frequency and the number of ports increase, however, the losses of signal traces and switches become overwhelming, and designing a printed circuit board (PCB) layout with global interconnections and with minimal signal integrity issues is no easy task.

\begin{figure}
    \centering
    \input{fig9}
\includegraphics[width=0.9\linewidth]{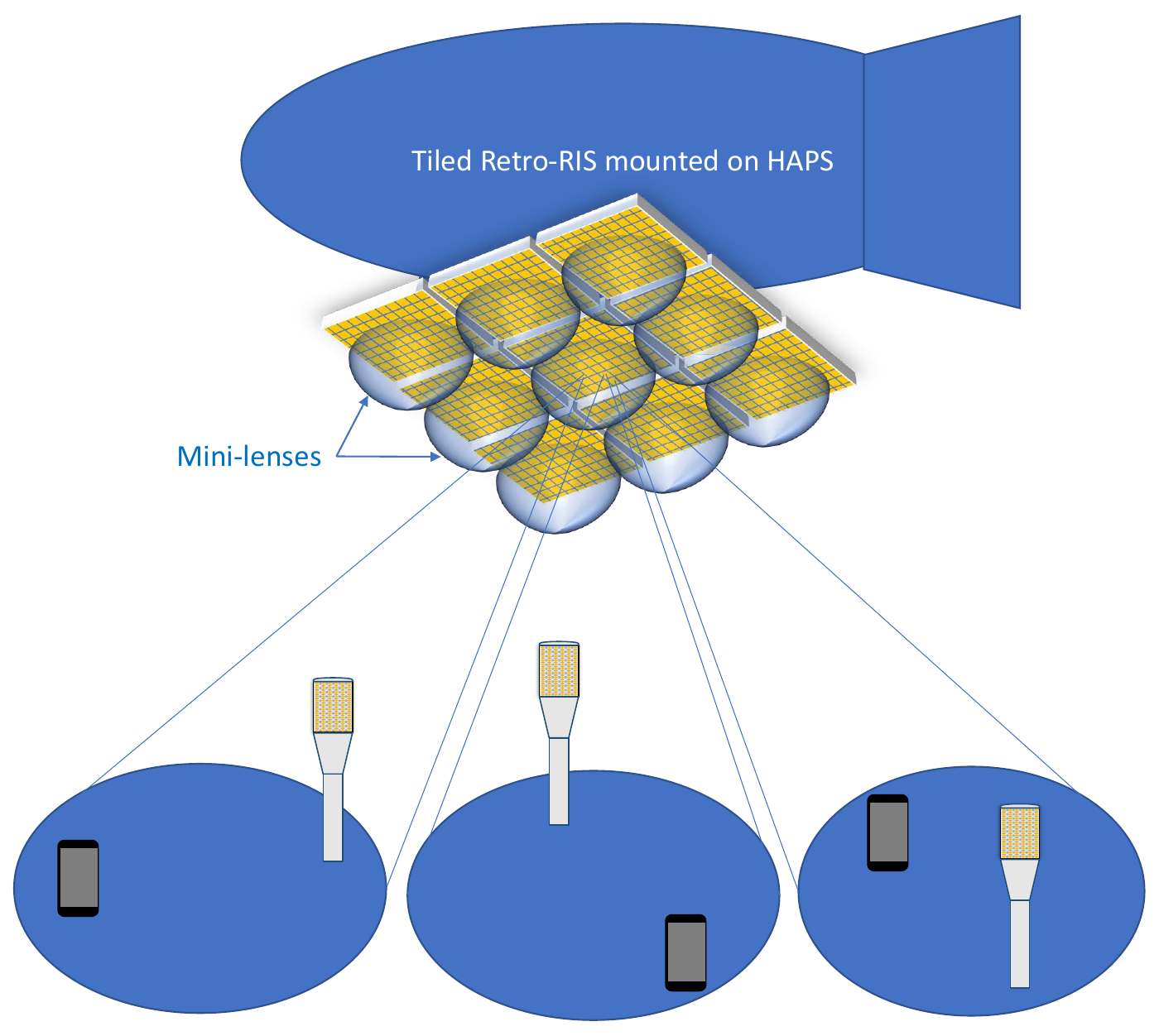}
\caption{Scalable and modular  multi-lens-based RedRIS array $\B{S}_L=(\B{F} \otimes {\bf I})\B{\Phi} (\B{F} \otimes {\bf I})$. 
}
\label{RIS_multilens}
\end{figure}

\begin{figure}
    \centering
    \includegraphics[width=0.9\linewidth]{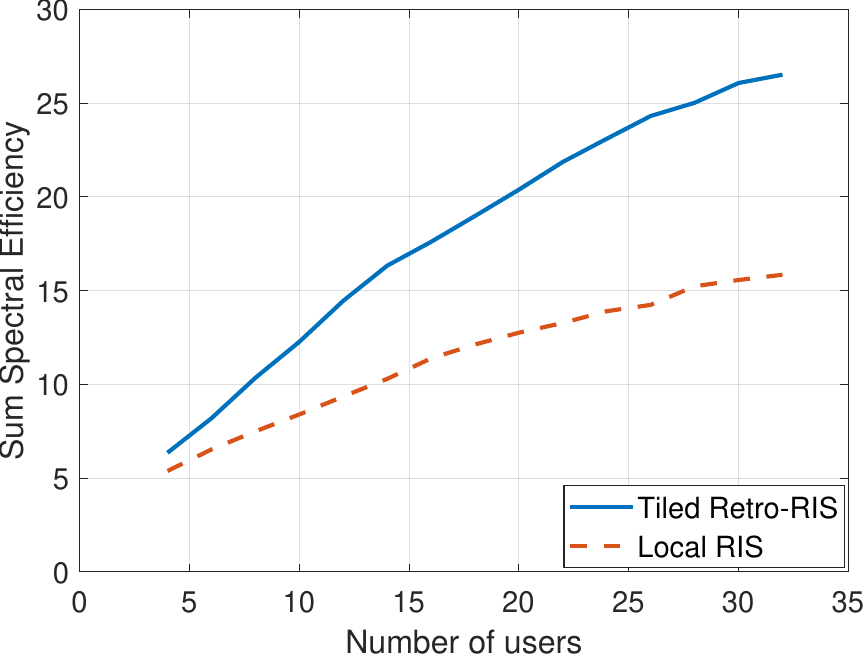}
\caption{Performance of local and nonlocal tiled Retro-RIS in a cell-free scenario with 36 access points. The tiled retro-RIS total size is $M=24\times 24$ elements partitioned into $3\times3$ tiles. $P/\sigma_n^2=5$.}
\label{rate_vs_users5}
\end{figure}

These problems may be manageable for RedRIS configuration with 2D beamforming, particularly when using distributed amplification to offset switching insertion losses. However, 3D redirective RIS implementation is likely to be highly challenging, and eventually, a hybrid approach combining both phased and lens-based RIS types would be meaningful \cite{Fikes_2021}, trading the wavefront selectivity for scalability. Such a salable hybrid implementation is shown in Fig.~\ref{RIS_multilens}. The total tile-based periodic aperture consists of partitioned small RedRIS apertures. The distributed and fractionated implementation can reduce the cost, size, and complexity significantly. The wave processing is quasi-local with short-range interactions since there are no interconnections between the RedRIS tiles. The beam port-to-port connections in each small RedRIS can also include a phase or delay shifter that can be optimized to achieve the desired phase/delay coherence across the array. In Fig.~\ref{rate_vs_users5}, we simulate a cell-free scenario with 36 access points a tiled retroreflective RIS, with $\B{S}_L=(\B{F} \otimes {\bf I}_3)\B{\Phi} (\B{F} \otimes {\bf I}_3)$,   at a high altitude is used as a relay to the users. The individual ports are connected to phase shifters (i.e., no port switching). We observe, as we increase that number of users, that performance is better than the one achievable with a conventional RIS with the same aperture and the same number of phase shifters. This can be explained by the angular sectorization capability of the retroreflective RIS that reduces the interference between the users.  

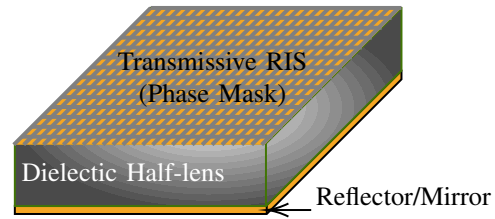
\begin{figure}
    \centering
    \input{fig10}
\caption{Strongly non-local (layered/two-plane) RIS with $\B{S}_L=\B{\Phi}\B{F}\B{\Phi} $. The strong non-local behavior is enabled through wave propagation inside the cavity inducing notable interactions between distant RIS elements.
}
\label{RIS_backed}
\end{figure}

\begin{figure}
    \centering
    \includegraphics[width=0.9\linewidth]{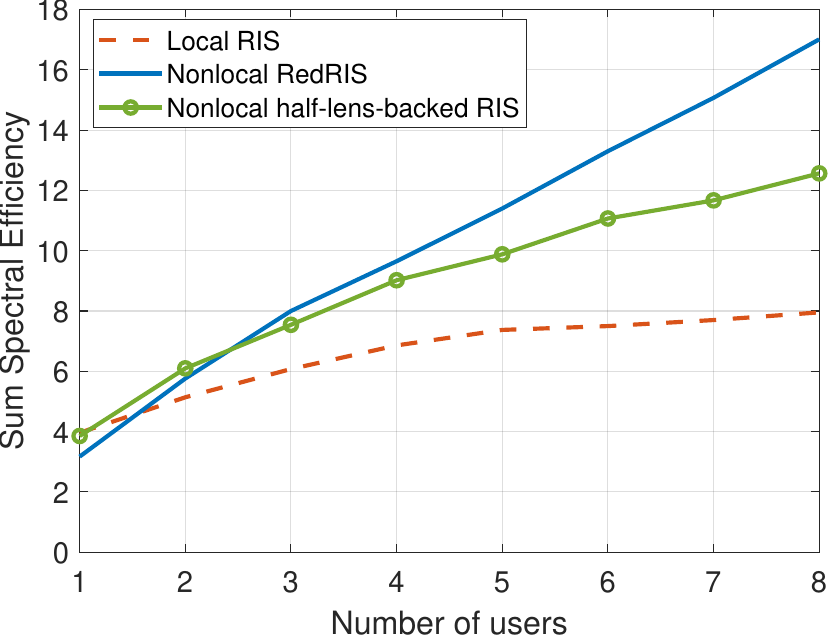}
\caption{Performance of local and nonlocal (redirective and backed) RIS in $K$-user interference channels with LOS channels. For the RedRIS, the number of switched connections is equal to the number of users. $P/\sigma_n^2=20$.}
\label{rate_vs_users}
\end{figure}

\begin{figure}
    \centering
    \includegraphics[width=0.9\linewidth]{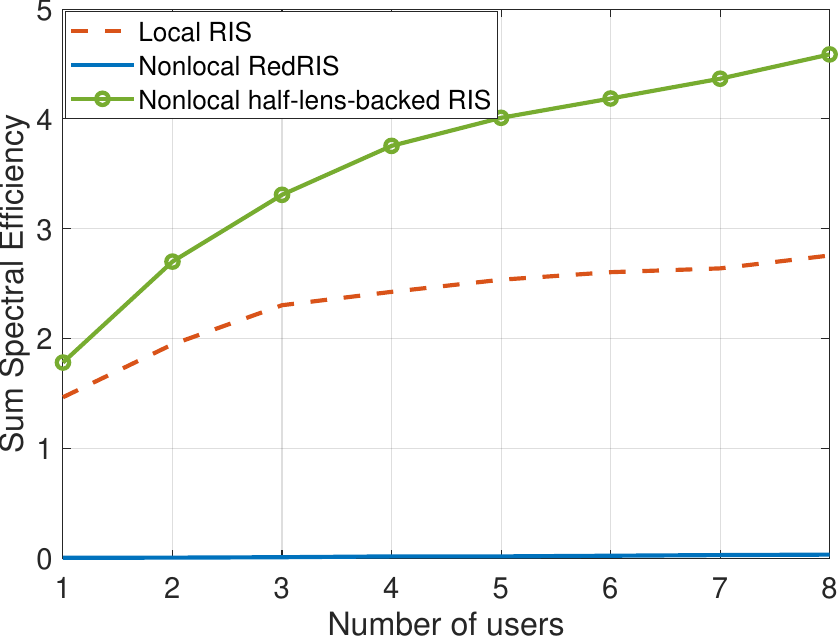}
\caption{Performance of local and nonlocal (redirective and backed) RIS in $K$-user interference channels with IID channels. Due to the lack of directional propagation, RedRIS performs poorly while nonlocal backed RIS still performs properly. $P/\sigma_n^2=5$.}
\label{rate_vs_users4_1}
\end{figure}

Another possible salable design is to use a two-pass cavity RIS which is composed of a transmissive RIS, i.e. phase-shift mask $\B{\Phi}$,  and a mirror as shown in Fig.~\ref{RIS_backed}. This results in a second-order wave conversion with $\B{S}_L=\B{\Phi}\B{F}\B{\Phi}$ where $\B{F}$ represents the two-way wave propagation inside the cavity (e.g. an optical Fourier transform with a parabolic mirror). Despite having the same size and the same number of phase-shifters as a conventional RIS, this type of wave conversion provides better performance for structured LOS channels (c.f. Fig.\ref{rate_vs_users}) as well as for random IID Rayleigh fading channels (c.f. Fig.\ref{rate_vs_users4_1}). This can be explained by the fact that this dense conversion matrix can approximate the optimal unitary transformation better than a diagonal matrix. 

\section{Redirective RIS: Other Enabled Functionaries}
In this section, we discuss some further advanced functionalities provided by the use of RedRIS in wireless networks. While some of these functionalities might be 
possible with a reflective RIS configuration (usually at the cost of spurious reflections\cite{Asadchy_2016}), their implementation becomes quite natural and with no side effects in the context of RedRIS due to the resulting directional propagation.

\subsection{Retrodirective RIS and Explicit Channel Estimation}
\label{estimation}

The channel estimation task in RIS-based systems is challenging due to the lack of digital processing capability at the RIS node. Several algorithms and techniques have been developed in the literature to obtain the channel state information (CSI) of the two-hop channel where pilot symbols are transmitted by the users and are received by the central node via the reflective RIS \cite{Chen_2019,Zheng_2022,Swindlehurst_2022}. The channel estimation task is performed at the central node to determine either the cascaded (two-hop) channel or even the separate channels between the reflective RIS on the one hand and the transmitter and receiver, on the other hand, using factorization techniques. These channel estimation methods usually suffer from extremely low SNR (prior to beamforming) and challenging synchronization due to the multiplicative path loss and since the high beamforming gain of the reflective RIS node is not effective during this channel estimation phase. 

In the following, we highlight the fact that this issue can be mitigated by using a passive \textit{retrodirective} RIS functionality and operating the base station or the UEs as a monostatic radar. Importantly, the two channels involving the retrodirective RIS on both sides of the communication link can be estimated separately, thereby taking into account their different dynamics and required CSI quality that is dictated by the required beamforming gain.         

Consider for simplicity the case of a single antenna at the base station. The channel between the base station and the RIS node is denoted by $\B{h}_{\rm BS-RIS}$ having the sparse representation $\B{h}_{\rm BS-RIS}= \sqrt{M} \B{Fs}_\textrm{BS}$ in the angular domain (recall here that $\B{F}$ is the normalized DFT matrix). The base station transmits a pilot signal and the incident wave impinges on the lens-based RIS where all the beam ports are modulated using a sequence of short or open circuits $\B{S}_L^{(t)}=\B{F}\B{S}_{L}'^{(t)}\B{F}$ with a diagonal matrix $\B{S}_{L}'^{(t)}$ at each time index $t$, having $\pm 1$ diagonal entries.
The modulated signals at the beam ports are therefore retro-reflected toward the base station taking advantage of the full gain $M$ of the RIS area. At time index $t$, the received retro-reflected signal at the base station can be written as 
\begin{equation}
\begin{aligned}
y_{\rm retro}^{(t)}&= \B{h}_{\rm BS-RIS}^{\mathsf{T}}  \B{S}_L^{(t)}\B{h}_{\rm BS-RIS}+n^{(t)} \\
&=M \B{s}_{\rm BS}^{\mathsf{T}} \B{FF} \B{S}_L'^{(t)} \B{FF}  \B{s}_{\rm BS}+n^{(t)} \\
&=M \B{s}_{\rm BS}^{\mathsf{T}} \B{S}_L'^{(t)}   \B{s}_{\rm BS}+n^{(t)}, ~(\B{FF}\textrm{ is anti-diagonal identity}) \\
&= M ({\bf diag}(\B{S}_L'^{(1)}))^{\mathsf{T}} \left( \B{s}_{\rm BS} \odot \B{s}_{\rm BS}\right)+n^{(t)},
\end{aligned}    
\nonumber
\end{equation}
where ${\bf diag}(\B{S}_L'^{(t)})$ is a column vector containing the diagonal elements of $\B{S}_L'^{(t)}$. 
Stacking $T$ samples in a vector, we get  
\begin{equation}
\begin{aligned}
\B{y}_{\rm retro}\,=\,M \begin{bmatrix} 
({\bf diag}(\B{S}_L'^{(1)}))^{\mathsf{T}} \\
({\bf diag}(\B{S}_L'^{(2)}))^{\mathsf{T}} \\
\vdots \\
({\bf diag}(\B{S}_L'^{(T)}))^{\mathsf{T}} 
  \end{bmatrix}   \left( \B{s}_{\rm BS} \odot \B{s}_{\rm BS}\right) +\B{n},
\end{aligned}   \label{retroreflective_channel} 
\end{equation}
where the gain of the RIS node is operational thereby significantly benefiting the channel estimation task.  
This radar-inspired technique can be applied on the UE side and can also be used for localization purposes during the training phase without requiring communication with the central base station.

Consider instead conventional direct estimation of the two-hop channel:
\begin{equation}
\B{h}_{\rm cascaded}\,=\,\B{h}_{\rm BS-RIS}\odot\B{h}_{\rm RIS-UE}\,=\,\sqrt{M} \B{F} \B{s}_{\rm cascaded},
\end{equation}
at a non-colocated receiver (i.e., the UE side) under phased RIS control without the analog DFT.  The corresponding received signal is given by 

\begin{equation}
\begin{aligned}
\B{y}_{\rm cascaded}\,=\,\sqrt{M} \begin{bmatrix} 
({\bf diag}(\B{S}_L^{(1)}))^{\mathsf{T}} \\
({\bf diag}(\B{S}_L^{(2)}))^{\mathsf{T}} \\
\vdots \\
({\bf diag}(\B{S}_L^{(T)}))^{\mathsf{T}} 
  \end{bmatrix}  \B{F} \B{s}_{\rm cascaded}  +\B{n}. 
\end{aligned}  \label{cascaded_channel}  
\end{equation}
As opposed to  (\ref{retroreflective_channel}),  the multiplicative gain/factor of the traditional phased RIS is only $\sqrt{M}$,  thereby revealing the clear advantage of retroreflective explicit channel estimation.  

The proposed scattering-based channel estimation method is particularly suitable for the discovery of new RIS nodes by the central node upon installation. The routine can be invoked periodically or automatically at power-on and simplifies the RIS installation by the technicians since no accurate manual steering of the RedRIS towards the BS is required. 
\subsection{Redirective RIS for Integrated Fronthaul and Access:  Edge network Inspired by passive optical networks}
The use of distributed antennas is a common approach to address the coverage issue. The fronthaul connection that is needed between the central node and the remote radio heads is highly challenging due to its high bandwidth and stringent latency requirements. It is generally implemented by an optical network. RIS-based wireless networks can be regarded as a more cost-effective alternative for implementing a distributed antenna system with integrated access and fronthaul. This can be enabled by the following distributed network components. 

\subsubsection{Wave/beam splitting and combining} Power splitter can be 
used to allow the same link from the central node to be passively shared between users simultaneously (for instance using OFDMA instead of TDMA) or between the access link and the fronthaul link without electronics and power at the cost of 3dB attenuation per binary splitting/combining, similar to passive optical networks (or with zero loss using active splitters). In a multi-hop setting, this broadcast architecture will allow the fronthaul configuration to be kept static and the access configuration alone at the endpoints to be updated at each time slot thereby reducing the control overhead.

\subsubsection{Beam widening and narrowing} for the access link using two separate small and large lens-type antenna arrays. Due to the challenging tasks of beam training/tracking and node handover or beam recovery triggered by blockages/abrupt changes under mobility with too narrow beams. Therefore the control overhead, as well as channel estimation overhead, only scales with the antenna gain of the access link.   

\subsubsection{Semi-static configuration}  Redirective RIS-based networks can be operated under two distinct configuration modes running at two different time scales. Indeed, a quasi-static configuration is used for the fronthaul link and is adapted periodically at a large time scale while a dynamic and fast configuration is adopted for the access links. This reduces the signaling overhead appreciably as discussed in Section~\ref{control_overhead}.


\subsection{Redirective RIS for Enhanced Physical Layer Security}

Due to the directional retransmission and confined wave propagation (via ``wireless waveguides''), the directive RIS network configuration is expected to have a higher security level \cite{Elayan_2020},  comparable to hard-wired systems (physical security or layer-0 security). Eavesdropping becomes harder as signals can only be received along well-confined wave trajectories. This is different from conventional physical layer security methods, where channel state information of the attacker as well as a larger coding block length, i.e. higher latency, is needed. In addition, any attempt to intercept the transmission, by placing for instance an object in the first Fresnel zone, will lead to a channel distortion/disruption or a detectable change of the spatial channel experienced by the receiver. Such propagation anomaly can be detected during the channel estimation procedure. 
This concept for secure communication can be regarded as an intrusion-alarmed quasi-passive wireless network where eavesdropping attacks can be detected. In case of an anomaly, the central node might decide to redirect the path over a different RIS route. 
Another approach is to intentionally use broken (zig-zag) multi-hop trajectories to mislead the attacker or avoid risk areas. 

\section{Conclusion}

We introduced the concept of nonlocal or redirective reconfigurable surfaces with low-rank scattering as an artificial wave-guiding structure for wireless wave propagation at high frequencies. We showed multiple functionalities that can be implemented, including beam bending, multi-beam data forwarding, wave amplification, routing, splitting, and combining. The observed results indicate that transformation-based intelligent surfaces can make mmWave and THz networks more salable, secure, flexible, and robust while being energy, cost, and spectrally efficient.  Mitigating the coverage issue of these frequency bands can be considered a critical milestone in the evolution of terrestrial wireless mobile access. Other than the improved coverage, RIS-based remote nodes can also improve the network capacity due to the extremely high directional propagation and the possibility for massive spatial multiplexing with massive MIMO at the central macro baseband node. This enables tens or even hundreds of bits per hertz and kilometer square area spectral efficiency for mmWaves at low cost and high coverage. While lens-based RIS offers much better performance in terms of signal processing efficiency, its bulkiness (particularly in the case of 3D beamforming) and scalability issues (due to the longer RF interconnections and switching implementation) might be disadvantageous 
as compared to planar and local phased RIS.
Fractionated and distributed RedRIS or mirror-backed RIS structures provide potentially the right remedy. 
In general, there are several challenges for scaling reconfigurable surfaces on many fronts: $i$) element count and node density required to break the frequency barrier $ii$) bandwidth limitations $iii$) channel estimation $iv$) control overhead and $v$) RF switching/tuning network complexity. This requires further extensive research and innovations for designing quasi-passive directional mobile networks and low-cost high-density communication nodes dedicated to mmWave/THz frequencies which will represent the mobile counterpart of passive optical networks (PON).

\bibliographystyle{IEEEtran}
\bibliography{IEEEabrv,references.bib}
 
\end{document}

%% file: fig5.tex
\usetikzlibrary{patterns}

 
\tikzset{
pattern size/.store in=\mcSize, 
pattern size = 5pt,
pattern thickness/.store in=\mcThickness, 
pattern thickness = 0.3pt,
pattern radius/.store in=\mcRadius, 
pattern radius = 1pt}\makeatletter
\pgfutil@ifundefined{pgf@pattern@name@_33oxsip9h}{
\pgfdeclarepatternformonly[\mcThickness,\mcSize]{_33oxsip9h}
{\pgfqpoint{-\mcThickness}{-\mcThickness}}
{\pgfpoint{\mcSize}{\mcSize}}
{\pgfpoint{\mcSize}{\mcSize}}
{\pgfsetcolor{\tikz@pattern@color}
\pgfsetlinewidth{\mcThickness}
\pgfpathmoveto{\pgfpointorigin}
\pgfpathlineto{\pgfpoint{\mcSize}{0}}
\pgfpathmoveto{\pgfpointorigin}
\pgfpathlineto{\pgfpoint{0}{\mcSize}}
\pgfusepath{stroke}}}
\makeatother
\tikzset{every picture/.style={line width=0.75pt}} 

\begin{tikzpicture}[x=0.75pt,y=0.75pt,yscale=-1,xscale=1]

\draw  [fill={rgb, 255:red, 0; green, 0; blue, 0 }  ,fill opacity=1 ] (304.3,55) -- (305.9,15) -- (308.03,15) -- (309.63,55) -- cycle ;
\draw  [fill={rgb, 255:red, 74; green, 144; blue, 226 }  ,fill opacity=1 ] (460,95.83) -- (441.17,77) -- (385,77) -- (385,178.17) -- (403.83,197) -- (460,197) -- cycle ; \draw   (385,77) -- (403.83,95.83) -- (460,95.83) ; \draw   (403.83,95.83) -- (403.83,197) ;
\draw  [draw opacity=0][fill={rgb, 255:red, 155; green, 155; blue, 155 }  ,fill opacity=1 ] (245,202) -- (468.75,202) -- (493.75,222) -- (270,222) -- cycle ;
\draw  [draw opacity=0][fill={rgb, 255:red, 155; green, 155; blue, 155 }  ,fill opacity=1 ] (457.96,258.14) -- (418.8,258.86) -- (312,152) -- (351.15,151.28) -- cycle ;
\draw  [draw opacity=0][fill={rgb, 255:red, 255; green, 255; blue, 255 }  ,fill opacity=1 ] (361.28,204.08) -- (335.33,204.08) -- (337.1,205.69) -- (363.04,205.69) -- cycle ;
\draw  [draw opacity=0][fill={rgb, 255:red, 255; green, 255; blue, 255 }  ,fill opacity=1 ] (364.99,207.46) -- (339.05,207.46) -- (340.82,209.08) -- (366.76,209.08) -- cycle ;
\draw  [draw opacity=0][fill={rgb, 255:red, 255; green, 255; blue, 255 }  ,fill opacity=1 ] (368.49,210.92) -- (342.55,210.92) -- (344.32,212.54) -- (370.26,212.54) -- cycle ;
\draw  [draw opacity=0][fill={rgb, 255:red, 255; green, 255; blue, 255 }  ,fill opacity=1 ] (371.99,214.38) -- (346.05,214.38) -- (347.81,216) -- (373.76,216) -- cycle ;
\draw  [draw opacity=0][fill={rgb, 255:red, 255; green, 255; blue, 255 }  ,fill opacity=1 ] (375.91,218) -- (349.97,218) -- (351.73,219.62) -- (377.67,219.62) -- cycle ;

\draw  [fill={rgb, 255:red, 208; green, 2; blue, 27 }  ,fill opacity=1 ] (275.53,209.29) -- (280.11,202) -- (307.62,202) -- (312.2,209.29) -- cycle ;
\draw  [color={rgb, 255:red, 208; green, 2; blue, 27 }  ,draw opacity=1 ][fill={rgb, 255:red, 208; green, 2; blue, 27 }  ,fill opacity=1 ] (270,210.81) .. controls (270,209.97) and (270.68,209.29) .. (271.53,209.29) -- (323.47,209.29) .. controls (324.32,209.29) and (325,209.97) .. (325,210.81) -- (325,215.39) .. controls (325,216.23) and (324.32,216.92) .. (323.47,216.92) -- (271.53,216.92) .. controls (270.68,216.92) and (270,216.23) .. (270,215.39) -- cycle ;
\draw  [fill={rgb, 255:red, 0; green, 0; blue, 0 }  ,fill opacity=1 ] (273.84,217.51) .. controls (273.84,215.03) and (275.91,213.02) .. (278.47,213.02) .. controls (281.02,213.02) and (283.1,215.03) .. (283.1,217.51) .. controls (283.1,219.99) and (281.02,222) .. (278.47,222) .. controls (275.91,222) and (273.84,219.99) .. (273.84,217.51) -- cycle ;
\draw  [fill={rgb, 255:red, 0; green, 0; blue, 0 }  ,fill opacity=1 ] (311.9,217.51) .. controls (311.9,215.03) and (313.98,213.02) .. (316.53,213.02) .. controls (319.09,213.02) and (321.16,215.03) .. (321.16,217.51) .. controls (321.16,219.99) and (319.09,222) .. (316.53,222) .. controls (313.98,222) and (311.9,219.99) .. (311.9,217.51) -- cycle ;

\draw  [fill={rgb, 255:red, 255; green, 255; blue, 255 }  ,fill opacity=1 ] (412,102) -- (422,102) -- (422,112) -- (412,112) -- cycle ;
\draw  [fill={rgb, 255:red, 255; green, 255; blue, 255 }  ,fill opacity=1 ] (442,102) -- (452,102) -- (452,112) -- (442,112) -- cycle ;
\draw  [fill={rgb, 255:red, 255; green, 255; blue, 255 }  ,fill opacity=1 ] (427,102) -- (437,102) -- (437,112) -- (427,112) -- cycle ;
\draw  [fill={rgb, 255:red, 255; green, 255; blue, 255 }  ,fill opacity=1 ] (412,117) -- (422,117) -- (422,127) -- (412,127) -- cycle ;
\draw  [fill={rgb, 255:red, 255; green, 255; blue, 255 }  ,fill opacity=1 ] (442,117) -- (452,117) -- (452,127) -- (442,127) -- cycle ;
\draw  [fill={rgb, 255:red, 255; green, 255; blue, 255 }  ,fill opacity=1 ] (427,117) -- (437,117) -- (437,127) -- (427,127) -- cycle ;
\draw  [fill={rgb, 255:red, 255; green, 255; blue, 255 }  ,fill opacity=1 ] (412,132) -- (422,132) -- (422,142) -- (412,142) -- cycle ;
\draw  [fill={rgb, 255:red, 255; green, 255; blue, 255 }  ,fill opacity=1 ] (442,132) -- (452,132) -- (452,142) -- (442,142) -- cycle ;
\draw  [fill={rgb, 255:red, 255; green, 255; blue, 255 }  ,fill opacity=1 ] (427,132) -- (437,132) -- (437,142) -- (427,142) -- cycle ;
\draw  [fill={rgb, 255:red, 255; green, 255; blue, 255 }  ,fill opacity=1 ] (412,147) -- (422,147) -- (422,157) -- (412,157) -- cycle ;
\draw  [fill={rgb, 255:red, 255; green, 255; blue, 255 }  ,fill opacity=1 ] (442,147) -- (452,147) -- (452,157) -- (442,157) -- cycle ;
\draw  [fill={rgb, 255:red, 255; green, 255; blue, 255 }  ,fill opacity=1 ] (427,147) -- (437,147) -- (437,157) -- (427,157) -- cycle ;
\draw  [fill={rgb, 255:red, 255; green, 255; blue, 255 }  ,fill opacity=1 ] (412,162) -- (422,162) -- (422,172) -- (412,172) -- cycle ;
\draw  [fill={rgb, 255:red, 255; green, 255; blue, 255 }  ,fill opacity=1 ] (442,162) -- (452,162) -- (452,172) -- (442,172) -- cycle ;
\draw  [fill={rgb, 255:red, 255; green, 255; blue, 255 }  ,fill opacity=1 ] (427,162) -- (437,162) -- (437,172) -- (427,172) -- cycle ;
\draw  [fill={rgb, 255:red, 255; green, 255; blue, 255 }  ,fill opacity=1 ] (412,177) -- (422,177) -- (422,187) -- (412,187) -- cycle ;
\draw  [fill={rgb, 255:red, 255; green, 255; blue, 255 }  ,fill opacity=1 ] (442,177) -- (452,177) -- (452,187) -- (442,187) -- cycle ;
\draw  [fill={rgb, 255:red, 255; green, 255; blue, 255 }  ,fill opacity=1 ] (427,177) -- (437,177) -- (437,187) -- (427,187) -- cycle ;

\draw  [fill={rgb, 255:red, 74; green, 144; blue, 226 }  ,fill opacity=1 ] (505,142.83) -- (486.17,124) -- (430,124) -- (430,225.17) -- (448.83,244) -- (505,244) -- cycle ; \draw   (430,124) -- (448.83,142.83) -- (505,142.83) ; \draw   (448.83,142.83) -- (448.83,244) ;
\draw  [fill={rgb, 255:red, 255; green, 255; blue, 255 }  ,fill opacity=1 ] (457,149) -- (467,149) -- (467,159) -- (457,159) -- cycle ;
\draw  [fill={rgb, 255:red, 255; green, 255; blue, 255 }  ,fill opacity=1 ] (487,149) -- (497,149) -- (497,159) -- (487,159) -- cycle ;
\draw  [fill={rgb, 255:red, 255; green, 255; blue, 255 }  ,fill opacity=1 ] (472,149) -- (482,149) -- (482,159) -- (472,159) -- cycle ;
\draw  [fill={rgb, 255:red, 255; green, 255; blue, 255 }  ,fill opacity=1 ] (457,164) -- (467,164) -- (467,174) -- (457,174) -- cycle ;
\draw  [fill={rgb, 255:red, 255; green, 255; blue, 255 }  ,fill opacity=1 ] (487,164) -- (497,164) -- (497,174) -- (487,174) -- cycle ;
\draw  [fill={rgb, 255:red, 255; green, 255; blue, 255 }  ,fill opacity=1 ] (472,164) -- (482,164) -- (482,174) -- (472,174) -- cycle ;
\draw  [fill={rgb, 255:red, 255; green, 255; blue, 255 }  ,fill opacity=1 ] (457,179) -- (467,179) -- (467,189) -- (457,189) -- cycle ;
\draw  [fill={rgb, 255:red, 255; green, 255; blue, 255 }  ,fill opacity=1 ] (487,179) -- (497,179) -- (497,189) -- (487,189) -- cycle ;
\draw  [fill={rgb, 255:red, 255; green, 255; blue, 255 }  ,fill opacity=1 ] (472,179) -- (482,179) -- (482,189) -- (472,189) -- cycle ;
\draw  [fill={rgb, 255:red, 255; green, 255; blue, 255 }  ,fill opacity=1 ] (457,194) -- (467,194) -- (467,204) -- (457,204) -- cycle ;
\draw  [fill={rgb, 255:red, 255; green, 255; blue, 255 }  ,fill opacity=1 ] (487,194) -- (497,194) -- (497,204) -- (487,204) -- cycle ;
\draw  [fill={rgb, 255:red, 255; green, 255; blue, 255 }  ,fill opacity=1 ] (472,194) -- (482,194) -- (482,204) -- (472,204) -- cycle ;
\draw  [fill={rgb, 255:red, 255; green, 255; blue, 255 }  ,fill opacity=1 ] (457,209) -- (467,209) -- (467,219) -- (457,219) -- cycle ;
\draw  [fill={rgb, 255:red, 255; green, 255; blue, 255 }  ,fill opacity=1 ] (487,209) -- (497,209) -- (497,219) -- (487,219) -- cycle ;
\draw  [fill={rgb, 255:red, 255; green, 255; blue, 255 }  ,fill opacity=1 ] (472,209) -- (482,209) -- (482,219) -- (472,219) -- cycle ;
\draw  [fill={rgb, 255:red, 255; green, 255; blue, 255 }  ,fill opacity=1 ] (457,224) -- (467,224) -- (467,234) -- (457,234) -- cycle ;
\draw  [fill={rgb, 255:red, 255; green, 255; blue, 255 }  ,fill opacity=1 ] (487,224) -- (497,224) -- (497,234) -- (487,234) -- cycle ;
\draw  [fill={rgb, 255:red, 255; green, 255; blue, 255 }  ,fill opacity=1 ] (472,224) -- (482,224) -- (482,234) -- (472,234) -- cycle ;

\draw  [draw opacity=0][fill={rgb, 255:red, 255; green, 255; blue, 255 }  ,fill opacity=1 ] (388.39,224.36) -- (401.2,236.86) -- (404.56,236.92) -- (391.75,224.42) -- cycle ;
\draw  [draw opacity=0][fill={rgb, 255:red, 255; green, 255; blue, 255 }  ,fill opacity=1 ] (395.41,224.47) -- (408.22,236.97) -- (411.58,237.03) -- (398.77,224.52) -- cycle ;
\draw  [draw opacity=0][fill={rgb, 255:red, 255; green, 255; blue, 255 }  ,fill opacity=1 ] (402.75,224.72) -- (415.56,237.23) -- (418.91,237.28) -- (406.1,224.78) -- cycle ;
\draw  [draw opacity=0][fill={rgb, 255:red, 255; green, 255; blue, 255 }  ,fill opacity=1 ] (410.08,224.98) -- (422.89,237.48) -- (426.25,237.54) -- (413.44,225.04) -- cycle ;
\draw  [draw opacity=0][fill={rgb, 255:red, 255; green, 255; blue, 255 }  ,fill opacity=1 ] (417.61,225.12) -- (430.42,237.63) -- (433.78,237.68) -- (420.97,225.18) -- cycle ;

\draw  [fill={rgb, 255:red, 245; green, 166; blue, 35 }  ,fill opacity=1 ] (337.11,53.62) .. controls (350.94,79.19) and (362.98,104.55) .. (368.29,124.43) .. controls (373.6,144.32) and (366.33,148.07) .. (355.67,129.89) .. controls (345,111.72) and (337.36,85.4) .. (329.69,56.95) .. controls (322.01,28.51) and (323.27,28.04) .. (337.11,53.62) -- cycle ;
\draw  [fill={rgb, 255:red, 0; green, 0; blue, 0 }  ,fill opacity=1 ] (259.23,154.32) -- (263.16,163.57) -- (255,162.26) -- cycle ;
\draw  [draw opacity=0][fill={rgb, 255:red, 245; green, 166; blue, 35 }  ,fill opacity=1 ][line width=1.5]  (266.24,128.93) -- (274.69,135.76) -- (275,149.69) -- (266.55,142.85) -- cycle ; \draw  [color={rgb, 255:red, 128; green, 128; blue, 128 }  ,draw opacity=1 ][line width=1.5]  (266.24,128.93) -- (266.55,142.85)(268.89,131.07) -- (269.2,144.99)(271.54,133.21) -- (271.85,147.14)(274.18,135.35) -- (274.49,149.28) ; \draw  [color={rgb, 255:red, 128; green, 128; blue, 128 }  ,draw opacity=1 ][line width=1.5]  (266.24,128.93) -- (274.69,135.76)(266.34,133.39) -- (274.79,140.22)(266.44,137.85) -- (274.89,144.68)(266.54,142.31) -- (274.99,149.14) ; \draw  [color={rgb, 255:red, 128; green, 128; blue, 128 }  ,draw opacity=1 ][line width=1.5]   ;
\draw  [draw opacity=0][line width=3]  (259.15,158.62) .. controls (259.28,153.1) and (261.98,149.06) .. (265.23,149.57) .. controls (268.48,150.08) and (271.05,154.97) .. (270.99,160.52) -- (265.07,159.67) -- cycle ; \draw  [line width=3]  (259.15,158.62) .. controls (259.28,153.1) and (261.98,149.06) .. (265.23,149.57) .. controls (268.48,150.08) and (271.05,154.97) .. (270.99,160.52) ;  
\draw  [color={rgb, 255:red, 0; green, 0; blue, 0 }  ,draw opacity=1 ][line width=3]  (412.41,260) -- (410.65,237.82) -- (407.64,259.13) (410.03,259.57) -- (411.57,205.21) ;
\draw  [fill={rgb, 255:red, 0; green, 0; blue, 0 }  ,fill opacity=1 ] (399.23,213.32) -- (403.16,222.57) -- (395,221.26) -- cycle ;
\draw  [draw opacity=0][fill={rgb, 255:red, 245; green, 166; blue, 35 }  ,fill opacity=1 ][line width=1.5]  (406.24,187.93) -- (414.69,194.76) -- (415,208.69) -- (406.55,201.85) -- cycle ; \draw  [color={rgb, 255:red, 128; green, 128; blue, 128 }  ,draw opacity=1 ][line width=1.5]  (406.24,187.93) -- (406.55,201.85)(408.89,190.07) -- (409.2,203.99)(411.53,192.21) -- (411.85,206.14)(414.18,194.35) -- (414.49,208.28) ; \draw  [color={rgb, 255:red, 128; green, 128; blue, 128 }  ,draw opacity=1 ][line width=1.5]  (406.24,187.93) -- (414.69,194.76)(406.34,192.39) -- (414.79,199.22)(406.44,196.85) -- (414.89,203.68)(406.54,201.31) -- (414.99,208.14) ; \draw  [color={rgb, 255:red, 128; green, 128; blue, 128 }  ,draw opacity=1 ][line width=1.5]   ;
\draw  [draw opacity=0][line width=3]  (399.15,217.62) .. controls (399.27,212.1) and (401.98,208.06) .. (405.23,208.57) .. controls (408.48,209.08) and (411.05,213.97) .. (410.99,219.52) -- (405.07,218.67) -- cycle ; \draw  [line width=3]  (399.15,217.62) .. controls (399.27,212.1) and (401.98,208.06) .. (405.23,208.57) .. controls (408.48,209.08) and (411.05,213.97) .. (410.99,219.52) ;  
\draw  [color={rgb, 255:red, 0; green, 0; blue, 0 }  ,draw opacity=1 ][line width=3]  (384.41,228) -- (382.65,205.82) -- (379.64,227.13) (382.03,227.57) -- (383.57,173.21) ;
\draw  [fill={rgb, 255:red, 0; green, 0; blue, 0 }  ,fill opacity=1 ] (371.23,181.32) -- (375.16,190.57) -- (367,189.26) -- cycle ;
\draw  [draw opacity=0][fill={rgb, 255:red, 245; green, 166; blue, 35 }  ,fill opacity=1 ][line width=1.5]  (378.24,155.93) -- (386.69,162.76) -- (387,176.69) -- (378.55,169.85) -- cycle ; \draw  [color={rgb, 255:red, 128; green, 128; blue, 128 }  ,draw opacity=1 ][line width=1.5]  (378.24,155.93) -- (378.55,169.85)(380.89,158.07) -- (381.2,171.99)(383.53,160.21) -- (383.85,174.14)(386.18,162.35) -- (386.49,176.28) ; \draw  [color={rgb, 255:red, 128; green, 128; blue, 128 }  ,draw opacity=1 ][line width=1.5]  (378.24,155.93) -- (386.69,162.76)(378.34,160.39) -- (386.79,167.22)(378.44,164.85) -- (386.89,171.68)(378.54,169.31) -- (386.99,176.14) ; \draw  [color={rgb, 255:red, 128; green, 128; blue, 128 }  ,draw opacity=1 ][line width=1.5]   ;
\draw  [draw opacity=0][line width=3]  (371.15,185.62) .. controls (371.27,180.1) and (373.98,176.06) .. (377.23,176.57) .. controls (380.48,177.08) and (383.05,181.97) .. (382.99,187.52) -- (377.07,186.67) -- cycle ; \draw  [line width=3]  (371.15,185.62) .. controls (371.27,180.1) and (373.98,176.06) .. (377.23,176.57) .. controls (380.48,177.08) and (383.05,181.97) .. (382.99,187.52) ;  
\draw  [fill={rgb, 255:red, 80; green, 227; blue, 194 }  ,fill opacity=1 ] (344.98,60.18) .. controls (370.02,95.23) and (393.2,129.89) .. (407.32,156.87) .. controls (421.44,183.84) and (414.61,187.64) .. (398.87,166.08) .. controls (383.13,144.53) and (359.47,102.89) .. (339.2,64.29) .. controls (318.92,25.69) and (319.94,25.13) .. (344.98,60.18) -- cycle ;
\draw  [fill={rgb, 255:red, 80; green, 227; blue, 194 }  ,fill opacity=1 ] (434.91,193.71) .. controls (452.67,184.98) and (470.29,177.75) .. (484.14,175.58) .. controls (497.99,173.4) and (500.25,180.87) .. (489.34,186.54) .. controls (478.43,192.21) and (457.12,197.21) .. (437.31,200.38) .. controls (417.49,203.55) and (417.15,202.43) .. (434.91,193.71) -- cycle ;
\draw  [fill={rgb, 255:red, 0; green, 0; blue, 0 }  ,fill opacity=1 ] (494,163) .. controls (494.55,163) and (495,163.45) .. (495,164) -- (495,172) .. controls (495,172.55) and (494.55,173) .. (494,173) -- (491,173) .. controls (490.45,173) and (490,172.55) .. (490,172) -- (490,164) .. controls (490,163.45) and (490.45,163) .. (491,163) -- cycle ;
\draw  [color={rgb, 255:red, 155; green, 155; blue, 155 }  ,draw opacity=1 ][fill={rgb, 255:red, 155; green, 155; blue, 155 }  ,fill opacity=1 ] (495,164.29) -- (495,171.57) -- (490,171.57) -- (490,164.29) -- cycle ;

\draw  [fill={rgb, 255:red, 0; green, 0; blue, 0 }  ,fill opacity=1 ] (302,204) .. controls (302.55,204) and (303,204.45) .. (303,205) -- (303,213) .. controls (303,213.55) and (302.55,214) .. (302,214) -- (299,214) .. controls (298.45,214) and (298,213.55) .. (298,213) -- (298,205) .. controls (298,204.45) and (298.45,204) .. (299,204) -- cycle ;
\draw  [color={rgb, 255:red, 155; green, 155; blue, 155 }  ,draw opacity=1 ][fill={rgb, 255:red, 155; green, 155; blue, 155 }  ,fill opacity=1 ] (303,205.29) -- (303,212.57) -- (298,212.57) -- (298,205.29) -- cycle ;

\draw  [draw opacity=0][fill={rgb, 255:red, 74; green, 144; blue, 226 }  ,fill opacity=1 ] (294.5,185) -- (302,45) -- (312,45) -- (319.5,185) -- cycle ;
\draw  [draw opacity=0][fill={rgb, 255:red, 74; green, 144; blue, 226 }  ,fill opacity=1 ] (270,194) -- (345,194) -- (345,184) -- (270,184) -- cycle ;
\draw  [color={rgb, 255:red, 0; green, 0; blue, 0 }  ,draw opacity=1 ][fill={rgb, 255:red, 126; green, 211; blue, 33 }  ,fill opacity=1 ] (280,83) -- (287.5,108) -- (327.5,108) -- (335,83) -- cycle ;
\draw  [fill={rgb, 255:red, 155; green, 155; blue, 155 }  ,fill opacity=1 ] (281.96,91) -- (284.06,98) -- (330.94,98) -- (333.04,91) -- cycle ;

\draw  [fill={rgb, 255:red, 245; green, 166; blue, 35 }  ,fill opacity=1 ] (365.97,160.53) .. controls (354.75,176.64) and (342.81,191.23) .. (331.43,199.26) .. controls (320.04,207.28) and (313.75,200.92) .. (320.46,190.72) .. controls (327.18,180.52) and (343.75,166.54) .. (360,155) .. controls (376.25,143.46) and (377.19,144.41) .. (365.97,160.53) -- cycle ;
\draw  [color={rgb, 255:red, 0; green, 0; blue, 0 }  ,draw opacity=1 ][line width=3]  (272.42,201) -- (270.65,178.82) -- (267.64,200.13) (270.03,200.57) -- (271.57,146.21) ;
\draw  [fill={rgb, 255:red, 248; green, 231; blue, 28 }  ,fill opacity=1 ] (290,55) .. controls (284.7,77.19) and (278.6,98.39) .. (271.76,113.01) .. controls (264.91,127.63) and (259.22,125.47) .. (262.3,111.65) .. controls (265.37,97.84) and (274.94,74.38) .. (284.69,53.43) .. controls (294.45,32.48) and (295.3,32.81) .. (290,55) -- cycle ;
\draw  [fill={rgb, 255:red, 248; green, 231; blue, 28 }  ,fill opacity=1 ] (282.03,145.8) .. controls (289.99,153.02) and (297,160.61) .. (300.3,167.61) .. controls (303.6,174.62) and (299.23,178.08) .. (294.15,173.73) .. controls (289.06,169.39) and (282.94,159.1) .. (278.17,149.1) .. controls (273.41,139.1) and (274.06,138.58) .. (282.03,145.8) -- cycle ;
\draw  [fill={rgb, 255:red, 0; green, 0; blue, 0 }  ,fill opacity=1 ] (299,180) .. controls (299.55,180) and (300,180.45) .. (300,181) -- (300,189) .. controls (300,189.55) and (299.55,190) .. (299,190) -- (296,190) .. controls (295.45,190) and (295,189.55) .. (295,189) -- (295,181) .. controls (295,180.45) and (295.45,180) .. (296,180) -- cycle ;
\draw  [color={rgb, 255:red, 155; green, 155; blue, 155 }  ,draw opacity=1 ][fill={rgb, 255:red, 155; green, 155; blue, 155 }  ,fill opacity=1 ] (300,181.29) -- (300,188.57) -- (295,188.57) -- (295,181.29) -- cycle ;

\draw  [fill={rgb, 255:red, 245; green, 166; blue, 35 }  ,fill opacity=1 ] (317,5.5) -- (317,18.5) .. controls (317,20.43) and (312.52,22) .. (307,22) .. controls (301.48,22) and (297,20.43) .. (297,18.5) -- (297,5.5)(317,5.5) .. controls (317,7.43) and (312.52,9) .. (307,9) .. controls (301.48,9) and (297,7.43) .. (297,5.5) .. controls (297,3.57) and (301.48,2) .. (307,2) .. controls (312.52,2) and (317,3.57) .. (317,5.5) -- cycle ;
\draw  [pattern=_33oxsip9h,pattern size=3.75pt,pattern thickness=1.5pt,pattern radius=0pt, pattern color={rgb, 255:red, 65; green, 117; blue, 5}] (317,5.5) -- (317,18.5) .. controls (317,20.43) and (312.52,22) .. (307,22) .. controls (301.48,22) and (297,20.43) .. (297,18.5) -- (297,5.5)(317,5.5) .. controls (317,7.43) and (312.52,9) .. (307,9) .. controls (301.48,9) and (297,7.43) .. (297,5.5) .. controls (297,3.57) and (301.48,2) .. (307,2) .. controls (312.52,2) and (317,3.57) .. (317,5.5) -- cycle ;

\draw (319,2.5) node [anchor=north west][inner sep=0.75pt]  [font=\normalsize] [align=left] {\begin{minipage}[lt]{107.88pt}\setlength\topsep{0pt}
\begin{center}
{\scriptsize High site/Gateway (multifaceted) }\\{\scriptsize Ultra-high capacity}
\end{center}

\end{minipage}};
\draw (244,127) node [anchor=north west][inner sep=0.75pt]   [align=left] {{\scriptsize RIS}};

\end{tikzpicture}

%% file: fig8.tex
  
\tikzset {_3niybcaub/.code = {\pgfsetadditionalshadetransform{ \pgftransformshift{\pgfpoint{0 bp } { 0 bp }  }  \pgftransformrotate{0 }  \pgftransformscale{2 }  }}}
\pgfdeclarehorizontalshading{_y41gbpwdn}{150bp}{rgb(0bp)=(1,1,0);
rgb(37.5bp)=(1,1,0);
rgb(62.5bp)=(0,0.5,0.5);
rgb(100bp)=(0,0.5,0.5)}

 
\tikzset{
pattern size/.store in=\mcSize, 
pattern size = 5pt,
pattern thickness/.store in=\mcThickness, 
pattern thickness = 0.3pt,
pattern radius/.store in=\mcRadius, 
pattern radius = 1pt}\makeatletter
\pgfutil@ifundefined{pgf@pattern@name@_0nc0am1sm}{
\pgfdeclarepatternformonly[\mcThickness,\mcSize]{_0nc0am1sm}
{\pgfqpoint{-\mcThickness}{-\mcThickness}}
{\pgfpoint{\mcSize}{\mcSize}}
{\pgfpoint{\mcSize}{\mcSize}}
{\pgfsetcolor{\tikz@pattern@color}
\pgfsetlinewidth{\mcThickness}
\pgfpathmoveto{\pgfpointorigin}
\pgfpathlineto{\pgfpoint{\mcSize}{0}}
\pgfpathmoveto{\pgfpointorigin}
\pgfpathlineto{\pgfpoint{0}{\mcSize}}
\pgfusepath{stroke}}}
\makeatother

 
\tikzset{
pattern size/.store in=\mcSize, 
pattern size = 5pt,
pattern thickness/.store in=\mcThickness, 
pattern thickness = 0.3pt,
pattern radius/.store in=\mcRadius, 
pattern radius = 1pt}\makeatletter
\pgfutil@ifundefined{pgf@pattern@name@_y5wt77mwz}{
\pgfdeclarepatternformonly[\mcThickness,\mcSize]{_y5wt77mwz}
{\pgfqpoint{-\mcThickness}{-\mcThickness}}
{\pgfpoint{\mcSize}{\mcSize}}
{\pgfpoint{\mcSize}{\mcSize}}
{\pgfsetcolor{\tikz@pattern@color}
\pgfsetlinewidth{\mcThickness}
\pgfpathmoveto{\pgfpointorigin}
\pgfpathlineto{\pgfpoint{\mcSize}{0}}
\pgfpathmoveto{\pgfpointorigin}
\pgfpathlineto{\pgfpoint{0}{\mcSize}}
\pgfusepath{stroke}}}
\makeatother
\tikzset{every picture/.style={line width=0.75pt}} 

\!\!\!\begin{tikzpicture}[x=0.75pt,y=0.75pt,yscale=-1,xscale=0.6]

\path  [shading=_y41gbpwdn,_3niybcaub] (356.42,54.11) .. controls (366.42,50.69) and (370.42,52.52) .. (381.42,74.11) .. controls (392.42,95.69) and (430.42,88.52) .. (396.42,117.11) .. controls (362.42,145.69) and (334.92,122.69) .. (328.92,114.11) .. controls (322.92,105.52) and (346.42,57.52) .. (356.42,54.11) -- cycle ; 
 \draw   (356.42,54.11) .. controls (366.42,50.69) and (370.42,52.52) .. (381.42,74.11) .. controls (392.42,95.69) and (430.42,88.52) .. (396.42,117.11) .. controls (362.42,145.69) and (334.92,122.69) .. (328.92,114.11) .. controls (322.92,105.52) and (346.42,57.52) .. (356.42,54.11) -- cycle ; 

\draw  [draw opacity=0][fill={rgb, 255:red, 245; green, 166; blue, 35 }  ,fill opacity=1 ][line width=1.5]  (351.42,44.11) -- (371.87,44.11) -- (371.87,64.37) -- (351.42,64.37) -- cycle ; \draw  [color={rgb, 255:red, 128; green, 128; blue, 128 }  ,draw opacity=1 ][line width=1.5]  (351.42,44.11) -- (351.42,64.37)(356.42,44.11) -- (356.42,64.37)(361.42,44.11) -- (361.42,64.37)(366.42,44.11) -- (366.42,64.37)(371.42,44.11) -- (371.42,64.37) ; \draw  [color={rgb, 255:red, 128; green, 128; blue, 128 }  ,draw opacity=1 ][line width=1.5]  (351.42,44.11) -- (371.87,44.11)(351.42,49.11) -- (371.87,49.11)(351.42,54.11) -- (371.87,54.11)(351.42,59.11) -- (371.87,59.11)(351.42,64.11) -- (371.87,64.11) ; \draw  [color={rgb, 255:red, 128; green, 128; blue, 128 }  ,draw opacity=1 ][line width=1.5]   ;

\draw  [fill={rgb, 255:red, 0; green, 0; blue, 0 }  ,fill opacity=1 ] (401.96,96) .. controls (402.51,96) and (402.96,96.45) .. (402.96,97) -- (402.96,105) .. controls (402.96,105.55) and (402.51,106) .. (401.96,106) -- (398.96,106) .. controls (398.4,106) and (397.96,105.55) .. (397.96,105) -- (397.96,97) .. controls (397.96,96.45) and (398.4,96) .. (398.96,96) -- cycle ;
\draw  [color={rgb, 255:red, 155; green, 155; blue, 155 }  ,draw opacity=1 ][fill={rgb, 255:red, 155; green, 155; blue, 155 }  ,fill opacity=1 ] (402.96,97.29) -- (402.96,104.57) -- (397.96,104.57) -- (397.96,97.29) -- cycle ;

\draw  [fill={rgb, 255:red, 0; green, 0; blue, 0 }  ,fill opacity=1 ] (323.96,96) .. controls (323.4,96) and (322.96,96.45) .. (322.96,97) -- (322.96,105) .. controls (322.96,105.55) and (323.4,106) .. (323.96,106) -- (326.96,106) .. controls (327.51,106) and (327.96,105.55) .. (327.96,105) -- (327.96,97) .. controls (327.96,96.45) and (327.51,96) .. (326.96,96) -- cycle ;
\draw  [color={rgb, 255:red, 155; green, 155; blue, 155 }  ,draw opacity=1 ][fill={rgb, 255:red, 155; green, 155; blue, 155 }  ,fill opacity=1 ] (322.96,97.29) -- (322.96,104.57) -- (327.96,104.57) -- (327.96,97.29) -- cycle ;

\draw [color={rgb, 255:red, 144; green, 19; blue, 254 }  ,draw opacity=1 ][line width=3]    (392.96,96) -- (371.42,54.11) -- (602.96,71) ;
\draw [shift={(377.43,65.8)}, rotate = 62.8] [fill={rgb, 255:red, 144; green, 19; blue, 254 }  ,fill opacity=1 ][line width=0.08]  [draw opacity=0] (18.75,-9.01) -- (0,0) -- (18.75,9.01) -- (12.45,0) -- cycle    ;
\draw [shift={(497.56,63.31)}, rotate = 184.17] [fill={rgb, 255:red, 144; green, 19; blue, 254 }  ,fill opacity=1 ][line width=0.08]  [draw opacity=0] (18.75,-9.01) -- (0,0) -- (18.75,9.01) -- (12.45,0) -- cycle    ;
\draw [color={rgb, 255:red, 208; green, 2; blue, 27 }  ,draw opacity=1 ][line width=3]    (332.96,96) -- (351.42,54.11) -- (122.96,71) ;
\draw [shift={(346.38,65.54)}, rotate = 113.79] [fill={rgb, 255:red, 208; green, 2; blue, 27 }  ,fill opacity=1 ][line width=0.08]  [draw opacity=0] (18.75,-9.01) -- (0,0) -- (18.75,9.01) -- (12.45,0) -- cycle    ;
\draw [shift={(226.82,63.32)}, rotate = 355.77] [fill={rgb, 255:red, 208; green, 2; blue, 27 }  ,fill opacity=1 ][line width=0.08]  [draw opacity=0] (18.75,-9.01) -- (0,0) -- (18.75,9.01) -- (12.45,0) -- cycle    ;
\draw  [fill={rgb, 255:red, 0; green, 0; blue, 0 }  ,fill opacity=1 ] (102.08,121) -- (104.08,81.75) -- (106.75,81.75) -- (108.75,121) -- cycle ;
\draw  [fill={rgb, 255:red, 245; green, 166; blue, 35 }  ,fill opacity=1 ] (117.96,72.43) -- (117.96,85.19) .. controls (117.96,87.09) and (112.36,88.62) .. (105.46,88.62) .. controls (98.55,88.62) and (92.96,87.09) .. (92.96,85.19) -- (92.96,72.43)(117.96,72.43) .. controls (117.96,74.33) and (112.36,75.87) .. (105.46,75.87) .. controls (98.55,75.87) and (92.96,74.33) .. (92.96,72.43) .. controls (92.96,70.54) and (98.55,69) .. (105.46,69) .. controls (112.36,69) and (117.96,70.54) .. (117.96,72.43) -- cycle ;
\draw  [pattern=_0nc0am1sm,pattern size=3.75pt,pattern thickness=1.5pt,pattern radius=0pt, pattern color={rgb, 255:red, 65; green, 117; blue, 5}] (117.96,72.43) -- (117.96,85.19) .. controls (117.96,87.09) and (112.36,88.62) .. (105.46,88.62) .. controls (98.55,88.62) and (92.96,87.09) .. (92.96,85.19) -- (92.96,72.43)(117.96,72.43) .. controls (117.96,74.33) and (112.36,75.87) .. (105.46,75.87) .. controls (98.55,75.87) and (92.96,74.33) .. (92.96,72.43) .. controls (92.96,70.54) and (98.55,69) .. (105.46,69) .. controls (112.36,69) and (117.96,70.54) .. (117.96,72.43) -- cycle ;

\draw  [fill={rgb, 255:red, 139; green, 87; blue, 42 }  ,fill opacity=1 ] (304.71,96.54) -- (304.71,118.71) .. controls (304.71,119.42) and (302.8,120) .. (300.43,120) .. controls (298.06,120) and (296.14,119.42) .. (296.14,118.71) -- (296.14,96.54) .. controls (296.14,95.83) and (298.06,95.25) .. (300.43,95.25) .. controls (302.8,95.25) and (304.71,95.83) .. (304.71,96.54) .. controls (304.71,97.25) and (302.8,97.82) .. (300.43,97.82) .. controls (298.06,97.82) and (296.14,97.25) .. (296.14,96.54) ;
\draw  [fill={rgb, 255:red, 65; green, 117; blue, 5 }  ,fill opacity=1 ] (287.73,85.28) .. controls (287.49,83.01) and (288.28,80.76) .. (289.77,79.49) .. controls (291.26,78.21) and (293.19,78.14) .. (294.74,79.3) .. controls (295.29,77.98) and (296.29,77.07) .. (297.44,76.84) .. controls (298.6,76.61) and (299.77,77.1) .. (300.6,78.15) .. controls (301.07,76.95) and (301.98,76.14) .. (303.02,76.02) .. controls (304.06,75.89) and (305.08,76.46) .. (305.72,77.53) .. controls (306.56,76.26) and (307.9,75.72) .. (309.16,76.15) .. controls (310.42,76.59) and (311.37,77.91) .. (311.6,79.55) .. controls (312.63,79.91) and (313.49,80.82) .. (313.96,82.06) .. controls (314.42,83.3) and (314.45,84.73) .. (314.03,85.99) .. controls (315.04,87.69) and (315.28,89.94) .. (314.65,91.92) .. controls (314.02,93.9) and (312.62,95.3) .. (310.97,95.61) .. controls (310.95,97.46) and (310.16,99.17) .. (308.89,100.06) .. controls (307.61,100.96) and (306.06,100.9) .. (304.83,99.92) .. controls (304.31,102.14) and (302.83,103.78) .. (301.04,104.12) .. controls (299.25,104.46) and (297.46,103.45) .. (296.45,101.52) .. controls (295.22,102.47) and (293.74,102.74) .. (292.35,102.28) .. controls (290.95,101.81) and (289.77,100.65) .. (289.05,99.04) .. controls (287.79,99.23) and (286.57,98.4) .. (286,96.95) .. controls (285.43,95.5) and (285.62,93.75) .. (286.49,92.57) .. controls (285.37,91.72) and (284.79,90.04) .. (285.07,88.4) .. controls (285.34,86.77) and (286.41,85.54) .. (287.7,85.37) ; \draw   (286.49,92.57) .. controls (287.02,92.97) and (287.63,93.15) .. (288.25,93.09)(289.05,99.04) .. controls (289.31,99) and (289.57,98.92) .. (289.82,98.79)(296.45,101.52) .. controls (296.27,101.16) and (296.11,100.78) .. (295.99,100.38)(304.83,99.92) .. controls (304.93,99.51) and (304.99,99.1) .. (305.01,98.67)(310.97,95.61) .. controls (310.98,93.63) and (310.1,91.82) .. (308.71,90.95)(314.03,85.99) .. controls (313.8,86.66) and (313.46,87.26) .. (313.02,87.74)(311.6,79.55) .. controls (311.64,79.82) and (311.66,80.09) .. (311.65,80.37)(305.72,77.53) .. controls (305.5,77.84) and (305.33,78.2) .. (305.2,78.58)(300.6,78.15) .. controls (300.49,78.43) and (300.4,78.74) .. (300.35,79.05)(294.74,79.3) .. controls (295.07,79.55) and (295.37,79.84) .. (295.64,80.18)(287.73,85.28) .. controls (287.76,85.59) and (287.82,85.9) .. (287.89,86.21) ;

\draw  [fill={rgb, 255:red, 0; green, 0; blue, 0 }  ,fill opacity=1 ] (617.08,121) -- (619.08,81.75) -- (621.75,81.75) -- (623.75,121) -- cycle ;
\draw  [fill={rgb, 255:red, 245; green, 166; blue, 35 }  ,fill opacity=1 ] (632.96,72.43) -- (632.96,85.19) .. controls (632.96,87.09) and (627.36,88.62) .. (620.46,88.62) .. controls (613.55,88.62) and (607.96,87.09) .. (607.96,85.19) -- (607.96,72.43)(632.96,72.43) .. controls (632.96,74.33) and (627.36,75.87) .. (620.46,75.87) .. controls (613.55,75.87) and (607.96,74.33) .. (607.96,72.43) .. controls (607.96,70.54) and (613.55,69) .. (620.46,69) .. controls (627.36,69) and (632.96,70.54) .. (632.96,72.43) -- cycle ;
\draw  [pattern=_y5wt77mwz,pattern size=3.75pt,pattern thickness=1.5pt,pattern radius=0pt, pattern color={rgb, 255:red, 65; green, 117; blue, 5}] (632.96,72.43) -- (632.96,85.19) .. controls (632.96,87.09) and (627.36,88.62) .. (620.46,88.62) .. controls (613.55,88.62) and (607.96,87.09) .. (607.96,85.19) -- (607.96,72.43)(632.96,72.43) .. controls (632.96,74.33) and (627.36,75.87) .. (620.46,75.87) .. controls (613.55,75.87) and (607.96,74.33) .. (607.96,72.43) .. controls (607.96,70.54) and (613.55,69) .. (620.46,69) .. controls (627.36,69) and (632.96,70.54) .. (632.96,72.43) -- cycle ;

\draw  [fill={rgb, 255:red, 139; green, 87; blue, 42 }  ,fill opacity=1 ] (434.71,96.54) -- (434.71,118.71) .. controls (434.71,119.42) and (432.8,120) .. (430.43,120) .. controls (428.06,120) and (426.14,119.42) .. (426.14,118.71) -- (426.14,96.54) .. controls (426.14,95.83) and (428.06,95.25) .. (430.43,95.25) .. controls (432.8,95.25) and (434.71,95.83) .. (434.71,96.54) .. controls (434.71,97.25) and (432.8,97.82) .. (430.43,97.82) .. controls (428.06,97.82) and (426.14,97.25) .. (426.14,96.54) ;
\draw  [fill={rgb, 255:red, 65; green, 117; blue, 5 }  ,fill opacity=1 ] (417.73,85.28) .. controls (417.49,83.01) and (418.28,80.76) .. (419.77,79.49) .. controls (421.26,78.21) and (423.19,78.14) .. (424.74,79.3) .. controls (425.29,77.98) and (426.29,77.07) .. (427.44,76.84) .. controls (428.6,76.61) and (429.77,77.1) .. (430.6,78.15) .. controls (431.07,76.95) and (431.98,76.14) .. (433.02,76.02) .. controls (434.06,75.89) and (435.08,76.46) .. (435.72,77.53) .. controls (436.56,76.26) and (437.9,75.72) .. (439.16,76.15) .. controls (440.42,76.59) and (441.37,77.91) .. (441.6,79.55) .. controls (442.63,79.91) and (443.49,80.82) .. (443.96,82.06) .. controls (444.42,83.3) and (444.45,84.73) .. (444.03,85.99) .. controls (445.04,87.69) and (445.28,89.94) .. (444.65,91.92) .. controls (444.02,93.9) and (442.62,95.3) .. (440.97,95.61) .. controls (440.95,97.46) and (440.16,99.17) .. (438.89,100.06) .. controls (437.61,100.96) and (436.06,100.9) .. (434.83,99.92) .. controls (434.31,102.14) and (432.83,103.78) .. (431.04,104.12) .. controls (429.25,104.46) and (427.46,103.45) .. (426.45,101.52) .. controls (425.22,102.47) and (423.74,102.74) .. (422.35,102.28) .. controls (420.95,101.81) and (419.77,100.65) .. (419.05,99.04) .. controls (417.79,99.23) and (416.57,98.4) .. (416,96.95) .. controls (415.43,95.5) and (415.62,93.75) .. (416.49,92.57) .. controls (415.37,91.72) and (414.79,90.04) .. (415.07,88.4) .. controls (415.34,86.77) and (416.41,85.54) .. (417.7,85.37) ; \draw   (416.49,92.57) .. controls (417.02,92.97) and (417.63,93.15) .. (418.25,93.09)(419.05,99.04) .. controls (419.31,99) and (419.57,98.92) .. (419.82,98.79)(426.45,101.52) .. controls (426.27,101.16) and (426.11,100.78) .. (425.99,100.38)(434.83,99.92) .. controls (434.93,99.51) and (434.99,99.1) .. (435.01,98.67)(440.97,95.61) .. controls (440.98,93.63) and (440.1,91.82) .. (438.71,90.95)(444.03,85.99) .. controls (443.8,86.66) and (443.46,87.26) .. (443.02,87.74)(441.6,79.55) .. controls (441.64,79.82) and (441.66,80.09) .. (441.65,80.37)(435.72,77.53) .. controls (435.5,77.84) and (435.33,78.2) .. (435.2,78.58)(430.6,78.15) .. controls (430.49,78.43) and (430.4,78.74) .. (430.35,79.05)(424.74,79.3) .. controls (425.07,79.55) and (425.37,79.84) .. (425.64,80.18)(417.73,85.28) .. controls (417.76,85.59) and (417.82,85.9) .. (417.89,86.21) ;

\draw  [fill={rgb, 255:red, 139; green, 87; blue, 42 }  ,fill opacity=1 ] (469.71,96.54) -- (469.71,118.71) .. controls (469.71,119.42) and (467.8,120) .. (465.43,120) .. controls (463.06,120) and (461.14,119.42) .. (461.14,118.71) -- (461.14,96.54) .. controls (461.14,95.83) and (463.06,95.25) .. (465.43,95.25) .. controls (467.8,95.25) and (469.71,95.83) .. (469.71,96.54) .. controls (469.71,97.25) and (467.8,97.82) .. (465.43,97.82) .. controls (463.06,97.82) and (461.14,97.25) .. (461.14,96.54) ;
\draw  [fill={rgb, 255:red, 65; green, 117; blue, 5 }  ,fill opacity=1 ] (452.73,85.28) .. controls (452.49,83.01) and (453.28,80.76) .. (454.77,79.49) .. controls (456.26,78.21) and (458.19,78.14) .. (459.74,79.3) .. controls (460.29,77.98) and (461.29,77.07) .. (462.44,76.84) .. controls (463.6,76.61) and (464.77,77.1) .. (465.6,78.15) .. controls (466.07,76.95) and (466.98,76.14) .. (468.02,76.02) .. controls (469.06,75.89) and (470.08,76.46) .. (470.72,77.53) .. controls (471.56,76.26) and (472.9,75.72) .. (474.16,76.15) .. controls (475.42,76.59) and (476.37,77.91) .. (476.6,79.55) .. controls (477.63,79.91) and (478.49,80.82) .. (478.96,82.06) .. controls (479.42,83.3) and (479.45,84.73) .. (479.03,85.99) .. controls (480.04,87.69) and (480.28,89.94) .. (479.65,91.92) .. controls (479.02,93.9) and (477.62,95.3) .. (475.97,95.61) .. controls (475.95,97.46) and (475.16,99.17) .. (473.89,100.06) .. controls (472.61,100.96) and (471.06,100.9) .. (469.83,99.92) .. controls (469.31,102.14) and (467.83,103.78) .. (466.04,104.12) .. controls (464.25,104.46) and (462.46,103.45) .. (461.45,101.52) .. controls (460.22,102.47) and (458.74,102.74) .. (457.35,102.28) .. controls (455.95,101.81) and (454.77,100.65) .. (454.05,99.04) .. controls (452.79,99.23) and (451.57,98.4) .. (451,96.95) .. controls (450.43,95.5) and (450.62,93.75) .. (451.49,92.57) .. controls (450.37,91.72) and (449.79,90.04) .. (450.07,88.4) .. controls (450.34,86.77) and (451.41,85.54) .. (452.7,85.37) ; \draw   (451.49,92.57) .. controls (452.02,92.97) and (452.63,93.15) .. (453.25,93.09)(454.05,99.04) .. controls (454.31,99) and (454.57,98.92) .. (454.82,98.79)(461.45,101.52) .. controls (461.27,101.16) and (461.11,100.78) .. (460.99,100.38)(469.83,99.92) .. controls (469.93,99.51) and (469.99,99.1) .. (470.01,98.67)(475.97,95.61) .. controls (475.98,93.63) and (475.1,91.82) .. (473.71,90.95)(479.03,85.99) .. controls (478.8,86.66) and (478.46,87.26) .. (478.02,87.74)(476.6,79.55) .. controls (476.64,79.82) and (476.66,80.09) .. (476.65,80.37)(470.72,77.53) .. controls (470.5,77.84) and (470.33,78.2) .. (470.2,78.58)(465.6,78.15) .. controls (465.49,78.43) and (465.4,78.74) .. (465.35,79.05)(459.74,79.3) .. controls (460.07,79.55) and (460.37,79.84) .. (460.64,80.18)(452.73,85.28) .. controls (452.76,85.59) and (452.82,85.9) .. (452.89,86.21) ;

\draw  [fill={rgb, 255:red, 74; green, 144; blue, 226 }  ,fill opacity=1 ] (177.46,91.04) -- (167.41,81) -- (127.96,81) -- (127.96,110.96) -- (138,121) -- (177.46,121) -- cycle ; \draw   (127.96,81) -- (138,91.04) -- (177.46,91.04) ; \draw   (138,91.04) -- (138,121) ;
\draw  [fill={rgb, 255:red, 255; green, 255; blue, 255 }  ,fill opacity=1 ] (143.51,95.55) -- (150.58,95.55) -- (150.58,102.82) -- (143.51,102.82) -- cycle ;
\draw  [fill={rgb, 255:red, 255; green, 255; blue, 255 }  ,fill opacity=1 ] (164.73,95.55) -- (171.8,95.55) -- (171.8,102.82) -- (164.73,102.82) -- cycle ;
\draw  [fill={rgb, 255:red, 255; green, 255; blue, 255 }  ,fill opacity=1 ] (154.12,95.55) -- (161.19,95.55) -- (161.19,102.82) -- (154.12,102.82) -- cycle ;
\draw  [fill={rgb, 255:red, 255; green, 255; blue, 255 }  ,fill opacity=1 ] (143.51,106.45) -- (150.58,106.45) -- (150.58,113.73) -- (143.51,113.73) -- cycle ;
\draw  [fill={rgb, 255:red, 255; green, 255; blue, 255 }  ,fill opacity=1 ] (164.73,106.45) -- (171.8,106.45) -- (171.8,113.73) -- (164.73,113.73) -- cycle ;
\draw  [fill={rgb, 255:red, 255; green, 255; blue, 255 }  ,fill opacity=1 ] (154.12,106.45) -- (161.19,106.45) -- (161.19,113.73) -- (154.12,113.73) -- cycle ;

\draw  [fill={rgb, 255:red, 74; green, 144; blue, 226 }  ,fill opacity=1 ] (602.96,91.04) -- (592.91,81) -- (553.46,81) -- (553.46,110.96) -- (563.5,121) -- (602.96,121) -- cycle ; \draw   (553.46,81) -- (563.5,91.04) -- (602.96,91.04) ; \draw   (563.5,91.04) -- (563.5,121) ;
\draw  [fill={rgb, 255:red, 255; green, 255; blue, 255 }  ,fill opacity=1 ] (569.01,95.55) -- (576.08,95.55) -- (576.08,102.82) -- (569.01,102.82) -- cycle ;
\draw  [fill={rgb, 255:red, 255; green, 255; blue, 255 }  ,fill opacity=1 ] (590.23,95.55) -- (597.3,95.55) -- (597.3,102.82) -- (590.23,102.82) -- cycle ;
\draw  [fill={rgb, 255:red, 255; green, 255; blue, 255 }  ,fill opacity=1 ] (579.62,95.55) -- (586.69,95.55) -- (586.69,102.82) -- (579.62,102.82) -- cycle ;
\draw  [fill={rgb, 255:red, 255; green, 255; blue, 255 }  ,fill opacity=1 ] (569.01,106.45) -- (576.08,106.45) -- (576.08,113.73) -- (569.01,113.73) -- cycle ;
\draw  [fill={rgb, 255:red, 255; green, 255; blue, 255 }  ,fill opacity=1 ] (590.23,106.45) -- (597.3,106.45) -- (597.3,113.73) -- (590.23,113.73) -- cycle ;
\draw  [fill={rgb, 255:red, 255; green, 255; blue, 255 }  ,fill opacity=1 ] (579.62,106.45) -- (586.69,106.45) -- (586.69,113.73) -- (579.62,113.73) -- cycle ;

\draw  [fill={rgb, 255:red, 139; green, 87; blue, 42 }  ,fill opacity=1 ] (332.96,106) -- (387.96,106) -- (387.96,109) -- (332.96,109) -- cycle ;
\draw  [fill={rgb, 255:red, 74; green, 144; blue, 226 }  ,fill opacity=1 ] (332.96,109) -- (387.96,109) -- (387.96,130) -- (332.96,130) -- cycle ;
\draw  [color={rgb, 255:red, 0; green, 0; blue, 0 }  ,draw opacity=1 ][fill={rgb, 255:red, 74; green, 144; blue, 226 }  ,fill opacity=1 ] (348.96,109) -- (370.96,109) -- (370.96,131) -- (348.96,131) -- cycle ;
\draw  [color={rgb, 255:red, 0; green, 0; blue, 0 }  ,draw opacity=1 ][fill={rgb, 255:red, 139; green, 87; blue, 42 }  ,fill opacity=1 ] (360.46,101) -- (377.96,109) -- (342.96,109) -- cycle ;

\draw  [color={rgb, 255:red, 0; green, 0; blue, 0 }  ,draw opacity=1 ][fill={rgb, 255:red, 255; green, 255; blue, 255 }  ,fill opacity=1 ] (357.96,119) -- (362.96,119) -- (362.96,129) -- (357.96,129) -- cycle ;

\draw    (70,125) .. controls (84.96,107) and (89,140.93) .. (105.08,121) ;
\draw    (621.08,121) .. controls (640,136.93) and (636,102.93) .. (655,120) ;

\draw (340.96,27) node [anchor=north west][inner sep=0.75pt]   [align=left] {\begin{minipage}[lt]{19.73pt}\setlength\topsep{0pt}
\begin{center}
RIS
\end{center}

\end{minipage}};
\draw (328.96,132) node [anchor=north west][inner sep=0.75pt]   [align=left] {Rural area};
\draw (88,131) node [anchor=north west][inner sep=0.75pt]   [align=left] {Suburban area};
\draw (503,132) node [anchor=north west][inner sep=0.75pt]   [align=left] {Suburban area};

\end{tikzpicture}

%% file: fig1.tex
\tikzset{every picture/.style={line width=0.75pt}} 

\begin{tikzpicture}[x=0.75pt,y=0.75pt,yscale=-0.5,xscale=0.5]

\draw   (190.5,105) -- (106.5,105) .. controls (105.67,105) and (105,102.76) .. (105,100) .. controls (105,97.24) and (105.67,95) .. (106.5,95) -- (190.5,95) .. controls (191.33,95) and (192,97.24) .. (192,100) .. controls (192,102.76) and (191.33,105) .. (190.5,105) .. controls (189.67,105) and (189,102.76) .. (189,100) .. controls (189,97.24) and (189.67,95) .. (190.5,95) ;
\draw    (190.5,105) -- (182,115) ;
\draw   (179.5,117.5) .. controls (179.5,116.12) and (180.62,115) .. (182,115) .. controls (183.38,115) and (184.5,116.12) .. (184.5,117.5) .. controls (184.5,118.88) and (183.38,120) .. (182,120) .. controls (180.62,120) and (179.5,118.88) .. (179.5,117.5) -- cycle ;
\draw   (280.5,105) -- (196.5,105) .. controls (195.67,105) and (195,102.76) .. (195,100) .. controls (195,97.24) and (195.67,95) .. (196.5,95) -- (280.5,95) .. controls (281.33,95) and (282,97.24) .. (282,100) .. controls (282,102.76) and (281.33,105) .. (280.5,105) .. controls (279.67,105) and (279,102.76) .. (279,100) .. controls (279,97.24) and (279.67,95) .. (280.5,95) ;
\draw    (280.5,105) -- (272,115) ;
\draw    (196.5,105) -- (205,115) ;
\draw   (269.5,117.5) .. controls (269.5,116.12) and (270.62,115) .. (272,115) .. controls (273.38,115) and (274.5,116.12) .. (274.5,117.5) .. controls (274.5,118.88) and (273.38,120) .. (272,120) .. controls (270.62,120) and (269.5,118.88) .. (269.5,117.5) -- cycle ;
\draw   (202.5,117.5) .. controls (202.5,116.12) and (203.62,115) .. (205,115) .. controls (206.38,115) and (207.5,116.12) .. (207.5,117.5) .. controls (207.5,118.88) and (206.38,120) .. (205,120) .. controls (203.62,120) and (202.5,118.88) .. (202.5,117.5) -- cycle ;

\draw   (371.5,105) -- (287.5,105) .. controls (286.67,105) and (286,102.76) .. (286,100) .. controls (286,97.24) and (286.67,95) .. (287.5,95) -- (371.5,95) .. controls (372.33,95) and (373,97.24) .. (373,100) .. controls (373,102.76) and (372.33,105) .. (371.5,105) .. controls (370.67,105) and (370,102.76) .. (370,100) .. controls (370,97.24) and (370.67,95) .. (371.5,95) ;
\draw    (371.5,105) -- (363,115) ;
\draw    (287.5,105) -- (296,115) ;
\draw   (360.5,117.5) .. controls (360.5,116.12) and (361.62,115) .. (363,115) .. controls (364.38,115) and (365.5,116.12) .. (365.5,117.5) .. controls (365.5,118.88) and (364.38,120) .. (363,120) .. controls (361.62,120) and (360.5,118.88) .. (360.5,117.5) -- cycle ;
\draw   (293.5,117.5) .. controls (293.5,116.12) and (294.62,115) .. (296,115) .. controls (297.38,115) and (298.5,116.12) .. (298.5,117.5) .. controls (298.5,118.88) and (297.38,120) .. (296,120) .. controls (294.62,120) and (293.5,118.88) .. (293.5,117.5) -- cycle ;

\draw   (461.5,105) -- (377.5,105) .. controls (376.67,105) and (376,102.76) .. (376,100) .. controls (376,97.24) and (376.67,95) .. (377.5,95) -- (461.5,95) .. controls (462.33,95) and (463,97.24) .. (463,100) .. controls (463,102.76) and (462.33,105) .. (461.5,105) .. controls (460.67,105) and (460,102.76) .. (460,100) .. controls (460,97.24) and (460.67,95) .. (461.5,95) ;
\draw    (461.5,105) -- (453,115) ;
\draw    (377.5,105) -- (386,115) ;
\draw   (450.5,117.5) .. controls (450.5,116.12) and (451.62,115) .. (453,115) .. controls (454.38,115) and (455.5,116.12) .. (455.5,117.5) .. controls (455.5,118.88) and (454.38,120) .. (453,120) .. controls (451.62,120) and (450.5,118.88) .. (450.5,117.5) -- cycle ;
\draw   (383.5,117.5) .. controls (383.5,116.12) and (384.62,115) .. (386,115) .. controls (387.38,115) and (388.5,116.12) .. (388.5,117.5) .. controls (388.5,118.88) and (387.38,120) .. (386,120) .. controls (384.62,120) and (383.5,118.88) .. (383.5,117.5) -- cycle ;

\draw   (550.5,105) -- (466.5,105) .. controls (465.67,105) and (465,102.76) .. (465,100) .. controls (465,97.24) and (465.67,95) .. (466.5,95) -- (550.5,95) .. controls (551.33,95) and (552,97.24) .. (552,100) .. controls (552,102.76) and (551.33,105) .. (550.5,105) .. controls (549.67,105) and (549,102.76) .. (549,100) .. controls (549,97.24) and (549.67,95) .. (550.5,95) ;
\draw    (550.5,105) -- (542,115) ;
\draw    (466.5,105) -- (475,115) ;
\draw   (539.5,117.5) .. controls (539.5,116.12) and (540.62,115) .. (542,115) .. controls (543.38,115) and (544.5,116.12) .. (544.5,117.5) .. controls (544.5,118.88) and (543.38,120) .. (542,120) .. controls (540.62,120) and (539.5,118.88) .. (539.5,117.5) -- cycle ;
\draw   (472.5,117.5) .. controls (472.5,116.12) and (473.62,115) .. (475,115) .. controls (476.38,115) and (477.5,116.12) .. (477.5,117.5) .. controls (477.5,118.88) and (476.38,120) .. (475,120) .. controls (473.62,120) and (472.5,118.88) .. (472.5,117.5) -- cycle ;

\draw   (640.5,105) -- (556.5,105) .. controls (555.67,105) and (555,102.76) .. (555,100) .. controls (555,97.24) and (555.67,95) .. (556.5,95) -- (640.5,95) .. controls (641.33,95) and (642,97.24) .. (642,100) .. controls (642,102.76) and (641.33,105) .. (640.5,105) .. controls (639.67,105) and (639,102.76) .. (639,100) .. controls (639,97.24) and (639.67,95) .. (640.5,95) ;
\draw    (556.5,105) -- (565,115) ;
\draw   (562.5,117.5) .. controls (562.5,116.12) and (563.62,115) .. (565,115) .. controls (566.38,115) and (567.5,116.12) .. (567.5,117.5) .. controls (567.5,118.88) and (566.38,120) .. (565,120) .. controls (563.62,120) and (562.5,118.88) .. (562.5,117.5) -- cycle ;
\draw   (210.5,140) -- (126.5,140) .. controls (125.67,140) and (125,137.76) .. (125,135) .. controls (125,132.24) and (125.67,130) .. (126.5,130) -- (210.5,130) .. controls (211.33,130) and (212,132.24) .. (212,135) .. controls (212,137.76) and (211.33,140) .. (210.5,140) .. controls (209.67,140) and (209,137.76) .. (209,135) .. controls (209,132.24) and (209.67,130) .. (210.5,130) ;
\draw    (210.5,140) -- (202,150) ;
\draw   (199.5,152.5) .. controls (199.5,151.12) and (200.62,150) .. (202,150) .. controls (203.38,150) and (204.5,151.12) .. (204.5,152.5) .. controls (204.5,153.88) and (203.38,155) .. (202,155) .. controls (200.62,155) and (199.5,153.88) .. (199.5,152.5) -- cycle ;
\draw   (300.5,140) -- (216.5,140) .. controls (215.67,140) and (215,137.76) .. (215,135) .. controls (215,132.24) and (215.67,130) .. (216.5,130) -- (300.5,130) .. controls (301.33,130) and (302,132.24) .. (302,135) .. controls (302,137.76) and (301.33,140) .. (300.5,140) .. controls (299.67,140) and (299,137.76) .. (299,135) .. controls (299,132.24) and (299.67,130) .. (300.5,130) ;
\draw    (300.5,140) -- (292,150) ;
\draw    (216.5,140) -- (225,150) ;
\draw   (289.5,152.5) .. controls (289.5,151.12) and (290.62,150) .. (292,150) .. controls (293.38,150) and (294.5,151.12) .. (294.5,152.5) .. controls (294.5,153.88) and (293.38,155) .. (292,155) .. controls (290.62,155) and (289.5,153.88) .. (289.5,152.5) -- cycle ;
\draw   (222.5,152.5) .. controls (222.5,151.12) and (223.62,150) .. (225,150) .. controls (226.38,150) and (227.5,151.12) .. (227.5,152.5) .. controls (227.5,153.88) and (226.38,155) .. (225,155) .. controls (223.62,155) and (222.5,153.88) .. (222.5,152.5) -- cycle ;

\draw   (391.5,140) -- (307.5,140) .. controls (306.67,140) and (306,137.76) .. (306,135) .. controls (306,132.24) and (306.67,130) .. (307.5,130) -- (391.5,130) .. controls (392.33,130) and (393,132.24) .. (393,135) .. controls (393,137.76) and (392.33,140) .. (391.5,140) .. controls (390.67,140) and (390,137.76) .. (390,135) .. controls (390,132.24) and (390.67,130) .. (391.5,130) ;
\draw    (391.5,140) -- (383,150) ;
\draw    (307.5,140) -- (316,150) ;
\draw   (380.5,152.5) .. controls (380.5,151.12) and (381.62,150) .. (383,150) .. controls (384.38,150) and (385.5,151.12) .. (385.5,152.5) .. controls (385.5,153.88) and (384.38,155) .. (383,155) .. controls (381.62,155) and (380.5,153.88) .. (380.5,152.5) -- cycle ;
\draw   (313.5,152.5) .. controls (313.5,151.12) and (314.62,150) .. (316,150) .. controls (317.38,150) and (318.5,151.12) .. (318.5,152.5) .. controls (318.5,153.88) and (317.38,155) .. (316,155) .. controls (314.62,155) and (313.5,153.88) .. (313.5,152.5) -- cycle ;

\draw   (481.5,140) -- (397.5,140) .. controls (396.67,140) and (396,137.76) .. (396,135) .. controls (396,132.24) and (396.67,130) .. (397.5,130) -- (481.5,130) .. controls (482.33,130) and (483,132.24) .. (483,135) .. controls (483,137.76) and (482.33,140) .. (481.5,140) .. controls (480.67,140) and (480,137.76) .. (480,135) .. controls (480,132.24) and (480.67,130) .. (481.5,130) ;
\draw    (481.5,140) -- (473,150) ;
\draw    (397.5,140) -- (406,150) ;
\draw   (470.5,152.5) .. controls (470.5,151.12) and (471.62,150) .. (473,150) .. controls (474.38,150) and (475.5,151.12) .. (475.5,152.5) .. controls (475.5,153.88) and (474.38,155) .. (473,155) .. controls (471.62,155) and (470.5,153.88) .. (470.5,152.5) -- cycle ;
\draw   (403.5,152.5) .. controls (403.5,151.12) and (404.62,150) .. (406,150) .. controls (407.38,150) and (408.5,151.12) .. (408.5,152.5) .. controls (408.5,153.88) and (407.38,155) .. (406,155) .. controls (404.62,155) and (403.5,153.88) .. (403.5,152.5) -- cycle ;

\draw   (570.5,140) -- (486.5,140) .. controls (485.67,140) and (485,137.76) .. (485,135) .. controls (485,132.24) and (485.67,130) .. (486.5,130) -- (570.5,130) .. controls (571.33,130) and (572,132.24) .. (572,135) .. controls (572,137.76) and (571.33,140) .. (570.5,140) .. controls (569.67,140) and (569,137.76) .. (569,135) .. controls (569,132.24) and (569.67,130) .. (570.5,130) ;
\draw    (570.5,140) -- (562,150) ;
\draw    (486.5,140) -- (495,150) ;
\draw   (559.5,152.5) .. controls (559.5,151.12) and (560.62,150) .. (562,150) .. controls (563.38,150) and (564.5,151.12) .. (564.5,152.5) .. controls (564.5,153.88) and (563.38,155) .. (562,155) .. controls (560.62,155) and (559.5,153.88) .. (559.5,152.5) -- cycle ;
\draw   (492.5,152.5) .. controls (492.5,151.12) and (493.62,150) .. (495,150) .. controls (496.38,150) and (497.5,151.12) .. (497.5,152.5) .. controls (497.5,153.88) and (496.38,155) .. (495,155) .. controls (493.62,155) and (492.5,153.88) .. (492.5,152.5) -- cycle ;

\draw   (660.5,140) -- (576.5,140) .. controls (575.67,140) and (575,137.76) .. (575,135) .. controls (575,132.24) and (575.67,130) .. (576.5,130) -- (660.5,130) .. controls (661.33,130) and (662,132.24) .. (662,135) .. controls (662,137.76) and (661.33,140) .. (660.5,140) .. controls (659.67,140) and (659,137.76) .. (659,135) .. controls (659,132.24) and (659.67,130) .. (660.5,130) ;
\draw    (576.5,140) -- (585,150) ;
\draw   (582.5,152.5) .. controls (582.5,151.12) and (583.62,150) .. (585,150) .. controls (586.38,150) and (587.5,151.12) .. (587.5,152.5) .. controls (587.5,153.88) and (586.38,155) .. (585,155) .. controls (583.62,155) and (582.5,153.88) .. (582.5,152.5) -- cycle ;
\draw   (230.5,175) -- (146.5,175) .. controls (145.67,175) and (145,172.76) .. (145,170) .. controls (145,167.24) and (145.67,165) .. (146.5,165) -- (230.5,165) .. controls (231.33,165) and (232,167.24) .. (232,170) .. controls (232,172.76) and (231.33,175) .. (230.5,175) .. controls (229.67,175) and (229,172.76) .. (229,170) .. controls (229,167.24) and (229.67,165) .. (230.5,165) ;
\draw    (230.5,175) -- (222,185) ;
\draw   (219.5,187.5) .. controls (219.5,186.12) and (220.62,185) .. (222,185) .. controls (223.38,185) and (224.5,186.12) .. (224.5,187.5) .. controls (224.5,188.88) and (223.38,190) .. (222,190) .. controls (220.62,190) and (219.5,188.88) .. (219.5,187.5) -- cycle ;
\draw   (320.5,175) -- (236.5,175) .. controls (235.67,175) and (235,172.76) .. (235,170) .. controls (235,167.24) and (235.67,165) .. (236.5,165) -- (320.5,165) .. controls (321.33,165) and (322,167.24) .. (322,170) .. controls (322,172.76) and (321.33,175) .. (320.5,175) .. controls (319.67,175) and (319,172.76) .. (319,170) .. controls (319,167.24) and (319.67,165) .. (320.5,165) ;
\draw    (320.5,175) -- (312,185) ;
\draw    (236.5,175) -- (245,185) ;
\draw   (309.5,187.5) .. controls (309.5,186.12) and (310.62,185) .. (312,185) .. controls (313.38,185) and (314.5,186.12) .. (314.5,187.5) .. controls (314.5,188.88) and (313.38,190) .. (312,190) .. controls (310.62,190) and (309.5,188.88) .. (309.5,187.5) -- cycle ;
\draw   (242.5,187.5) .. controls (242.5,186.12) and (243.62,185) .. (245,185) .. controls (246.38,185) and (247.5,186.12) .. (247.5,187.5) .. controls (247.5,188.88) and (246.38,190) .. (245,190) .. controls (243.62,190) and (242.5,188.88) .. (242.5,187.5) -- cycle ;

\draw   (411.5,175) -- (327.5,175) .. controls (326.67,175) and (326,172.76) .. (326,170) .. controls (326,167.24) and (326.67,165) .. (327.5,165) -- (411.5,165) .. controls (412.33,165) and (413,167.24) .. (413,170) .. controls (413,172.76) and (412.33,175) .. (411.5,175) .. controls (410.67,175) and (410,172.76) .. (410,170) .. controls (410,167.24) and (410.67,165) .. (411.5,165) ;
\draw    (411.5,175) -- (403,185) ;
\draw    (327.5,175) -- (336,185) ;
\draw   (400.5,187.5) .. controls (400.5,186.12) and (401.62,185) .. (403,185) .. controls (404.38,185) and (405.5,186.12) .. (405.5,187.5) .. controls (405.5,188.88) and (404.38,190) .. (403,190) .. controls (401.62,190) and (400.5,188.88) .. (400.5,187.5) -- cycle ;
\draw   (333.5,187.5) .. controls (333.5,186.12) and (334.62,185) .. (336,185) .. controls (337.38,185) and (338.5,186.12) .. (338.5,187.5) .. controls (338.5,188.88) and (337.38,190) .. (336,190) .. controls (334.62,190) and (333.5,188.88) .. (333.5,187.5) -- cycle ;

\draw   (501.5,175) -- (417.5,175) .. controls (416.67,175) and (416,172.76) .. (416,170) .. controls (416,167.24) and (416.67,165) .. (417.5,165) -- (501.5,165) .. controls (502.33,165) and (503,167.24) .. (503,170) .. controls (503,172.76) and (502.33,175) .. (501.5,175) .. controls (500.67,175) and (500,172.76) .. (500,170) .. controls (500,167.24) and (500.67,165) .. (501.5,165) ;
\draw    (501.5,175) -- (493,185) ;
\draw    (417.5,175) -- (426,185) ;
\draw   (490.5,187.5) .. controls (490.5,186.12) and (491.62,185) .. (493,185) .. controls (494.38,185) and (495.5,186.12) .. (495.5,187.5) .. controls (495.5,188.88) and (494.38,190) .. (493,190) .. controls (491.62,190) and (490.5,188.88) .. (490.5,187.5) -- cycle ;
\draw   (423.5,187.5) .. controls (423.5,186.12) and (424.62,185) .. (426,185) .. controls (427.38,185) and (428.5,186.12) .. (428.5,187.5) .. controls (428.5,188.88) and (427.38,190) .. (426,190) .. controls (424.62,190) and (423.5,188.88) .. (423.5,187.5) -- cycle ;

\draw   (590.5,175) -- (506.5,175) .. controls (505.67,175) and (505,172.76) .. (505,170) .. controls (505,167.24) and (505.67,165) .. (506.5,165) -- (590.5,165) .. controls (591.33,165) and (592,167.24) .. (592,170) .. controls (592,172.76) and (591.33,175) .. (590.5,175) .. controls (589.67,175) and (589,172.76) .. (589,170) .. controls (589,167.24) and (589.67,165) .. (590.5,165) ;
\draw    (590.5,175) -- (582,185) ;
\draw    (506.5,175) -- (515,185) ;
\draw   (579.5,187.5) .. controls (579.5,186.12) and (580.62,185) .. (582,185) .. controls (583.38,185) and (584.5,186.12) .. (584.5,187.5) .. controls (584.5,188.88) and (583.38,190) .. (582,190) .. controls (580.62,190) and (579.5,188.88) .. (579.5,187.5) -- cycle ;
\draw   (512.5,187.5) .. controls (512.5,186.12) and (513.62,185) .. (515,185) .. controls (516.38,185) and (517.5,186.12) .. (517.5,187.5) .. controls (517.5,188.88) and (516.38,190) .. (515,190) .. controls (513.62,190) and (512.5,188.88) .. (512.5,187.5) -- cycle ;

\draw   (680.5,175) -- (596.5,175) .. controls (595.67,175) and (595,172.76) .. (595,170) .. controls (595,167.24) and (595.67,165) .. (596.5,165) -- (680.5,165) .. controls (681.33,165) and (682,167.24) .. (682,170) .. controls (682,172.76) and (681.33,175) .. (680.5,175) .. controls (679.67,175) and (679,172.76) .. (679,170) .. controls (679,167.24) and (679.67,165) .. (680.5,165) ;
\draw    (596.5,175) -- (605,185) ;
\draw   (602.5,187.5) .. controls (602.5,186.12) and (603.62,185) .. (605,185) .. controls (606.38,185) and (607.5,186.12) .. (607.5,187.5) .. controls (607.5,188.88) and (606.38,190) .. (605,190) .. controls (603.62,190) and (602.5,188.88) .. (602.5,187.5) -- cycle ;
\draw   (250.5,210) -- (166.5,210) .. controls (165.67,210) and (165,207.76) .. (165,205) .. controls (165,202.24) and (165.67,200) .. (166.5,200) -- (250.5,200) .. controls (251.33,200) and (252,202.24) .. (252,205) .. controls (252,207.76) and (251.33,210) .. (250.5,210) .. controls (249.67,210) and (249,207.76) .. (249,205) .. controls (249,202.24) and (249.67,200) .. (250.5,200) ;
\draw    (250.5,210) -- (242,220) ;
\draw   (239.5,222.5) .. controls (239.5,221.12) and (240.62,220) .. (242,220) .. controls (243.38,220) and (244.5,221.12) .. (244.5,222.5) .. controls (244.5,223.88) and (243.38,225) .. (242,225) .. controls (240.62,225) and (239.5,223.88) .. (239.5,222.5) -- cycle ;
\draw   (340.5,210) -- (256.5,210) .. controls (255.67,210) and (255,207.76) .. (255,205) .. controls (255,202.24) and (255.67,200) .. (256.5,200) -- (340.5,200) .. controls (341.33,200) and (342,202.24) .. (342,205) .. controls (342,207.76) and (341.33,210) .. (340.5,210) .. controls (339.67,210) and (339,207.76) .. (339,205) .. controls (339,202.24) and (339.67,200) .. (340.5,200) ;
\draw    (340.5,210) -- (332,220) ;
\draw    (256.5,210) -- (265,220) ;
\draw   (329.5,222.5) .. controls (329.5,221.12) and (330.62,220) .. (332,220) .. controls (333.38,220) and (334.5,221.12) .. (334.5,222.5) .. controls (334.5,223.88) and (333.38,225) .. (332,225) .. controls (330.62,225) and (329.5,223.88) .. (329.5,222.5) -- cycle ;
\draw   (262.5,222.5) .. controls (262.5,221.12) and (263.62,220) .. (265,220) .. controls (266.38,220) and (267.5,221.12) .. (267.5,222.5) .. controls (267.5,223.88) and (266.38,225) .. (265,225) .. controls (263.62,225) and (262.5,223.88) .. (262.5,222.5) -- cycle ;

\draw   (431.5,210) -- (347.5,210) .. controls (346.67,210) and (346,207.76) .. (346,205) .. controls (346,202.24) and (346.67,200) .. (347.5,200) -- (431.5,200) .. controls (432.33,200) and (433,202.24) .. (433,205) .. controls (433,207.76) and (432.33,210) .. (431.5,210) .. controls (430.67,210) and (430,207.76) .. (430,205) .. controls (430,202.24) and (430.67,200) .. (431.5,200) ;
\draw    (431.5,210) -- (423,220) ;
\draw    (347.5,210) -- (356,220) ;
\draw   (420.5,222.5) .. controls (420.5,221.12) and (421.62,220) .. (423,220) .. controls (424.38,220) and (425.5,221.12) .. (425.5,222.5) .. controls (425.5,223.88) and (424.38,225) .. (423,225) .. controls (421.62,225) and (420.5,223.88) .. (420.5,222.5) -- cycle ;
\draw   (353.5,222.5) .. controls (353.5,221.12) and (354.62,220) .. (356,220) .. controls (357.38,220) and (358.5,221.12) .. (358.5,222.5) .. controls (358.5,223.88) and (357.38,225) .. (356,225) .. controls (354.62,225) and (353.5,223.88) .. (353.5,222.5) -- cycle ;

\draw   (521.5,210) -- (437.5,210) .. controls (436.67,210) and (436,207.76) .. (436,205) .. controls (436,202.24) and (436.67,200) .. (437.5,200) -- (521.5,200) .. controls (522.33,200) and (523,202.24) .. (523,205) .. controls (523,207.76) and (522.33,210) .. (521.5,210) .. controls (520.67,210) and (520,207.76) .. (520,205) .. controls (520,202.24) and (520.67,200) .. (521.5,200) ;
\draw    (521.5,210) -- (513,220) ;
\draw    (437.5,210) -- (446,220) ;
\draw   (510.5,222.5) .. controls (510.5,221.12) and (511.62,220) .. (513,220) .. controls (514.38,220) and (515.5,221.12) .. (515.5,222.5) .. controls (515.5,223.88) and (514.38,225) .. (513,225) .. controls (511.62,225) and (510.5,223.88) .. (510.5,222.5) -- cycle ;
\draw   (443.5,222.5) .. controls (443.5,221.12) and (444.62,220) .. (446,220) .. controls (447.38,220) and (448.5,221.12) .. (448.5,222.5) .. controls (448.5,223.88) and (447.38,225) .. (446,225) .. controls (444.62,225) and (443.5,223.88) .. (443.5,222.5) -- cycle ;

\draw   (610.5,210) -- (526.5,210) .. controls (525.67,210) and (525,207.76) .. (525,205) .. controls (525,202.24) and (525.67,200) .. (526.5,200) -- (610.5,200) .. controls (611.33,200) and (612,202.24) .. (612,205) .. controls (612,207.76) and (611.33,210) .. (610.5,210) .. controls (609.67,210) and (609,207.76) .. (609,205) .. controls (609,202.24) and (609.67,200) .. (610.5,200) ;
\draw    (610.5,210) -- (602,220) ;
\draw    (526.5,210) -- (535,220) ;
\draw   (599.5,222.5) .. controls (599.5,221.12) and (600.62,220) .. (602,220) .. controls (603.38,220) and (604.5,221.12) .. (604.5,222.5) .. controls (604.5,223.88) and (603.38,225) .. (602,225) .. controls (600.62,225) and (599.5,223.88) .. (599.5,222.5) -- cycle ;
\draw   (532.5,222.5) .. controls (532.5,221.12) and (533.62,220) .. (535,220) .. controls (536.38,220) and (537.5,221.12) .. (537.5,222.5) .. controls (537.5,223.88) and (536.38,225) .. (535,225) .. controls (533.62,225) and (532.5,223.88) .. (532.5,222.5) -- cycle ;

\draw   (700.5,210) -- (616.5,210) .. controls (615.67,210) and (615,207.76) .. (615,205) .. controls (615,202.24) and (615.67,200) .. (616.5,200) -- (700.5,200) .. controls (701.33,200) and (702,202.24) .. (702,205) .. controls (702,207.76) and (701.33,210) .. (700.5,210) .. controls (699.67,210) and (699,207.76) .. (699,205) .. controls (699,202.24) and (699.67,200) .. (700.5,200) ;
\draw    (616.5,210) -- (625,220) ;
\draw   (622.5,222.5) .. controls (622.5,221.12) and (623.62,220) .. (625,220) .. controls (626.38,220) and (627.5,221.12) .. (627.5,222.5) .. controls (627.5,223.88) and (626.38,225) .. (625,225) .. controls (623.62,225) and (622.5,223.88) .. (622.5,222.5) -- cycle ;
\draw    (410,160) -- (503.85,26.64) ;
\draw [shift={(505,25)}, rotate = 125.13] [color={rgb, 255:red, 0; green, 0; blue, 0 }  ][line width=0.75]    (10.93,-3.29) .. controls (6.95,-1.4) and (3.31,-0.3) .. (0,0) .. controls (3.31,0.3) and (6.95,1.4) .. (10.93,3.29)   ;
\draw  [fill={rgb, 255:red, 0; green, 0; blue, 0 }  ,fill opacity=1 ] (290,35) .. controls (290,32.24) and (292.24,30) .. (295,30) .. controls (297.76,30) and (300,32.24) .. (300,35) .. controls (300,37.76) and (297.76,40) .. (295,40) .. controls (292.24,40) and (290,37.76) .. (290,35) -- cycle ;
\draw  [fill={rgb, 255:red, 0; green, 0; blue, 0 }  ,fill opacity=1 ] (300,55) .. controls (300,52.24) and (302.24,50) .. (305,50) .. controls (307.76,50) and (310,52.24) .. (310,55) .. controls (310,57.76) and (307.76,60) .. (305,60) .. controls (302.24,60) and (300,57.76) .. (300,55) -- cycle ;
\draw  [fill={rgb, 255:red, 0; green, 0; blue, 0 }  ,fill opacity=1 ] (310,75) .. controls (310,72.24) and (312.24,70) .. (315,70) .. controls (317.76,70) and (320,72.24) .. (320,75) .. controls (320,77.76) and (317.76,80) .. (315,80) .. controls (312.24,80) and (310,77.76) .. (310,75) -- cycle ;
\draw  [fill={rgb, 255:red, 0; green, 0; blue, 0 }  ,fill opacity=1 ] (390,225) .. controls (390,222.24) and (392.24,220) .. (395,220) .. controls (397.76,220) and (400,222.24) .. (400,225) .. controls (400,227.76) and (397.76,230) .. (395,230) .. controls (392.24,230) and (390,227.76) .. (390,225) -- cycle ;
\draw  [fill={rgb, 255:red, 0; green, 0; blue, 0 }  ,fill opacity=1 ] (400,245) .. controls (400,242.24) and (402.24,240) .. (405,240) .. controls (407.76,240) and (410,242.24) .. (410,245) .. controls (410,247.76) and (407.76,250) .. (405,250) .. controls (402.24,250) and (400,247.76) .. (400,245) -- cycle ;
\draw  [fill={rgb, 255:red, 0; green, 0; blue, 0 }  ,fill opacity=1 ] (410,265) .. controls (410,262.24) and (412.24,260) .. (415,260) .. controls (417.76,260) and (420,262.24) .. (420,265) .. controls (420,267.76) and (417.76,270) .. (415,270) .. controls (412.24,270) and (410,267.76) .. (410,265) -- cycle ;
\draw  [draw opacity=0][fill={rgb, 255:red, 74; green, 144; blue, 226 }  ,fill opacity=0.22 ] (454.54,96.81) .. controls (464.19,107.8) and (455.34,141.52) .. (433.37,175.23) .. controls (433.02,175.77) and (432.67,176.3) .. (432.32,176.84) -- (410,160) -- cycle ; \draw  [color={rgb, 255:red, 74; green, 144; blue, 226 }  ,draw opacity=1 ] (454.54,96.81) .. controls (464.19,107.8) and (455.34,141.52) .. (433.37,175.23) .. controls (433.02,175.77) and (432.67,176.3) .. (432.32,176.84) ;
\draw [color={rgb, 255:red, 208; green, 2; blue, 27 }  ,draw opacity=1 ]   (410,120) .. controls (311.49,121.99) and (339.04,186.19) .. (430.93,176.98) ;
\draw [shift={(432.32,176.84)}, rotate = 173.86] [color={rgb, 255:red, 208; green, 2; blue, 27 }  ,draw opacity=1 ][line width=0.75]    (10.93,-3.29) .. controls (6.95,-1.4) and (3.31,-0.3) .. (0,0) .. controls (3.31,0.3) and (6.95,1.4) .. (10.93,3.29)   ;
\draw  [fill={rgb, 255:red, 0; green, 0; blue, 0 }  ,fill opacity=1 ] (110,155) .. controls (110,152.24) and (112.24,150) .. (115,150) .. controls (117.76,150) and (120,152.24) .. (120,155) .. controls (120,157.76) and (117.76,160) .. (115,160) .. controls (112.24,160) and (110,157.76) .. (110,155) -- cycle ;
\draw  [fill={rgb, 255:red, 0; green, 0; blue, 0 }  ,fill opacity=1 ] (90,155) .. controls (90,152.24) and (92.24,150) .. (95,150) .. controls (97.76,150) and (100,152.24) .. (100,155) .. controls (100,157.76) and (97.76,160) .. (95,160) .. controls (92.24,160) and (90,157.76) .. (90,155) -- cycle ;
\draw  [fill={rgb, 255:red, 0; green, 0; blue, 0 }  ,fill opacity=1 ] (70,155) .. controls (70,152.24) and (72.24,150) .. (75,150) .. controls (77.76,150) and (80,152.24) .. (80,155) .. controls (80,157.76) and (77.76,160) .. (75,160) .. controls (72.24,160) and (70,157.76) .. (70,155) -- cycle ;
\draw  [fill={rgb, 255:red, 0; green, 0; blue, 0 }  ,fill opacity=1 ] (720,150) .. controls (720,147.24) and (722.24,145) .. (725,145) .. controls (727.76,145) and (730,147.24) .. (730,150) .. controls (730,152.76) and (727.76,155) .. (725,155) .. controls (722.24,155) and (720,152.76) .. (720,150) -- cycle ;
\draw  [fill={rgb, 255:red, 0; green, 0; blue, 0 }  ,fill opacity=1 ] (700,150) .. controls (700,147.24) and (702.24,145) .. (705,145) .. controls (707.76,145) and (710,147.24) .. (710,150) .. controls (710,152.76) and (707.76,155) .. (705,155) .. controls (702.24,155) and (700,152.76) .. (700,150) -- cycle ;
\draw  [fill={rgb, 255:red, 0; green, 0; blue, 0 }  ,fill opacity=1 ] (680,150) .. controls (680,147.24) and (682.24,145) .. (685,145) .. controls (687.76,145) and (690,147.24) .. (690,150) .. controls (690,152.76) and (687.76,155) .. (685,155) .. controls (682.24,155) and (680,152.76) .. (680,150) -- cycle ;
\draw    (192,85) -- (283,85) ;
\draw [shift={(285,85)}, rotate = 180] [color={rgb, 255:red, 0; green, 0; blue, 0 }  ][line width=0.75]    (10.93,-3.29) .. controls (6.95,-1.4) and (3.31,-0.3) .. (0,0) .. controls (3.31,0.3) and (6.95,1.4) .. (10.93,3.29)   ;
\draw [shift={(190,85)}, rotate = 0] [color={rgb, 255:red, 0; green, 0; blue, 0 }  ][line width=0.75]    (10.93,-3.29) .. controls (6.95,-1.4) and (3.31,-0.3) .. (0,0) .. controls (3.31,0.3) and (6.95,1.4) .. (10.93,3.29)   ;
\draw    (95.89,101.79) -- (114.11,138.21) ;
\draw [shift={(115,140)}, rotate = 243.43] [color={rgb, 255:red, 0; green, 0; blue, 0 }  ][line width=0.75]    (10.93,-3.29) .. controls (6.95,-1.4) and (3.31,-0.3) .. (0,0) .. controls (3.31,0.3) and (6.95,1.4) .. (10.93,3.29)   ;
\draw [shift={(95,100)}, rotate = 63.43] [color={rgb, 255:red, 0; green, 0; blue, 0 }  ][line width=0.75]    (10.93,-3.29) .. controls (6.95,-1.4) and (3.31,-0.3) .. (0,0) .. controls (3.31,0.3) and (6.95,1.4) .. (10.93,3.29)   ;
\draw    (240,265) -- (240,237) ;
\draw [shift={(240,235)}, rotate = 90] [color={rgb, 255:red, 0; green, 0; blue, 0 }  ][line width=0.75]    (10.93,-3.29) .. controls (6.95,-1.4) and (3.31,-0.3) .. (0,0) .. controls (3.31,0.3) and (6.95,1.4) .. (10.93,3.29)   ;
\draw    (265,263) -- (265,235) ;
\draw [shift={(265,265)}, rotate = 270] [color={rgb, 255:red, 0; green, 0; blue, 0 }  ][line width=0.75]    (10.93,-3.29) .. controls (6.95,-1.4) and (3.31,-0.3) .. (0,0) .. controls (3.31,0.3) and (6.95,1.4) .. (10.93,3.29)   ;
\draw    (600,265) -- (600,237) ;
\draw [shift={(600,235)}, rotate = 90] [color={rgb, 255:red, 0; green, 0; blue, 0 }  ][line width=0.75]    (10.93,-3.29) .. controls (6.95,-1.4) and (3.31,-0.3) .. (0,0) .. controls (3.31,0.3) and (6.95,1.4) .. (10.93,3.29)   ;
\draw    (625,263) -- (625,235) ;
\draw [shift={(625,265)}, rotate = 270] [color={rgb, 255:red, 0; green, 0; blue, 0 }  ][line width=0.75]    (10.93,-3.29) .. controls (6.95,-1.4) and (3.31,-0.3) .. (0,0) .. controls (3.31,0.3) and (6.95,1.4) .. (10.93,3.29)   ;
\draw   (326.3,-14.75) -- (372.38,31.33) -- (383.15,20.16) -- (392.33,73.22) -- (340.07,64.83) -- (350.84,53.66) -- (304.76,7.58) -- cycle ;
\draw   (394.03,53.55) -- (442.47,5.12) -- (430.12,-5.53) -- (487.11,-16.52) -- (479.52,37.08) -- (467.17,26.42) -- (418.73,74.86) -- cycle ;

\draw (415.8,151.12) node [anchor=north west][inner sep=0.75pt]  [rotate=-301.76]  {$\textcolor[rgb]{0.29,0.56,0.89}{\frac{\pi }{2}}\textcolor[rgb]{0.29,0.56,0.89}{-\theta }$};
\draw (355,137.4) node [anchor=north west][inner sep=0.75pt]    {$\textcolor[rgb]{0.82,0.01,0.11}{\varphi }$};
\draw (236,67.4) node [anchor=north west][inner sep=0.75pt]    {$a$};
\draw (78,112.4) node [anchor=north west][inner sep=0.75pt]    {$a$};
\draw (272,242.4) node [anchor=north west][inner sep=0.75pt]    {$b_{\alpha ,1}$};
\draw (186,240.4) node [anchor=north west][inner sep=0.75pt]    {$a_{\alpha ,1}$};
\draw (631,240.4) node [anchor=north west][inner sep=0.75pt]    {$b_{\alpha ,m}$};
\draw (535,237.4) node [anchor=north west][inner sep=0.75pt]    {$a_{\alpha ,m}$};
\draw (325.39,-15.09) node [anchor=north west][inner sep=0.75pt]  [rotate=-45]  {$a_{\beta }( \theta ,\varphi )$};
\draw (397.42,50.45) node [anchor=north west][inner sep=0.75pt]  [rotate=-315]  {$b_{\beta }( \theta ,\varphi )$};

\end{tikzpicture}

%% file: fig4.tex
\tikzset{every picture/.style={line width=0.75pt}} 

\begin{tikzpicture}[x=0.75pt,y=0.75pt,yscale=-0.75,xscale=0.75]

\draw [line width=3]    (350,10) -- (350.5,140) ;
\draw  [fill={rgb, 255:red, 0; green, 0; blue, 0 }  ,fill opacity=1 ] (340,15) -- (360,15) -- (360,35) -- (340,35) -- cycle ;
\draw  [fill={rgb, 255:red, 0; green, 0; blue, 0 }  ,fill opacity=1 ] (340.25,47.5) -- (360.25,47.5) -- (360.25,67.5) -- (340.25,67.5) -- cycle ;
\draw  [fill={rgb, 255:red, 0; green, 0; blue, 0 }  ,fill opacity=1 ] (340,80) -- (360,80) -- (360,100) -- (340,100) -- cycle ;
\draw  [fill={rgb, 255:red, 0; green, 0; blue, 0 }  ,fill opacity=1 ] (340,112) -- (360,112) -- (360,132) -- (340,132) -- cycle ;

\draw [line width=3]    (394.75,10) -- (395.25,140) ;
\draw  [fill={rgb, 255:red, 0; green, 0; blue, 0 }  ,fill opacity=1 ] (384.75,15) -- (404.75,15) -- (404.75,35) -- (384.75,35) -- cycle ;
\draw  [fill={rgb, 255:red, 0; green, 0; blue, 0 }  ,fill opacity=1 ] (385,47.5) -- (405,47.5) -- (405,67.5) -- (385,67.5) -- cycle ;
\draw  [fill={rgb, 255:red, 0; green, 0; blue, 0 }  ,fill opacity=1 ] (384.75,80) -- (404.75,80) -- (404.75,100) -- (384.75,100) -- cycle ;
\draw  [fill={rgb, 255:red, 0; green, 0; blue, 0 }  ,fill opacity=1 ] (384.75,112) -- (404.75,112) -- (404.75,132) -- (384.75,132) -- cycle ;

\draw [line width=3]    (439.75,10) -- (440.25,140) ;
\draw  [fill={rgb, 255:red, 0; green, 0; blue, 0 }  ,fill opacity=1 ] (429.75,15) -- (449.75,15) -- (449.75,35) -- (429.75,35) -- cycle ;
\draw  [fill={rgb, 255:red, 0; green, 0; blue, 0 }  ,fill opacity=1 ] (430,47.5) -- (450,47.5) -- (450,67.5) -- (430,67.5) -- cycle ;
\draw  [fill={rgb, 255:red, 0; green, 0; blue, 0 }  ,fill opacity=1 ] (429.75,80) -- (449.75,80) -- (449.75,100) -- (429.75,100) -- cycle ;
\draw  [fill={rgb, 255:red, 0; green, 0; blue, 0 }  ,fill opacity=1 ] (429.75,112) -- (449.75,112) -- (449.75,132) -- (429.75,132) -- cycle ;

\draw [line width=3]    (305,10) -- (305.5,140) ;
\draw  [fill={rgb, 255:red, 0; green, 0; blue, 0 }  ,fill opacity=1 ] (295,15) -- (315,15) -- (315,35) -- (295,35) -- cycle ;
\draw  [fill={rgb, 255:red, 0; green, 0; blue, 0 }  ,fill opacity=1 ] (295.25,47.5) -- (315.25,47.5) -- (315.25,67.5) -- (295.25,67.5) -- cycle ;
\draw  [fill={rgb, 255:red, 0; green, 0; blue, 0 }  ,fill opacity=1 ] (295,80) -- (315,80) -- (315,100) -- (295,100) -- cycle ;
\draw  [fill={rgb, 255:red, 0; green, 0; blue, 0 }  ,fill opacity=1 ] (295,112) -- (315,112) -- (315,132) -- (295,132) -- cycle ;

\draw [line width=3]    (259.75,10) -- (260.25,140) ;
\draw  [fill={rgb, 255:red, 0; green, 0; blue, 0 }  ,fill opacity=1 ] (249.75,15) -- (269.75,15) -- (269.75,35) -- (249.75,35) -- cycle ;
\draw  [fill={rgb, 255:red, 0; green, 0; blue, 0 }  ,fill opacity=1 ] (250,47.5) -- (270,47.5) -- (270,67.5) -- (250,67.5) -- cycle ;
\draw  [fill={rgb, 255:red, 0; green, 0; blue, 0 }  ,fill opacity=1 ] (249.75,80) -- (269.75,80) -- (269.75,100) -- (249.75,100) -- cycle ;
\draw  [fill={rgb, 255:red, 0; green, 0; blue, 0 }  ,fill opacity=1 ] (249.75,112) -- (269.75,112) -- (269.75,132) -- (249.75,132) -- cycle ;

\draw   (325,247) -- (375,247) -- (375,297) -- (325,297) -- cycle ;
\draw [line width=2.25]    (315,252) -- (325,252) ;
\draw [line width=2.25]    (315,262) -- (325,262) ;
\draw [line width=2.25]    (315,272) -- (325,272) ;
\draw [line width=2.25]    (315,282) -- (325,282) ;
\draw [line width=2.25]    (315,292) -- (325,292) ;
\draw [line width=2.25]    (375,252) -- (385,252) ;
\draw [line width=2.25]    (375,262) -- (385,262) ;
\draw [line width=2.25]    (375,282) -- (385,282) ;
\draw [line width=2.25]    (375,292) -- (385,292) ;
\draw [line width=2.25]    (375,272) -- (385,272) ;

\draw [line width=2.25]    (330,237) -- (330,247) ;
\draw [line width=2.25]    (330,297) -- (330,307) ;
\draw [line width=2.25]    (340,237) -- (340,247) ;
\draw [line width=2.25]    (350,237) -- (350,247) ;
\draw [line width=2.25]    (360,237) -- (360,247) ;
\draw [line width=2.25]    (370,237) -- (370,247) ;
\draw [line width=2.25]    (340,297) -- (340,307) ;
\draw [line width=2.25]    (350,297) -- (350,307) ;
\draw [line width=2.25]    (360,297) -- (360,307) ;
\draw [line width=2.25]    (370,297) -- (370,307) ;
\draw [line width=2.25]    (335,297) -- (335,307) ;
\draw [line width=2.25]    (345,297) -- (345,307) ;
\draw [line width=2.25]    (355,297) -- (355,307) ;
\draw [line width=2.25]    (365,297) -- (365,307) ;
\draw [line width=2.25]    (335,237) -- (335,247) ;
\draw [line width=2.25]    (345,237) -- (345,247) ;
\draw [line width=2.25]    (355,237) -- (355,247) ;
\draw [line width=2.25]    (365,237) -- (365,247) ;
\draw [line width=2.25]    (325,267) -- (315,267) ;
\draw [line width=2.25]    (325,257) -- (315,257) ;
\draw [line width=2.25]    (325,277) -- (315,277) ;
\draw [line width=2.25]    (325,287) -- (315,287) ;
\draw [line width=2.25]    (385,257) -- (375,257) ;
\draw [line width=2.25]    (385,267) -- (375,267) ;
\draw [line width=2.25]    (385,277) -- (375,277) ;
\draw [line width=2.25]    (385,287) -- (375,287) ;

\draw  [dash pattern={on 0.84pt off 2.51pt}]  (260.25,140) -- (275,140) -- (275,288) -- (315,287) ;
\draw  [dash pattern={on 0.84pt off 2.51pt}]  (305.71,140.17) -- (290,140) -- (290,266) -- (316,267) ;
\draw   (253.12,161.4) -- (260.45,149.1) -- (268.33,161.06) -- (253.12,161.4) -- cycle (260.93,170.32) -- (260.73,161.23) (252.85,149.27) -- (268.06,148.93) (260.45,149.1) -- (260.25,140) ;
\draw   (255.15,173.55) -- (260.93,180) -- (266.72,173.55) -- (255.15,173.55) -- cycle (260.93,170.32) -- (260.93,173.55) ;

\draw   (298.59,161.57) -- (305.92,149.27) -- (313.8,161.23) -- (298.59,161.57) -- cycle (306.39,170.49) -- (306.19,161.4) (298.31,149.44) -- (313.52,149.1) (305.92,149.27) -- (305.71,140.17) ;
\draw   (300.61,173.72) -- (306.39,180.17) -- (312.18,173.72) -- (300.61,173.72) -- cycle (306.39,170.49) -- (306.39,173.72) ;

\draw   (343.59,161.57) -- (350.92,149.27) -- (358.8,161.23) -- (343.59,161.57) -- cycle (351.39,170.49) -- (351.19,161.4) (343.31,149.44) -- (358.52,149.1) (350.92,149.27) -- (350.71,140.17) ;
\draw   (345.61,173.72) -- (351.39,180.17) -- (357.18,173.72) -- (345.61,173.72) -- cycle (351.39,170.49) -- (351.39,173.72) ;

\draw   (388.59,161.57) -- (395.92,149.27) -- (403.8,161.23) -- (388.59,161.57) -- cycle (396.39,170.49) -- (396.19,161.4) (388.31,149.44) -- (403.52,149.1) (395.92,149.27) -- (395.71,140.17) ;
\draw   (390.61,173.72) -- (396.39,180.17) -- (402.18,173.72) -- (390.61,173.72) -- cycle (396.39,170.49) -- (396.39,173.72) ;

\draw   (433.59,161.57) -- (440.92,149.27) -- (448.8,161.23) -- (433.59,161.57) -- cycle (441.39,170.49) -- (441.19,161.4) (433.31,149.44) -- (448.52,149.1) (440.92,149.27) -- (440.71,140.17) ;
\draw   (435.61,173.72) -- (441.39,180.17) -- (447.18,173.72) -- (435.61,173.72) -- cycle (441.39,170.49) -- (441.39,173.72) ;

\draw (340,259.4) node [anchor=north west][inner sep=0.75pt]    {$\!\mu$C};
\draw (301,187) node [anchor=north west][inner sep=0.75pt]   [align=left] {Phase shifters};

\end{tikzpicture}

%% file: fig3.tex
\tikzset{every picture/.style={line width=0.75pt}} 

\begin{tikzpicture}[x=0.75pt,y=0.75pt,yscale=-0.75,xscale=0.75]

\draw    (398,208) -- (423,185) -- (428,214) ;
\draw    (368,205) -- (393,175) -- (398,208) ;
\draw    (333,205) -- (350.5,170) -- (368,205) ;
\draw    (428,214) -- (448,201) -- (468,240) ;

\draw    (303,208) -- (278,185) -- (273,214) ;
\draw    (333,205) -- (308,175) -- (303,208) ;
\draw    (273,214) -- (253,201) -- (233,240) ;

\draw    (398,272) -- (423,295) -- (428,266) ;
\draw    (368,275) -- (393,305) -- (398,272) ;
\draw    (333,275) -- (350.5,310) -- (368,275) ;
\draw    (428,266) -- (448,279) -- (468,240) ;

\draw    (303,272) -- (278,295) -- (273,266) ;
\draw    (333,275) -- (308,305) -- (303,272) ;
\draw    (273,266) -- (253,279) -- (233,240) ;

\draw [line width=3]    (350,10) -- (350.5,140) ;
\draw  [fill={rgb, 255:red, 0; green, 0; blue, 0 }  ,fill opacity=1 ] (340,15) -- (360,15) -- (360,35) -- (340,35) -- cycle ;
\draw  [fill={rgb, 255:red, 0; green, 0; blue, 0 }  ,fill opacity=1 ] (340.25,47.5) -- (360.25,47.5) -- (360.25,67.5) -- (340.25,67.5) -- cycle ;
\draw  [fill={rgb, 255:red, 0; green, 0; blue, 0 }  ,fill opacity=1 ] (340,80) -- (360,80) -- (360,100) -- (340,100) -- cycle ;
\draw  [fill={rgb, 255:red, 0; green, 0; blue, 0 }  ,fill opacity=1 ] (340,112) -- (360,112) -- (360,132) -- (340,132) -- cycle ;

\draw [line width=3]    (394.75,10) -- (395.25,140) ;
\draw  [fill={rgb, 255:red, 0; green, 0; blue, 0 }  ,fill opacity=1 ] (384.75,15) -- (404.75,15) -- (404.75,35) -- (384.75,35) -- cycle ;
\draw  [fill={rgb, 255:red, 0; green, 0; blue, 0 }  ,fill opacity=1 ] (385,47.5) -- (405,47.5) -- (405,67.5) -- (385,67.5) -- cycle ;
\draw  [fill={rgb, 255:red, 0; green, 0; blue, 0 }  ,fill opacity=1 ] (384.75,80) -- (404.75,80) -- (404.75,100) -- (384.75,100) -- cycle ;
\draw  [fill={rgb, 255:red, 0; green, 0; blue, 0 }  ,fill opacity=1 ] (384.75,112) -- (404.75,112) -- (404.75,132) -- (384.75,132) -- cycle ;

\draw [line width=3]    (439.75,10) -- (440.25,140) ;
\draw  [fill={rgb, 255:red, 0; green, 0; blue, 0 }  ,fill opacity=1 ] (429.75,15) -- (449.75,15) -- (449.75,35) -- (429.75,35) -- cycle ;
\draw  [fill={rgb, 255:red, 0; green, 0; blue, 0 }  ,fill opacity=1 ] (430,47.5) -- (450,47.5) -- (450,67.5) -- (430,67.5) -- cycle ;
\draw  [fill={rgb, 255:red, 0; green, 0; blue, 0 }  ,fill opacity=1 ] (429.75,80) -- (449.75,80) -- (449.75,100) -- (429.75,100) -- cycle ;
\draw  [fill={rgb, 255:red, 0; green, 0; blue, 0 }  ,fill opacity=1 ] (429.75,112) -- (449.75,112) -- (449.75,132) -- (429.75,132) -- cycle ;

\draw [line width=3]    (350.5,140) .. controls (393,154.25) and (308,158.58) .. (350.5,170) ;
\draw [line width=3]    (395.25,140) .. controls (379,148.58) and (382,170.58) .. (393,175) ;
\draw [line width=3]    (440.25,140) -- (423,185) ;
\draw [line width=3]    (305,10) -- (305.5,140) ;
\draw  [fill={rgb, 255:red, 0; green, 0; blue, 0 }  ,fill opacity=1 ] (295,15) -- (315,15) -- (315,35) -- (295,35) -- cycle ;
\draw  [fill={rgb, 255:red, 0; green, 0; blue, 0 }  ,fill opacity=1 ] (295.25,47.5) -- (315.25,47.5) -- (315.25,67.5) -- (295.25,67.5) -- cycle ;
\draw  [fill={rgb, 255:red, 0; green, 0; blue, 0 }  ,fill opacity=1 ] (295,80) -- (315,80) -- (315,100) -- (295,100) -- cycle ;
\draw  [fill={rgb, 255:red, 0; green, 0; blue, 0 }  ,fill opacity=1 ] (295,112) -- (315,112) -- (315,132) -- (295,132) -- cycle ;

\draw [line width=3]    (305.5,140) .. controls (329,145.58) and (316,169.58) .. (308,175) ;
\draw [line width=3]    (259.75,10) -- (260.25,140) ;
\draw  [fill={rgb, 255:red, 0; green, 0; blue, 0 }  ,fill opacity=1 ] (249.75,15) -- (269.75,15) -- (269.75,35) -- (249.75,35) -- cycle ;
\draw  [fill={rgb, 255:red, 0; green, 0; blue, 0 }  ,fill opacity=1 ] (250,47.5) -- (270,47.5) -- (270,67.5) -- (250,67.5) -- cycle ;
\draw  [fill={rgb, 255:red, 0; green, 0; blue, 0 }  ,fill opacity=1 ] (249.75,80) -- (269.75,80) -- (269.75,100) -- (249.75,100) -- cycle ;
\draw  [fill={rgb, 255:red, 0; green, 0; blue, 0 }  ,fill opacity=1 ] (249.75,112) -- (269.75,112) -- (269.75,132) -- (249.75,132) -- cycle ;

\draw [line width=3]    (260.25,140) -- (278,185) ;
\draw    (405.11,354.91) -- (390.5,354.91) ;
\draw    (405.11,352.73) -- (390.5,352.73) ;
\draw    (390.5,352.73) -- (390.5,345.45) ;
\draw    (405.11,352.73) -- (405.11,345.45) ;
\draw    (397.81,354.91) -- (397.81,360) ;
\draw    (383.19,345.45) -- (390.5,345.45) ;
\draw    (405.11,345.45) -- (412.42,345.45) ;

\draw    (409.68,321.57) -- (416.99,334.17) ;
\draw    (407.79,322.66) -- (415.09,335.26) ;
\draw    (415.09,335.26) -- (408.77,338.9) ;
\draw    (407.79,322.66) -- (401.46,326.3) ;
\draw    (413.34,327.87) -- (417.77,325.32) ;
\draw    (412.42,345.19) -- (408.77,338.9) ;
\draw    (401.46,326.3) -- (397.81,320) ;

\draw    (378.62,334.43) -- (385.93,321.83) ;
\draw    (380.52,335.52) -- (387.82,322.92) ;
\draw    (387.82,322.92) -- (394.15,326.56) ;
\draw    (380.52,335.52) -- (386.84,339.16) ;
\draw    (382.27,328.13) -- (377.84,325.59) ;
\draw    (397.81,320.26) -- (394.15,326.56) ;
\draw    (386.84,339.16) -- (383.19,345.45) ;

\draw    (357.62,354.91) -- (343,354.91) ;
\draw    (357.62,352.73) -- (343,352.73) ;
\draw    (343,352.73) -- (343,345.45) ;
\draw    (357.62,352.73) -- (357.62,345.45) ;
\draw    (350.31,354.91) -- (350.31,360) ;
\draw    (335.7,345.45) -- (343,345.45) ;
\draw    (357.62,345.45) -- (364.92,345.45) ;

\draw    (362.19,321.57) -- (369.5,334.17) ;
\draw    (360.29,322.66) -- (367.6,335.26) ;
\draw    (367.6,335.26) -- (361.27,338.9) ;
\draw    (360.29,322.66) -- (353.96,326.3) ;
\draw    (365.84,327.87) -- (370.27,325.32) ;
\draw    (364.92,345.19) -- (361.27,338.9) ;
\draw    (353.96,326.3) -- (350.31,320) ;

\draw    (331.12,334.43) -- (338.43,321.83) ;
\draw    (333.02,335.52) -- (340.33,322.92) ;
\draw    (340.33,322.92) -- (346.66,326.56) ;
\draw    (333.02,335.52) -- (339.35,339.16) ;
\draw    (334.78,328.13) -- (330.35,325.59) ;
\draw    (350.31,320.26) -- (346.66,326.56) ;
\draw    (339.35,339.16) -- (335.7,345.45) ;

\draw    (310.12,354.91) -- (295.51,354.91) ;
\draw    (310.12,352.73) -- (295.51,352.73) ;
\draw    (295.51,352.73) -- (295.51,345.45) ;
\draw    (310.12,352.73) -- (310.12,345.45) ;
\draw    (302.82,354.91) -- (302.82,360) ;
\draw    (288.2,345.45) -- (295.51,345.45) ;
\draw    (310.12,345.45) -- (317.43,345.45) ;

\draw    (314.7,321.57) -- (322,334.17) ;
\draw    (312.8,322.66) -- (320.1,335.26) ;
\draw    (320.1,335.26) -- (313.78,338.9) ;
\draw    (312.8,322.66) -- (306.47,326.3) ;
\draw    (318.35,327.87) -- (322.78,325.32) ;
\draw    (317.43,345.19) -- (313.78,338.9) ;
\draw    (306.47,326.3) -- (302.82,320) ;

\draw    (283.63,334.43) -- (290.94,321.83) ;
\draw    (285.53,335.52) -- (292.83,322.92) ;
\draw    (292.83,322.92) -- (299.16,326.56) ;
\draw    (285.53,335.52) -- (291.86,339.16) ;
\draw    (287.28,328.13) -- (282.85,325.59) ;
\draw    (302.82,320.26) -- (299.16,326.56) ;
\draw    (291.86,339.16) -- (288.2,345.45) ;

\draw [line width=3]    (382.93,345.19) -- (364.92,345.19) ;
\draw [line width=3]    (335.43,345.45) -- (317.43,345.45) ;
\draw [line width=3]    (350.5,310) -- (350.31,320) ;
\draw [line width=3]    (397.81,320.26) -- (393,305) ;
\draw [line width=3]    (412.42,345.45) -- (435,330) -- (423,295) ;
\draw [line width=3]    (302.82,320.26) -- (308,305) ;
\draw [line width=3]    (278,295) -- (265,330) -- (288.2,345.45) ;
\draw   (325,395) -- (375,395) -- (375,445) -- (325,445) -- cycle ;
\draw [line width=2.25]    (315,400) -- (325,400) ;
\draw [line width=2.25]    (315,410) -- (325,410) ;
\draw [line width=2.25]    (315,420) -- (325,420) ;
\draw [line width=2.25]    (315,430) -- (325,430) ;
\draw [line width=2.25]    (315,440) -- (325,440) ;
\draw [line width=2.25]    (375,400) -- (385,400) ;
\draw [line width=2.25]    (375,410) -- (385,410) ;
\draw [line width=2.25]    (375,430) -- (385,430) ;
\draw [line width=2.25]    (375,440) -- (385,440) ;
\draw [line width=2.25]    (375,420) -- (385,420) ;

\draw [line width=2.25]    (330,385) -- (330,395) ;
\draw [line width=2.25]    (330,445) -- (330,455) ;
\draw [line width=2.25]    (340,385) -- (340,395) ;
\draw [line width=2.25]    (350,385) -- (350,395) ;
\draw [line width=2.25]    (360,385) -- (360,395) ;
\draw [line width=2.25]    (370,385) -- (370,395) ;
\draw [line width=2.25]    (340,445) -- (340,455) ;
\draw [line width=2.25]    (350,445) -- (350,455) ;
\draw [line width=2.25]    (360,445) -- (360,455) ;
\draw [line width=2.25]    (370,445) -- (370,455) ;
\draw [line width=2.25]    (335,445) -- (335,455) ;
\draw [line width=2.25]    (345,445) -- (345,455) ;
\draw [line width=2.25]    (355,445) -- (355,455) ;
\draw [line width=2.25]    (365,445) -- (365,455) ;
\draw [line width=2.25]    (335,385) -- (335,395) ;
\draw [line width=2.25]    (345,385) -- (345,395) ;
\draw [line width=2.25]    (355,385) -- (355,395) ;
\draw [line width=2.25]    (365,385) -- (365,395) ;
\draw [line width=2.25]    (325,415) -- (315,415) ;
\draw [line width=2.25]    (325,405) -- (315,405) ;
\draw [line width=2.25]    (325,425) -- (315,425) ;
\draw [line width=2.25]    (325,435) -- (315,435) ;
\draw [line width=2.25]    (385,405) -- (375,405) ;
\draw [line width=2.25]    (385,415) -- (375,415) ;
\draw [line width=2.25]    (385,425) -- (375,425) ;
\draw [line width=2.25]    (385,435) -- (375,435) ;

\draw  [dash pattern={on 0.84pt off 2.51pt}]  (302.82,360) -- (303,425) -- (320,425) ;
\draw  [dash pattern={on 0.84pt off 2.51pt}]  (282.85,325.59) -- (300,335) -- (300,435) -- (315,435) ;
\draw  [dash pattern={on 0.84pt off 2.51pt}]  (322.78,325.32) -- (306,335) -- (306,415) -- (316,415) ;

\draw (306,231) node [anchor=north west][inner sep=0.75pt]   [align=left] {\hspace{-1.2cm}Rotman Lens (Analog DFT)};
\draw (316,362) node [anchor=north west][inner sep=0.75pt]   [align=left] {Transfer Switch Line\\ ~~~~~~~~~~~(or Matrix)};
\draw (340,407.4) node [anchor=north west][inner sep=0.75pt]    {$\!\!\mu$C};

\end{tikzpicture}

%% file: fig7.tex
\usetikzlibrary{patterns}

\tikzset{every picture/.style={line width=0.75pt}} 

\begin{tikzpicture}[x=0.75pt,y=0.75pt,yscale=-1,xscale=1]

\draw  [fill={rgb, 255:red, 0; green, 0; blue, 0 }  ,fill opacity=1 ] (355,10) -- (375,10) -- (375,30) -- (355,30) -- cycle ;
\draw  [fill={rgb, 255:red, 0; green, 0; blue, 0 }  ,fill opacity=1 ] (399.75,10) -- (419.75,10) -- (419.75,30) -- (399.75,30) -- cycle ;
\draw  [fill={rgb, 255:red, 0; green, 0; blue, 0 }  ,fill opacity=1 ] (444.75,10) -- (464.75,10) -- (464.75,30) -- (444.75,30) -- cycle ;
\draw  [fill={rgb, 255:red, 0; green, 0; blue, 0 }  ,fill opacity=1 ] (310,10) -- (330,10) -- (330,30) -- (310,30) -- cycle ;
\draw  [fill={rgb, 255:red, 0; green, 0; blue, 0 }  ,fill opacity=1 ] (264.75,10) -- (284.75,10) -- (284.75,30) -- (264.75,30) -- cycle ;
\draw [line width=2.25]    (330,151) -- (340,151) ;
\draw [line width=2.25]    (330,161) -- (340,161) ;
\draw [line width=2.25]    (330,171) -- (340,171) ;
\draw [line width=2.25]    (330,181) -- (340,181) ;
\draw [line width=2.25]    (330,191) -- (340,191) ;
\draw [line width=2.25]    (390,151) -- (400,151) ;
\draw [line width=2.25]    (390,161) -- (400,161) ;
\draw [line width=2.25]    (390,181) -- (400,181) ;
\draw [line width=2.25]    (390,191) -- (400,191) ;
\draw [line width=2.25]    (390,171) -- (400,171) ;
\draw [line width=2.25]    (345,136) -- (345,146) ;
\draw [line width=2.25]    (345,196) -- (345,206) ;
\draw [line width=2.25]    (355,136) -- (355,146) ;
\draw [line width=2.25]    (365,136) -- (365,146) ;
\draw [line width=2.25]    (375,136) -- (375,146) ;
\draw [line width=2.25]    (385,136) -- (385,146) ;
\draw [line width=2.25]    (355,196) -- (355,206) ;
\draw [line width=2.25]    (365,196) -- (365,206) ;
\draw [line width=2.25]    (375,196) -- (375,206) ;
\draw [line width=2.25]    (385,196) -- (385,206) ;
\draw [line width=2.25]    (350,196) -- (350,206) ;
\draw [line width=2.25]    (360,196) -- (360,206) ;
\draw [line width=2.25]    (370,196) -- (370,206) ;
\draw [line width=2.25]    (380,196) -- (380,206) ;
\draw [line width=2.25]    (350,136) -- (350,146) ;
\draw [line width=2.25]    (360,136) -- (360,146) ;
\draw [line width=2.25]    (370,136) -- (370,146) ;
\draw [line width=2.25]    (380,136) -- (380,146) ;
\draw [line width=2.25]    (340,166) -- (330,166) ;
\draw [line width=2.25]    (340,156) -- (330,156) ;
\draw [line width=2.25]    (340,176) -- (330,176) ;
\draw [line width=2.25]    (340,186) -- (330,186) ;
\draw [line width=2.25]    (400,156) -- (390,156) ;
\draw [line width=2.25]    (400,166) -- (390,166) ;
\draw [line width=2.25]    (400,176) -- (390,176) ;
\draw [line width=2.25]    (400,186) -- (390,186) ;
\draw   (340,149.5) .. controls (340,147.29) and (341.79,145.5) .. (344,145.5) -- (386,145.5) .. controls (388.21,145.5) and (390,147.29) .. (390,149.5) -- (390,191.5) .. controls (390,193.71) and (388.21,195.5) .. (386,195.5) -- (344,195.5) .. controls (341.79,195.5) and (340,193.71) .. (340,191.5) -- cycle ;

\draw  [dash pattern={on 0.84pt off 2.51pt}]  (252.25,38) -- (267,38) -- (267,186) -- (330,186) ;
\draw  [dash pattern={on 0.84pt off 2.51pt}]  (297.71,38.17) -- (282,38) -- (282,165) -- (330,166) ;
\draw   (245.12,59.4) -- (252.45,47.1) -- (260.33,59.06) -- (245.12,59.4) -- cycle (252.93,68.32) -- (252.73,59.23) (244.85,47.27) -- (260.06,46.93) (252.45,47.1) -- (252.25,38) ;
\draw   (247.15,71.55) -- (252.93,78) -- (258.72,71.55) -- (247.15,71.55) -- cycle (252.93,68.32) -- (252.93,71.55) ;

\draw   (290.59,59.57) -- (297.92,47.27) -- (305.8,59.23) -- (290.59,59.57) -- cycle (298.39,68.49) -- (298.19,59.4) (290.31,47.44) -- (305.52,47.1) (297.92,47.27) -- (297.71,38.17) ;
\draw   (292.61,71.72) -- (298.39,78.17) -- (304.18,71.72) -- (292.61,71.72) -- cycle (298.39,68.49) -- (298.39,71.72) ;

\draw   (335.59,59.57) -- (342.92,47.27) -- (350.8,59.23) -- (335.59,59.57) -- cycle (343.39,68.49) -- (343.19,59.4) (335.31,47.44) -- (350.52,47.1) (342.92,47.27) -- (342.71,38.17) ;
\draw   (337.61,71.72) -- (343.39,78.17) -- (349.18,71.72) -- (337.61,71.72) -- cycle (343.39,68.49) -- (343.39,71.72) ;

\draw   (380.59,59.57) -- (387.92,47.27) -- (395.8,59.23) -- (380.59,59.57) -- cycle (388.39,68.49) -- (388.19,59.4) (380.31,47.44) -- (395.52,47.1) (387.92,47.27) -- (387.71,38.17) ;
\draw   (382.61,71.72) -- (388.39,78.17) -- (394.18,71.72) -- (382.61,71.72) -- cycle (388.39,68.49) -- (388.39,71.72) ;

\draw   (425.59,59.57) -- (432.92,47.27) -- (440.8,59.23) -- (425.59,59.57) -- cycle (433.39,68.49) -- (433.19,59.4) (425.31,47.44) -- (440.52,47.1) (432.92,47.27) -- (432.71,38.17) ;
\draw   (427.61,71.72) -- (433.39,78.17) -- (439.18,71.72) -- (427.61,71.72) -- cycle (433.39,68.49) -- (433.39,71.72) ;

\draw [line width=3]    (454.64,28) -- (455,98) ;
\draw [line width=3]    (409.64,28) -- (410,98) ;
\draw [line width=3]    (364.64,28) -- (365,98) ;
\draw [line width=3]    (319.64,28) -- (320,98) ;
\draw [line width=3]    (274.64,28) -- (275,98) ;
\draw    (275.25,38) -- (252.25,38) ;
\draw    (320.71,38.17) -- (297.71,38.17) ;
\draw    (365.71,38.17) -- (342.71,38.17) ;
\draw    (410.25,38) -- (387.25,38) ;
\draw    (455.71,38.17) -- (432.71,38.17) ;
\draw  [color={rgb, 255:red, 208; green, 2; blue, 27 }  ,draw opacity=1 ][line width=2.25]  (380,124.75) -- (355,133) -- (355,116.5) -- cycle ;
\draw [color={rgb, 255:red, 208; green, 2; blue, 27 }  ,draw opacity=1 ][line width=2.25]    (300,99.75) -- (300,124.75) -- (355,124.75) ;
\draw [color={rgb, 255:red, 208; green, 2; blue, 27 }  ,draw opacity=1 ][line width=2.25]    (435,99.75) -- (435,124.75) -- (380,124.75) ;
\draw  [color={rgb, 255:red, 74; green, 144; blue, 226 }  ,draw opacity=1 ][line width=2.25]  (240,86) -- (480,86) -- (480,111) -- (240,111) -- cycle ;

\draw (298,97) node [anchor=north west][inner sep=0.75pt]  [font=\small,color={rgb, 255:red, 74; green, 144; blue, 226 }  ,opacity=1 ] [align=left] {\textbf{{\footnotesize  ~~resonant cavity combiner}}};
\draw (355,158.4) node [anchor=north west][inner sep=0.75pt]    {$\mu \text{C}$};

\end{tikzpicture}

%% file: fig6.tex
 
\tikzset{
pattern size/.store in=\mcSize, 
pattern size = 5pt,
pattern thickness/.store in=\mcThickness, 
pattern thickness = 0.3pt,
pattern radius/.store in=\mcRadius, 
pattern radius = 1pt}\makeatletter
\pgfutil@ifundefined{pgf@pattern@name@_siqb6sr8t}{
\pgfdeclarepatternformonly[\mcThickness,\mcSize]{_siqb6sr8t}
{\pgfqpoint{-\mcThickness}{-\mcThickness}}
{\pgfpoint{\mcSize}{\mcSize}}
{\pgfpoint{\mcSize}{\mcSize}}
{\pgfsetcolor{\tikz@pattern@color}
\pgfsetlinewidth{\mcThickness}
\pgfpathmoveto{\pgfpointorigin}
\pgfpathlineto{\pgfpoint{\mcSize}{0}}
\pgfpathmoveto{\pgfpointorigin}
\pgfpathlineto{\pgfpoint{0}{\mcSize}}
\pgfusepath{stroke}}}
\makeatother

 
\tikzset{
pattern size/.store in=\mcSize, 
pattern size = 5pt,
pattern thickness/.store in=\mcThickness, 
pattern thickness = 0.3pt,
pattern radius/.store in=\mcRadius, 
pattern radius = 1pt}\makeatletter
\pgfutil@ifundefined{pgf@pattern@name@_34qbuvvyy}{
\pgfdeclarepatternformonly[\mcThickness,\mcSize]{_34qbuvvyy}
{\pgfqpoint{-\mcThickness}{-\mcThickness}}
{\pgfpoint{\mcSize}{\mcSize}}
{\pgfpoint{\mcSize}{\mcSize}}
{\pgfsetcolor{\tikz@pattern@color}
\pgfsetlinewidth{\mcThickness}
\pgfpathmoveto{\pgfpointorigin}
\pgfpathlineto{\pgfpoint{\mcSize}{0}}
\pgfpathmoveto{\pgfpointorigin}
\pgfpathlineto{\pgfpoint{0}{\mcSize}}
\pgfusepath{stroke}}}
\makeatother
\tikzset{every picture/.style={line width=0.75pt}} 

\begin{tikzpicture}[x=0.75pt,y=0.75pt,yscale=-1,xscale=0.95]

\draw  [fill={rgb, 255:red, 0; green, 0; blue, 0 }  ,fill opacity=1 ] (389,80) .. controls (389.55,80) and (390,80.45) .. (390,81) -- (390,89) .. controls (390,89.55) and (389.55,90) .. (389,90) -- (386,90) .. controls (385.45,90) and (385,89.55) .. (385,89) -- (385,81) .. controls (385,80.45) and (385.45,80) .. (386,80) -- cycle ;
\draw  [color={rgb, 255:red, 155; green, 155; blue, 155 }  ,draw opacity=1 ][fill={rgb, 255:red, 155; green, 155; blue, 155 }  ,fill opacity=1 ] (390,81.29) -- (390,88.57) -- (385,88.57) -- (385,81.29) -- cycle ;

\draw  [fill={rgb, 255:red, 0; green, 0; blue, 0 }  ,fill opacity=1 ] (311,80) .. controls (310.45,80) and (310,80.45) .. (310,81) -- (310,89) .. controls (310,89.55) and (310.45,90) .. (311,90) -- (314,90) .. controls (314.55,90) and (315,89.55) .. (315,89) -- (315,81) .. controls (315,80.45) and (314.55,80) .. (314,80) -- cycle ;
\draw  [color={rgb, 255:red, 155; green, 155; blue, 155 }  ,draw opacity=1 ][fill={rgb, 255:red, 155; green, 155; blue, 155 }  ,fill opacity=1 ] (310,81.29) -- (310,88.57) -- (315,88.57) -- (315,81.29) -- cycle ;

\draw  [color={rgb, 255:red, 74; green, 74; blue, 74 }  ,draw opacity=1 ][line width=3]  (344.89,61.23) -- (348.08,74.74) -- (351.28,61.23) (348.08,61.23) -- (348.08,95) ;
\draw  [draw opacity=0][fill={rgb, 255:red, 245; green, 166; blue, 35 }  ,fill opacity=1 ][line width=1.5]  (339.14,40.96) -- (359.58,40.96) -- (359.58,61.23) -- (339.14,61.23) -- cycle ; \draw  [color={rgb, 255:red, 128; green, 128; blue, 128 }  ,draw opacity=1 ][line width=1.5]  (339.14,40.96) -- (339.14,61.23)(344.14,40.96) -- (344.14,61.23)(349.14,40.96) -- (349.14,61.23)(354.14,40.96) -- (354.14,61.23)(359.14,40.96) -- (359.14,61.23) ; \draw  [color={rgb, 255:red, 128; green, 128; blue, 128 }  ,draw opacity=1 ][line width=1.5]  (339.14,40.96) -- (359.58,40.96)(339.14,45.96) -- (359.58,45.96)(339.14,50.96) -- (359.58,50.96)(339.14,55.96) -- (359.58,55.96)(339.14,60.96) -- (359.58,60.96) ; \draw  [color={rgb, 255:red, 128; green, 128; blue, 128 }  ,draw opacity=1 ][line width=1.5]   ;

\draw [color={rgb, 255:red, 144; green, 19; blue, 254 }  ,draw opacity=1 ][line width=3]    (380,80) -- (359.14,45.96) -- (490,25) ;
\draw [shift={(364.13,54.12)}, rotate = 58.49] [fill={rgb, 255:red, 144; green, 19; blue, 254 }  ,fill opacity=1 ][line width=0.08]  [draw opacity=0] (18.75,-9.01) -- (0,0) -- (18.75,9.01) -- (12.45,0) -- cycle    ;
\draw [shift={(434.84,33.84)}, rotate = 170.9] [fill={rgb, 255:red, 144; green, 19; blue, 254 }  ,fill opacity=1 ][line width=0.08]  [draw opacity=0] (18.75,-9.01) -- (0,0) -- (18.75,9.01) -- (12.45,0) -- cycle    ;
\draw [color={rgb, 255:red, 208; green, 2; blue, 27 }  ,draw opacity=1 ][line width=3]    (320,80) -- (339.14,45.96) -- (210,25) ;
\draw [shift={(334.67,53.92)}, rotate = 119.35] [fill={rgb, 255:red, 208; green, 2; blue, 27 }  ,fill opacity=1 ][line width=0.08]  [draw opacity=0] (18.75,-9.01) -- (0,0) -- (18.75,9.01) -- (12.45,0) -- cycle    ;
\draw [shift={(264.3,33.82)}, rotate = 9.22] [fill={rgb, 255:red, 208; green, 2; blue, 27 }  ,fill opacity=1 ][line width=0.08]  [draw opacity=0] (18.75,-9.01) -- (0,0) -- (18.75,9.01) -- (12.45,0) -- cycle    ;
\draw  [fill={rgb, 255:red, 0; green, 0; blue, 0 }  ,fill opacity=1 ] (505.95,105) -- (508.35,37.08) -- (511.55,37.08) -- (513.95,105) -- cycle ;
\draw  [fill={rgb, 255:red, 245; green, 166; blue, 35 }  ,fill opacity=1 ] (525,20.94) -- (525,43.02) .. controls (525,46.3) and (518.28,48.96) .. (510,48.96) .. controls (501.72,48.96) and (495,46.3) .. (495,43.02) -- (495,20.94)(525,20.94) .. controls (525,24.23) and (518.28,26.89) .. (510,26.89) .. controls (501.72,26.89) and (495,24.23) .. (495,20.94) .. controls (495,17.66) and (501.72,15) .. (510,15) .. controls (518.28,15) and (525,17.66) .. (525,20.94) -- cycle ;
\draw  [pattern=_siqb6sr8t,pattern size=3.75pt,pattern thickness=1.5pt,pattern radius=0pt, pattern color={rgb, 255:red, 65; green, 117; blue, 5}] (525,20.94) -- (525,43.02) .. controls (525,46.3) and (518.28,48.96) .. (510,48.96) .. controls (501.72,48.96) and (495,46.3) .. (495,43.02) -- (495,20.94)(525,20.94) .. controls (525,24.23) and (518.28,26.89) .. (510,26.89) .. controls (501.72,26.89) and (495,24.23) .. (495,20.94) .. controls (495,17.66) and (501.72,15) .. (510,15) .. controls (518.28,15) and (525,17.66) .. (525,20.94) -- cycle ;

\draw  [fill={rgb, 255:red, 139; green, 87; blue, 42 }  ,fill opacity=1 ] (293,69.5) -- (293,102.5) .. controls (293,103.33) and (290.76,104) .. (288,104) .. controls (285.24,104) and (283,103.33) .. (283,102.5) -- (283,69.5) .. controls (283,68.67) and (285.24,68) .. (288,68) .. controls (290.76,68) and (293,68.67) .. (293,69.5) .. controls (293,70.33) and (290.76,71) .. (288,71) .. controls (285.24,71) and (283,70.33) .. (283,69.5) ;
\draw  [fill={rgb, 255:red, 65; green, 117; blue, 5 }  ,fill opacity=1 ] (273.19,53.5) .. controls (272.9,50.19) and (273.83,46.92) .. (275.57,45.07) .. controls (277.31,43.22) and (279.56,43.11) .. (281.36,44.8) .. controls (282,42.88) and (283.17,41.55) .. (284.52,41.22) .. controls (285.86,40.89) and (287.23,41.6) .. (288.2,43.12) .. controls (288.74,41.38) and (289.81,40.21) .. (291.03,40.03) .. controls (292.24,39.84) and (293.43,40.67) .. (294.17,42.22) .. controls (295.15,40.37) and (296.72,39.59) .. (298.19,40.22) .. controls (299.65,40.85) and (300.76,42.77) .. (301.03,45.16) .. controls (302.24,45.68) and (303.24,47.02) .. (303.79,48.81) .. controls (304.33,50.61) and (304.36,52.7) .. (303.87,54.53) .. controls (305.05,57) and (305.33,60.28) .. (304.59,63.16) .. controls (303.86,66.04) and (302.22,68.08) .. (300.29,68.52) .. controls (300.28,71.22) and (299.35,73.7) .. (297.87,75) .. controls (296.38,76.3) and (294.57,76.22) .. (293.13,74.79) .. controls (292.52,78.03) and (290.8,80.41) .. (288.71,80.9) .. controls (286.62,81.4) and (284.54,79.93) .. (283.36,77.12) .. controls (281.92,78.5) and (280.2,78.9) .. (278.57,78.22) .. controls (276.95,77.55) and (275.56,75.85) .. (274.73,73.52) .. controls (273.26,73.79) and (271.83,72.57) .. (271.17,70.47) .. controls (270.5,68.37) and (270.73,65.82) .. (271.74,64.1) .. controls (270.43,62.87) and (269.76,60.43) .. (270.08,58.04) .. controls (270.4,55.66) and (271.64,53.88) .. (273.16,53.63) ; \draw   (271.74,64.1) .. controls (272.36,64.69) and (273.07,64.95) .. (273.79,64.86)(274.73,73.52) .. controls (275.03,73.46) and (275.34,73.34) .. (275.62,73.15)(283.36,77.12) .. controls (283.15,76.6) and (282.97,76.04) .. (282.82,75.46)(293.13,74.79) .. controls (293.25,74.2) and (293.32,73.59) .. (293.35,72.98)(300.29,68.52) .. controls (300.31,65.64) and (299.28,63.01) .. (297.66,61.75)(303.87,54.53) .. controls (303.6,55.51) and (303.2,56.38) .. (302.69,57.07)(301.03,45.16) .. controls (301.08,45.55) and (301.1,45.95) .. (301.09,46.36)(294.17,42.22) .. controls (293.92,42.68) and (293.72,43.2) .. (293.57,43.75)(288.2,43.12) .. controls (288.07,43.54) and (287.97,43.98) .. (287.91,44.44)(281.36,44.8) .. controls (281.74,45.16) and (282.1,45.59) .. (282.41,46.08)(273.19,53.5) .. controls (273.22,53.95) and (273.28,54.41) .. (273.37,54.84) ;

\draw  [fill={rgb, 255:red, 139; green, 87; blue, 42 }  ,fill opacity=1 ] (418,70.5) -- (418,103.5) .. controls (418,104.33) and (415.76,105) .. (413,105) .. controls (410.24,105) and (408,104.33) .. (408,103.5) -- (408,70.5) .. controls (408,69.67) and (410.24,69) .. (413,69) .. controls (415.76,69) and (418,69.67) .. (418,70.5) .. controls (418,71.33) and (415.76,72) .. (413,72) .. controls (410.24,72) and (408,71.33) .. (408,70.5) ;
\draw  [fill={rgb, 255:red, 65; green, 117; blue, 5 }  ,fill opacity=1 ] (398.19,54.5) .. controls (397.9,51.19) and (398.83,47.92) .. (400.57,46.07) .. controls (402.31,44.22) and (404.56,44.11) .. (406.36,45.8) .. controls (407,43.88) and (408.17,42.55) .. (409.52,42.22) .. controls (410.86,41.89) and (412.23,42.6) .. (413.2,44.12) .. controls (413.74,42.38) and (414.81,41.21) .. (416.03,41.03) .. controls (417.24,40.84) and (418.43,41.67) .. (419.17,43.22) .. controls (420.15,41.37) and (421.72,40.59) .. (423.19,41.22) .. controls (424.65,41.85) and (425.76,43.77) .. (426.03,46.16) .. controls (427.24,46.68) and (428.24,48.02) .. (428.79,49.81) .. controls (429.33,51.61) and (429.36,53.7) .. (428.87,55.53) .. controls (430.05,58) and (430.33,61.28) .. (429.59,64.16) .. controls (428.86,67.04) and (427.22,69.08) .. (425.29,69.52) .. controls (425.28,72.22) and (424.35,74.7) .. (422.87,76) .. controls (421.38,77.3) and (419.57,77.22) .. (418.13,75.79) .. controls (417.52,79.03) and (415.8,81.41) .. (413.71,81.9) .. controls (411.62,82.4) and (409.54,80.93) .. (408.36,78.12) .. controls (406.92,79.5) and (405.2,79.9) .. (403.57,79.22) .. controls (401.95,78.55) and (400.56,76.85) .. (399.73,74.52) .. controls (398.26,74.79) and (396.83,73.57) .. (396.17,71.47) .. controls (395.5,69.37) and (395.73,66.82) .. (396.74,65.1) .. controls (395.43,63.87) and (394.76,61.43) .. (395.08,59.04) .. controls (395.4,56.66) and (396.64,54.88) .. (398.16,54.63) ; \draw   (396.74,65.1) .. controls (397.36,65.69) and (398.07,65.95) .. (398.79,65.86)(399.73,74.52) .. controls (400.03,74.46) and (400.34,74.34) .. (400.62,74.15)(408.36,78.12) .. controls (408.15,77.6) and (407.97,77.04) .. (407.82,76.46)(418.13,75.79) .. controls (418.25,75.2) and (418.32,74.59) .. (418.35,73.98)(425.29,69.52) .. controls (425.31,66.64) and (424.28,64.01) .. (422.66,62.75)(428.87,55.53) .. controls (428.6,56.51) and (428.2,57.38) .. (427.69,58.07)(426.03,46.16) .. controls (426.08,46.55) and (426.1,46.95) .. (426.09,47.36)(419.17,43.22) .. controls (418.92,43.68) and (418.72,44.2) .. (418.57,44.75)(413.2,44.12) .. controls (413.07,44.54) and (412.97,44.98) .. (412.91,45.44)(406.36,45.8) .. controls (406.74,46.16) and (407.1,46.59) .. (407.41,47.08)(398.19,54.5) .. controls (398.22,54.95) and (398.28,55.41) .. (398.37,55.84) ;

\draw  [fill={rgb, 255:red, 0; green, 0; blue, 0 }  ,fill opacity=1 ] (185.95,105) -- (188.35,37.08) -- (191.55,37.08) -- (193.95,105) -- cycle ;
\draw  [fill={rgb, 255:red, 245; green, 166; blue, 35 }  ,fill opacity=1 ] (205,20.94) -- (205,43.02) .. controls (205,46.3) and (198.28,48.96) .. (190,48.96) .. controls (181.72,48.96) and (175,46.3) .. (175,43.02) -- (175,20.94)(205,20.94) .. controls (205,24.23) and (198.28,26.89) .. (190,26.89) .. controls (181.72,26.89) and (175,24.23) .. (175,20.94) .. controls (175,17.66) and (181.72,15) .. (190,15) .. controls (198.28,15) and (205,17.66) .. (205,20.94) -- cycle ;
\draw  [pattern=_34qbuvvyy,pattern size=3.75pt,pattern thickness=1.5pt,pattern radius=0pt, pattern color={rgb, 255:red, 65; green, 117; blue, 5}] (205,20.94) -- (205,43.02) .. controls (205,46.3) and (198.28,48.96) .. (190,48.96) .. controls (181.72,48.96) and (175,46.3) .. (175,43.02) -- (175,20.94)(205,20.94) .. controls (205,24.23) and (198.28,26.89) .. (190,26.89) .. controls (181.72,26.89) and (175,24.23) .. (175,20.94) .. controls (175,17.66) and (181.72,15) .. (190,15) .. controls (198.28,15) and (205,17.66) .. (205,20.94) -- cycle ;

\draw    (474.87,109) .. controls (489.82,91) and (493.87,124.93) .. (509.95,105) ;
\draw    (189.95,105) .. controls (208.87,120.93) and (204.87,86.93) .. (223.87,104) ;

\draw (336,26) node [anchor=north west][inner sep=0.75pt]   [align=left] {\begin{minipage}[lt]{19.73pt}\setlength\topsep{0pt}
\begin{center}
RIS
\end{center}

\end{minipage}};

\end{tikzpicture}

%% file: fig9.tex
  
\tikzset {_rjlfhqvi3/.code = {\pgfsetadditionalshadetransform{ \pgftransformshift{\pgfpoint{0 bp } { 0 bp }  }  \pgftransformscale{1 }  }}}
\pgfdeclareradialshading{_jj0twpct6}{\pgfpoint{0bp}{0bp}}{rgb(0bp)=(1,1,1);
rgb(0bp)=(1,1,1);
rgb(25bp)=(0.55,0.34,0.16);
rgb(400bp)=(0.55,0.34,0.16)}

  
\tikzset {_mvoekshjp/.code = {\pgfsetadditionalshadetransform{ \pgftransformshift{\pgfpoint{0 bp } { 0 bp }  }  \pgftransformscale{1 }  }}}
\pgfdeclareradialshading{_r6aq6ry1h}{\pgfpoint{0bp}{0bp}}{rgb(0bp)=(1,1,1);
rgb(0bp)=(1,1,1);
rgb(25bp)=(0.55,0.34,0.16);
rgb(400bp)=(0.55,0.34,0.16)}

  
\tikzset {_visngs7t7/.code = {\pgfsetadditionalshadetransform{ \pgftransformshift{\pgfpoint{0 bp } { 0 bp }  }  \pgftransformscale{1 }  }}}
\pgfdeclareradialshading{_5uffzvqya}{\pgfpoint{0bp}{0bp}}{rgb(0bp)=(1,1,1);
rgb(0bp)=(1,1,1);
rgb(25bp)=(0.55,0.34,0.16);
rgb(400bp)=(0.55,0.34,0.16)}

  
\tikzset {_5lhopmj94/.code = {\pgfsetadditionalshadetransform{ \pgftransformshift{\pgfpoint{0 bp } { 0 bp }  }  \pgftransformscale{1 }  }}}
\pgfdeclareradialshading{_te3e7yetc}{\pgfpoint{0bp}{0bp}}{rgb(0bp)=(1,1,1);
rgb(0bp)=(1,1,1);
rgb(25bp)=(0.55,0.34,0.16);
rgb(400bp)=(0.55,0.34,0.16)}

  
\tikzset {_jhl8op38e/.code = {\pgfsetadditionalshadetransform{ \pgftransformshift{\pgfpoint{0 bp } { 0 bp }  }  \pgftransformscale{1 }  }}}
\pgfdeclareradialshading{_olkle89n2}{\pgfpoint{0bp}{0bp}}{rgb(0bp)=(1,1,1);
rgb(0bp)=(1,1,1);
rgb(25bp)=(0.55,0.34,0.16);
rgb(400bp)=(0.55,0.34,0.16)}

  
\tikzset {_6cznxgjc9/.code = {\pgfsetadditionalshadetransform{ \pgftransformshift{\pgfpoint{0 bp } { 0 bp }  }  \pgftransformscale{1 }  }}}
\pgfdeclareradialshading{_9ngp94thp}{\pgfpoint{0bp}{0bp}}{rgb(0bp)=(1,1,1);
rgb(0bp)=(1,1,1);
rgb(25bp)=(0.55,0.34,0.16);
rgb(400bp)=(0.55,0.34,0.16)}

  
\tikzset {_c2x7dvq0w/.code = {\pgfsetadditionalshadetransform{ \pgftransformshift{\pgfpoint{0 bp } { 0 bp }  }  \pgftransformscale{1 }  }}}
\pgfdeclareradialshading{_nbcp8vyhp}{\pgfpoint{0bp}{0bp}}{rgb(0bp)=(1,1,1);
rgb(0bp)=(1,1,1);
rgb(25bp)=(0.55,0.34,0.16);
rgb(400bp)=(0.55,0.34,0.16)}

  
\tikzset {_cmu2dm98r/.code = {\pgfsetadditionalshadetransform{ \pgftransformshift{\pgfpoint{0 bp } { 0 bp }  }  \pgftransformscale{1 }  }}}
\pgfdeclareradialshading{_vwg353c8o}{\pgfpoint{0bp}{0bp}}{rgb(0bp)=(1,1,1);
rgb(0bp)=(1,1,1);
rgb(25bp)=(0.55,0.34,0.16);
rgb(400bp)=(0.55,0.34,0.16)}

  
\tikzset {_u1ick5kwd/.code = {\pgfsetadditionalshadetransform{ \pgftransformshift{\pgfpoint{0 bp } { 0 bp }  }  \pgftransformscale{1 }  }}}
\pgfdeclareradialshading{_9i0sfn6hq}{\pgfpoint{0bp}{0bp}}{rgb(0bp)=(1,1,1);
rgb(0bp)=(1,1,1);
rgb(25bp)=(0.55,0.34,0.16);
rgb(400bp)=(0.55,0.34,0.16)}

  
\tikzset {_wd8hquqkq/.code = {\pgfsetadditionalshadetransform{ \pgftransformshift{\pgfpoint{0 bp } { 0 bp }  }  \pgftransformscale{1 }  }}}
\pgfdeclareradialshading{_7mhry837e}{\pgfpoint{0bp}{0bp}}{rgb(0bp)=(1,1,1);
rgb(0bp)=(1,1,1);
rgb(25bp)=(0.55,0.34,0.16);
rgb(400bp)=(0.55,0.34,0.16)}

  
\tikzset {_eo8g2v1xj/.code = {\pgfsetadditionalshadetransform{ \pgftransformshift{\pgfpoint{0 bp } { 0 bp }  }  \pgftransformscale{1 }  }}}
\pgfdeclareradialshading{_f1tjfpc2p}{\pgfpoint{0bp}{0bp}}{rgb(0bp)=(1,1,1);
rgb(0bp)=(1,1,1);
rgb(25bp)=(0.55,0.34,0.16);
rgb(400bp)=(0.55,0.34,0.16)}

  
\tikzset {_51xfyfetl/.code = {\pgfsetadditionalshadetransform{ \pgftransformshift{\pgfpoint{0 bp } { 0 bp }  }  \pgftransformscale{1 }  }}}
\pgfdeclareradialshading{_56c1ushi2}{\pgfpoint{0bp}{0bp}}{rgb(0bp)=(1,1,1);
rgb(0bp)=(1,1,1);
rgb(25bp)=(0.55,0.34,0.16);
rgb(400bp)=(0.55,0.34,0.16)}

  
\tikzset {_av2nauclz/.code = {\pgfsetadditionalshadetransform{ \pgftransformshift{\pgfpoint{0 bp } { 0 bp }  }  \pgftransformscale{1 }  }}}
\pgfdeclareradialshading{_2260aj6xs}{\pgfpoint{0bp}{0bp}}{rgb(0bp)=(1,1,1);
rgb(0bp)=(1,1,1);
rgb(25bp)=(0.55,0.34,0.16);
rgb(400bp)=(0.55,0.34,0.16)}

  
\tikzset {_n8hlp70lc/.code = {\pgfsetadditionalshadetransform{ \pgftransformshift{\pgfpoint{0 bp } { 0 bp }  }  \pgftransformscale{1 }  }}}
\pgfdeclareradialshading{_m1t8zbz6x}{\pgfpoint{0bp}{0bp}}{rgb(0bp)=(1,1,1);
rgb(0bp)=(1,1,1);
rgb(25bp)=(0.55,0.34,0.16);
rgb(400bp)=(0.55,0.34,0.16)}

  
\tikzset {_yac683ncy/.code = {\pgfsetadditionalshadetransform{ \pgftransformshift{\pgfpoint{0 bp } { 0 bp }  }  \pgftransformscale{1 }  }}}
\pgfdeclareradialshading{_afuhexptm}{\pgfpoint{0bp}{0bp}}{rgb(0bp)=(1,1,1);
rgb(0bp)=(1,1,1);
rgb(25bp)=(0.55,0.34,0.16);
rgb(400bp)=(0.55,0.34,0.16)}

  
\tikzset {_ijosmhyoa/.code = {\pgfsetadditionalshadetransform{ \pgftransformshift{\pgfpoint{0 bp } { 0 bp }  }  \pgftransformscale{1 }  }}}
\pgfdeclareradialshading{_s22w9vran}{\pgfpoint{0bp}{0bp}}{rgb(0bp)=(1,1,1);
rgb(0bp)=(1,1,1);
rgb(25bp)=(0.55,0.34,0.16);
rgb(400bp)=(0.55,0.34,0.16)}

  
\tikzset {_jdm7vy99o/.code = {\pgfsetadditionalshadetransform{ \pgftransformshift{\pgfpoint{0 bp } { 0 bp }  }  \pgftransformscale{1 }  }}}
\pgfdeclareradialshading{_51j8fti4c}{\pgfpoint{0bp}{0bp}}{rgb(0bp)=(1,1,1);
rgb(0bp)=(1,1,1);
rgb(25bp)=(0.55,0.34,0.16);
rgb(400bp)=(0.55,0.34,0.16)}

  
\tikzset {_sgh0kqiqa/.code = {\pgfsetadditionalshadetransform{ \pgftransformshift{\pgfpoint{0 bp } { 0 bp }  }  \pgftransformscale{1 }  }}}
\pgfdeclareradialshading{_c372oph4r}{\pgfpoint{0bp}{0bp}}{rgb(0bp)=(1,1,1);
rgb(0bp)=(1,1,1);
rgb(25bp)=(0.55,0.34,0.16);
rgb(400bp)=(0.55,0.34,0.16)}

  
\tikzset {_mk2f2kmuq/.code = {\pgfsetadditionalshadetransform{ \pgftransformshift{\pgfpoint{0 bp } { 0 bp }  }  \pgftransformscale{1 }  }}}
\pgfdeclareradialshading{_a2h46gxxc}{\pgfpoint{0bp}{0bp}}{rgb(0bp)=(1,1,1);
rgb(0bp)=(1,1,1);
rgb(25bp)=(0.55,0.34,0.16);
rgb(400bp)=(0.55,0.34,0.16)}

  
\tikzset {_g5owlb4nt/.code = {\pgfsetadditionalshadetransform{ \pgftransformshift{\pgfpoint{0 bp } { 0 bp }  }  \pgftransformscale{1 }  }}}
\pgfdeclareradialshading{_q8ml3n1h1}{\pgfpoint{0bp}{0bp}}{rgb(0bp)=(1,1,1);
rgb(0bp)=(1,1,1);
rgb(25bp)=(0.55,0.34,0.16);
rgb(400bp)=(0.55,0.34,0.16)}

  
\tikzset {_0mg3osk66/.code = {\pgfsetadditionalshadetransform{ \pgftransformshift{\pgfpoint{0 bp } { 0 bp }  }  \pgftransformrotate{-73 }  \pgftransformscale{2 }  }}}
\pgfdeclarehorizontalshading{_gig91j6pn}{150bp}{rgb(0bp)=(0,1,0);
rgb(37.5bp)=(0,1,0);
rgb(47.32142857142857bp)=(1,1,0);
rgb(61.60714285714286bp)=(1,0,0);
rgb(100bp)=(1,0,0)}

  
\tikzset {_offw1b2wx/.code = {\pgfsetadditionalshadetransform{ \pgftransformshift{\pgfpoint{0 bp } { 0 bp }  }  \pgftransformrotate{-73 }  \pgftransformscale{2 }  }}}
\pgfdeclarehorizontalshading{_32mqwufzm}{150bp}{rgb(0bp)=(0,1,0);
rgb(37.5bp)=(0,1,0);
rgb(47.32142857142857bp)=(1,1,0);
rgb(61.60714285714286bp)=(1,0,0);
rgb(100bp)=(1,0,0)}

  
\tikzset {_md8v3l82l/.code = {\pgfsetadditionalshadetransform{ \pgftransformshift{\pgfpoint{0 bp } { 0 bp }  }  \pgftransformrotate{-73 }  \pgftransformscale{2 }  }}}
\pgfdeclarehorizontalshading{_smlfupdjv}{150bp}{rgb(0bp)=(0,1,0);
rgb(37.5bp)=(0,1,0);
rgb(47.32142857142857bp)=(1,1,0);
rgb(61.60714285714286bp)=(1,0,0);
rgb(100bp)=(1,0,0)}

  
\tikzset {_qutiidczg/.code = {\pgfsetadditionalshadetransform{ \pgftransformshift{\pgfpoint{0 bp } { 0 bp }  }  \pgftransformrotate{-73 }  \pgftransformscale{2 }  }}}
\pgfdeclarehorizontalshading{_5jxm9rz6k}{150bp}{rgb(0bp)=(0,1,0);
rgb(37.5bp)=(0,1,0);
rgb(47.32142857142857bp)=(1,1,0);
rgb(61.60714285714286bp)=(1,0,0);
rgb(100bp)=(1,0,0)}

  
\tikzset {_3hxpoh0x3/.code = {\pgfsetadditionalshadetransform{ \pgftransformshift{\pgfpoint{0 bp } { 0 bp }  }  \pgftransformrotate{-73 }  \pgftransformscale{2 }  }}}
\pgfdeclarehorizontalshading{_i47q77ym0}{150bp}{rgb(0bp)=(0,1,0);
rgb(37.5bp)=(0,1,0);
rgb(47.32142857142857bp)=(1,1,0);
rgb(61.60714285714286bp)=(1,0,0);
rgb(100bp)=(1,0,0)}

  
\tikzset {_54xptasjb/.code = {\pgfsetadditionalshadetransform{ \pgftransformshift{\pgfpoint{0 bp } { 0 bp }  }  \pgftransformrotate{-73 }  \pgftransformscale{2 }  }}}
\pgfdeclarehorizontalshading{_ku6oe10ic}{150bp}{rgb(0bp)=(0,1,0);
rgb(37.5bp)=(0,1,0);
rgb(47.32142857142857bp)=(1,1,0);
rgb(61.60714285714286bp)=(1,0,0);
rgb(100bp)=(1,0,0)}

  
\tikzset {_wcwrmiwnb/.code = {\pgfsetadditionalshadetransform{ \pgftransformshift{\pgfpoint{0 bp } { 0 bp }  }  \pgftransformrotate{-73 }  \pgftransformscale{2 }  }}}
\pgfdeclarehorizontalshading{_k28y3hq5h}{150bp}{rgb(0bp)=(0,1,0);
rgb(37.5bp)=(0,1,0);
rgb(47.32142857142857bp)=(1,1,0);
rgb(61.60714285714286bp)=(1,0,0);
rgb(100bp)=(1,0,0)}

  
\tikzset {_ar9gm5bln/.code = {\pgfsetadditionalshadetransform{ \pgftransformshift{\pgfpoint{0 bp } { 0 bp }  }  \pgftransformrotate{-73 }  \pgftransformscale{2 }  }}}
\pgfdeclarehorizontalshading{_o358n8jsf}{150bp}{rgb(0bp)=(0,1,0);
rgb(37.5bp)=(0,1,0);
rgb(47.32142857142857bp)=(1,1,0);
rgb(61.60714285714286bp)=(1,0,0);
rgb(100bp)=(1,0,0)}
\tikzset{every picture/.style={line width=0.75pt}} 

\begin{tikzpicture}[x=0.75pt,y=0.75pt,yscale=-1,xscale=1]

\draw  [fill={rgb, 255:red, 0; green, 0; blue, 0 }  ,fill opacity=1 ] (310.39,140.26) -- (319.91,130.74) -- (325.39,130.74) -- (325.39,150) -- (315.87,159.53) -- (310.39,159.53) -- cycle ; \draw   (325.39,130.74) -- (315.87,140.26) -- (310.39,140.26) ; \draw   (315.87,140.26) -- (315.87,159.53) ;
\draw  [fill={rgb, 255:red, 0; green, 0; blue, 0 }  ,fill opacity=1 ] (310.56,110.8) -- (320.09,101.27) -- (325.56,101.27) -- (325.56,120.54) -- (316.04,130.06) -- (310.56,130.06) -- cycle ; \draw   (325.56,101.27) -- (316.04,110.8) -- (310.56,110.8) ; \draw   (316.04,110.8) -- (316.04,130.06) ;
\draw  [fill={rgb, 255:red, 0; green, 0; blue, 0 }  ,fill opacity=1 ] (310.13,80.26) -- (319.66,70.74) -- (325.13,70.74) -- (325.13,90) -- (315.61,99.53) -- (310.13,99.53) -- cycle ; \draw   (325.13,70.74) -- (315.61,80.26) -- (310.13,80.26) ; \draw   (315.61,80.26) -- (315.61,99.53) ;
\draw  [fill={rgb, 255:red, 0; green, 0; blue, 0 }  ,fill opacity=1 ] (310,50.27) -- (319.52,40.74) -- (325,40.74) -- (325,60) -- (315.48,69.53) -- (310,69.53) -- cycle ; \draw   (325,40.74) -- (315.48,50.27) -- (310,50.27) ; \draw   (315.48,50.27) -- (315.48,69.53) ;
\draw  [fill={rgb, 255:red, 245; green, 166; blue, 35 }  ,fill opacity=1 ] (312.38,45.89) -- (353.15,5.12) -- (359.04,5.12) -- (359.04,124.23) -- (318.27,165) -- (312.38,165) -- cycle ; \draw   (359.04,5.12) -- (318.27,45.89) -- (312.38,45.89) ; \draw   (318.27,45.89) -- (318.27,165) ;
\draw  [fill={rgb, 255:red, 139; green, 87; blue, 42 }  ,fill opacity=1 ] (318.27,45.89) -- (359.04,5.12) -- (364.93,5.12) -- (364.93,124.23) -- (324.16,165) -- (318.27,165) -- cycle ; \draw   (364.93,5.12) -- (324.16,45.89) -- (318.27,45.89) ; \draw   (324.16,45.89) -- (324.16,165) ;
\draw  [draw opacity=0][shading=_jj0twpct6,_rjlfhqvi3] (364.38,125.12) .. controls (372.32,125.12) and (378.75,118.4) .. (378.76,110.12) .. controls (378.76,101.83) and (372.33,95.12) .. (364.38,95.12) .. controls (364.38,95.12) and (364.37,95.12) .. (364.37,95.12) -- (364.38,110.12) -- cycle ; \draw   (364.38,125.12) .. controls (372.32,125.12) and (378.75,118.4) .. (378.76,110.12) .. controls (378.76,101.83) and (372.33,95.12) .. (364.38,95.12) .. controls (364.38,95.12) and (364.37,95.12) .. (364.37,95.12) ;  
\draw  [draw opacity=0][shading=_r6aq6ry1h,_mvoekshjp] (364.38,95.12) .. controls (372.32,95.12) and (378.75,88.4) .. (378.76,80.12) .. controls (378.76,71.83) and (372.33,65.12) .. (364.38,65.12) .. controls (364.38,65.12) and (364.37,65.12) .. (364.37,65.12) -- (364.38,80.12) -- cycle ; \draw   (364.38,95.12) .. controls (372.32,95.12) and (378.75,88.4) .. (378.76,80.12) .. controls (378.76,71.83) and (372.33,65.12) .. (364.38,65.12) .. controls (364.38,65.12) and (364.37,65.12) .. (364.37,65.12) ;  
\draw  [draw opacity=0][shading=_5uffzvqya,_visngs7t7] (364.39,65.12) .. controls (372.33,65.12) and (378.77,58.4) .. (378.77,50.12) .. controls (378.78,41.83) and (372.34,35.12) .. (364.4,35.12) .. controls (364.4,35.12) and (364.39,35.12) .. (364.39,35.12) -- (364.39,50.12) -- cycle ; \draw   (364.39,65.12) .. controls (372.33,65.12) and (378.77,58.4) .. (378.77,50.12) .. controls (378.78,41.83) and (372.34,35.12) .. (364.4,35.12) .. controls (364.4,35.12) and (364.39,35.12) .. (364.39,35.12) ;  
\draw  [draw opacity=0][shading=_te3e7yetc,_5lhopmj94] (364.39,35.12) .. controls (372.33,35.12) and (378.77,28.4) .. (378.77,20.12) .. controls (378.77,11.84) and (372.34,5.12) .. (364.39,5.12) .. controls (364.39,5.12) and (364.39,5.12) .. (364.38,5.12) -- (364.39,20.12) -- cycle ; \draw   (364.39,35.12) .. controls (372.33,35.12) and (378.77,28.4) .. (378.77,20.12) .. controls (378.77,11.84) and (372.34,5.12) .. (364.39,5.12) .. controls (364.39,5.12) and (364.39,5.12) .. (364.38,5.12) ;  
\draw  [draw opacity=0][shading=_olkle89n2,_jhl8op38e] (354.28,135.12) .. controls (362.22,135.12) and (368.65,128.4) .. (368.66,120.12) .. controls (368.66,111.83) and (362.23,105.12) .. (354.28,105.12) .. controls (354.28,105.12) and (354.27,105.12) .. (354.27,105.12) -- (354.28,120.12) -- cycle ; \draw   (354.28,135.12) .. controls (362.22,135.12) and (368.65,128.4) .. (368.66,120.12) .. controls (368.66,111.83) and (362.23,105.12) .. (354.28,105.12) .. controls (354.28,105.12) and (354.27,105.12) .. (354.27,105.12) ;  
\draw  [draw opacity=0][shading=_9ngp94thp,_6cznxgjc9] (354.28,105.12) .. controls (362.22,105.12) and (368.65,98.4) .. (368.66,90.12) .. controls (368.66,81.83) and (362.23,75.12) .. (354.28,75.12) .. controls (354.28,75.12) and (354.27,75.12) .. (354.27,75.12) -- (354.28,90.12) -- cycle ; \draw   (354.28,105.12) .. controls (362.22,105.12) and (368.65,98.4) .. (368.66,90.12) .. controls (368.66,81.83) and (362.23,75.12) .. (354.28,75.12) .. controls (354.28,75.12) and (354.27,75.12) .. (354.27,75.12) ;  
\draw  [draw opacity=0][shading=_nbcp8vyhp,_c2x7dvq0w] (354.28,75.12) .. controls (362.22,75.12) and (368.65,68.4) .. (368.66,60.12) .. controls (368.66,51.83) and (362.23,45.12) .. (354.28,45.12) .. controls (354.28,45.12) and (354.27,45.12) .. (354.27,45.12) -- (354.28,60.12) -- cycle ; \draw   (354.28,75.12) .. controls (362.22,75.12) and (368.65,68.4) .. (368.66,60.12) .. controls (368.66,51.83) and (362.23,45.12) .. (354.28,45.12) .. controls (354.28,45.12) and (354.27,45.12) .. (354.27,45.12) ;  
\draw  [draw opacity=0][shading=_vwg353c8o,_cmu2dm98r] (354.26,45.12) .. controls (362.2,45.12) and (368.64,38.4) .. (368.64,30.12) .. controls (368.65,21.84) and (362.21,15.12) .. (354.27,15.12) .. controls (354.26,15.12) and (354.26,15.12) .. (354.26,15.12) -- (354.26,30.12) -- cycle ; \draw   (354.26,45.12) .. controls (362.2,45.12) and (368.64,38.4) .. (368.64,30.12) .. controls (368.65,21.84) and (362.21,15.12) .. (354.27,15.12) .. controls (354.26,15.12) and (354.26,15.12) .. (354.26,15.12) ;  
\draw  [draw opacity=0][shading=_9i0sfn6hq,_u1ick5kwd] (344.98,145.12) .. controls (352.92,145.12) and (359.35,138.4) .. (359.36,130.12) .. controls (359.36,121.83) and (352.93,115.12) .. (344.98,115.12) .. controls (344.98,115.12) and (344.97,115.12) .. (344.97,115.12) -- (344.98,130.12) -- cycle ; \draw   (344.98,145.12) .. controls (352.92,145.12) and (359.35,138.4) .. (359.36,130.12) .. controls (359.36,121.83) and (352.93,115.12) .. (344.98,115.12) .. controls (344.98,115.12) and (344.97,115.12) .. (344.97,115.12) ;  
\draw  [draw opacity=0][shading=_7mhry837e,_wd8hquqkq] (344.98,115.12) .. controls (352.92,115.12) and (359.35,108.4) .. (359.36,100.12) .. controls (359.36,91.83) and (352.93,85.12) .. (344.98,85.12) .. controls (344.98,85.12) and (344.97,85.12) .. (344.97,85.12) -- (344.98,100.12) -- cycle ; \draw   (344.98,115.12) .. controls (352.92,115.12) and (359.35,108.4) .. (359.36,100.12) .. controls (359.36,91.83) and (352.93,85.12) .. (344.98,85.12) .. controls (344.98,85.12) and (344.97,85.12) .. (344.97,85.12) ;  
\draw  [draw opacity=0][shading=_f1tjfpc2p,_eo8g2v1xj] (344.98,85.12) .. controls (352.92,85.12) and (359.35,78.4) .. (359.36,70.12) .. controls (359.36,61.83) and (352.93,55.12) .. (344.98,55.12) .. controls (344.98,55.12) and (344.97,55.12) .. (344.97,55.12) -- (344.98,70.12) -- cycle ; \draw   (344.98,85.12) .. controls (352.92,85.12) and (359.35,78.4) .. (359.36,70.12) .. controls (359.36,61.83) and (352.93,55.12) .. (344.98,55.12) .. controls (344.98,55.12) and (344.97,55.12) .. (344.97,55.12) ;  
\draw  [draw opacity=0][shading=_56c1ushi2,_51xfyfetl] (345,55.12) .. controls (352.94,55.12) and (359.38,48.4) .. (359.38,40.12) .. controls (359.38,31.84) and (352.95,25.12) .. (345.01,25.12) .. controls (345,25.12) and (345,25.12) .. (344.99,25.12) -- (345,40.12) -- cycle ; \draw   (345,55.12) .. controls (352.94,55.12) and (359.38,48.4) .. (359.38,40.12) .. controls (359.38,31.84) and (352.95,25.12) .. (345.01,25.12) .. controls (345,25.12) and (345,25.12) .. (344.99,25.12) ;  
\draw  [draw opacity=0][shading=_2260aj6xs,_av2nauclz] (334.98,155) .. controls (342.92,155) and (349.35,148.28) .. (349.36,140) .. controls (349.36,131.72) and (342.93,125) .. (334.98,125) .. controls (334.98,125) and (334.97,125) .. (334.97,125) -- (334.98,140) -- cycle ; \draw   (334.98,155) .. controls (342.92,155) and (349.35,148.28) .. (349.36,140) .. controls (349.36,131.72) and (342.93,125) .. (334.98,125) .. controls (334.98,125) and (334.97,125) .. (334.97,125) ;  
\draw  [draw opacity=0][shading=_m1t8zbz6x,_n8hlp70lc] (334.98,125) .. controls (342.92,125) and (349.35,118.28) .. (349.36,110) .. controls (349.36,101.72) and (342.93,95) .. (334.98,95) .. controls (334.98,95) and (334.97,95) .. (334.97,95) -- (334.98,110) -- cycle ; \draw   (334.98,125) .. controls (342.92,125) and (349.35,118.28) .. (349.36,110) .. controls (349.36,101.72) and (342.93,95) .. (334.98,95) .. controls (334.98,95) and (334.97,95) .. (334.97,95) ;  
\draw  [draw opacity=0][shading=_afuhexptm,_yac683ncy] (334.98,95) .. controls (342.92,95) and (349.35,88.28) .. (349.36,80) .. controls (349.36,71.72) and (342.93,65) .. (334.98,65) .. controls (334.98,65) and (334.97,65) .. (334.97,65) -- (334.98,80) -- cycle ; \draw   (334.98,95) .. controls (342.92,95) and (349.35,88.28) .. (349.36,80) .. controls (349.36,71.72) and (342.93,65) .. (334.98,65) .. controls (334.98,65) and (334.97,65) .. (334.97,65) ;  
\draw  [draw opacity=0][shading=_s22w9vran,_ijosmhyoa] (335,65) .. controls (342.94,65) and (349.38,58.29) .. (349.38,50) .. controls (349.38,41.72) and (342.95,35) .. (335.01,35) .. controls (335,35) and (335,35) .. (334.99,35) -- (335,50) -- cycle ; \draw   (335,65) .. controls (342.94,65) and (349.38,58.29) .. (349.38,50) .. controls (349.38,41.72) and (342.95,35) .. (335.01,35) .. controls (335,35) and (335,35) .. (334.99,35) ;  
\draw  [draw opacity=0][shading=_51j8fti4c,_jdm7vy99o] (325.01,165) .. controls (332.95,165) and (339.39,158.28) .. (339.4,150) .. controls (339.4,141.72) and (332.96,135) .. (325.02,135) .. controls (325.02,135) and (325.01,135) .. (325.01,135) -- (325.02,150) -- cycle ; \draw   (325.01,165) .. controls (332.95,165) and (339.39,158.28) .. (339.4,150) .. controls (339.4,141.72) and (332.96,135) .. (325.02,135) .. controls (325.02,135) and (325.01,135) .. (325.01,135) ;  
\draw  [draw opacity=0][shading=_c372oph4r,_sgh0kqiqa] (325.01,135) .. controls (332.95,135) and (339.39,128.28) .. (339.39,120) .. controls (339.39,111.72) and (332.96,105) .. (325.02,105) .. controls (325.01,105) and (325.01,105) .. (325,105) -- (325.01,120) -- cycle ; \draw   (325.01,135) .. controls (332.95,135) and (339.39,128.28) .. (339.39,120) .. controls (339.39,111.72) and (332.96,105) .. (325.02,105) .. controls (325.01,105) and (325.01,105) .. (325,105) ;  
\draw  [draw opacity=0][shading=_a2h46gxxc,_mk2f2kmuq] (325,105) .. controls (332.94,105) and (339.38,98.28) .. (339.38,90) .. controls (339.39,81.72) and (332.95,75) .. (325.01,75) .. controls (325.01,75) and (325,75) .. (325,75) -- (325.01,90) -- cycle ; \draw   (325,105) .. controls (332.94,105) and (339.38,98.28) .. (339.38,90) .. controls (339.39,81.72) and (332.95,75) .. (325.01,75) .. controls (325.01,75) and (325,75) .. (325,75) ;  
\draw  [draw opacity=0][shading=_q8ml3n1h1,_g5owlb4nt] (325,75) .. controls (332.94,75) and (339.38,68.29) .. (339.38,60) .. controls (339.38,51.72) and (332.95,45) .. (325.01,45) .. controls (325,45) and (325,45) .. (324.99,45) -- (325,60) -- cycle ; \draw   (325,75) .. controls (332.94,75) and (339.38,68.29) .. (339.38,60) .. controls (339.38,51.72) and (332.95,45) .. (325.01,45) .. controls (325,45) and (325,45) .. (324.99,45) ;  
\path  [shading=_gig91j6pn,_0mg3osk66] (388.21,23.2) .. controls (385.43,33.49) and (382.36,43.02) .. (379.2,48.85) .. controls (376.03,54.68) and (374.12,52.73) .. (376.14,45.54) .. controls (378.17,38.34) and (381.66,29.24) .. (386.17,20.87) .. controls (390.68,12.5) and (391,12.91) .. (388.21,23.2) -- cycle ; 
 \draw   (388.21,23.2) .. controls (385.43,33.49) and (382.36,43.02) .. (379.2,48.85) .. controls (376.03,54.68) and (374.12,52.73) .. (376.14,45.54) .. controls (378.17,38.34) and (381.66,29.24) .. (386.17,20.87) .. controls (390.68,12.5) and (391,12.91) .. (388.21,23.2) -- cycle ; 

\path  [shading=_32mqwufzm,_offw1b2wx] (388.21,23.2) .. controls (385.43,33.49) and (382.36,43.02) .. (379.2,48.85) .. controls (376.03,54.68) and (374.12,52.73) .. (376.14,45.54) .. controls (378.17,38.34) and (381.66,29.24) .. (386.17,20.87) .. controls (390.68,12.5) and (391,12.91) .. (388.21,23.2) -- cycle ; 
 \draw   (388.21,23.2) .. controls (385.43,33.49) and (382.36,43.02) .. (379.2,48.85) .. controls (376.03,54.68) and (374.12,52.73) .. (376.14,45.54) .. controls (378.17,38.34) and (381.66,29.24) .. (386.17,20.87) .. controls (390.68,12.5) and (391,12.91) .. (388.21,23.2) -- cycle ; 

\path  [shading=_smlfupdjv,_md8v3l82l] (381.75,55.37) .. controls (391.84,58.81) and (401.16,62.48) .. (406.78,66.01) .. controls (412.39,69.54) and (410.32,71.33) .. (403.27,68.85) .. controls (396.23,66.37) and (387.36,62.3) .. (379.3,57.27) .. controls (371.23,52.23) and (371.67,51.93) .. (381.75,55.37) -- cycle ; 
 \draw   (381.75,55.37) .. controls (391.84,58.81) and (401.16,62.48) .. (406.78,66.01) .. controls (412.39,69.54) and (410.32,71.33) .. (403.27,68.85) .. controls (396.23,66.37) and (387.36,62.3) .. (379.3,57.27) .. controls (371.23,52.23) and (371.67,51.93) .. (381.75,55.37) -- cycle ; 

\path  [shading=_5jxm9rz6k,_qutiidczg] (381.75,55.37) .. controls (391.84,58.81) and (401.16,62.48) .. (406.78,66.01) .. controls (412.39,69.54) and (410.32,71.33) .. (403.27,68.85) .. controls (396.23,66.37) and (387.36,62.3) .. (379.3,57.27) .. controls (371.23,52.23) and (371.67,51.93) .. (381.75,55.37) -- cycle ; 
 \draw   (381.75,55.37) .. controls (391.84,58.81) and (401.16,62.48) .. (406.78,66.01) .. controls (412.39,69.54) and (410.32,71.33) .. (403.27,68.85) .. controls (396.23,66.37) and (387.36,62.3) .. (379.3,57.27) .. controls (371.23,52.23) and (371.67,51.93) .. (381.75,55.37) -- cycle ; 

\path  [shading=_i47q77ym0,_3hxpoh0x3] (389.54,78.2) .. controls (386.75,88.49) and (383.68,98.02) .. (380.52,103.85) .. controls (377.36,109.68) and (375.44,107.73) .. (377.47,100.54) .. controls (379.49,93.34) and (382.98,84.24) .. (387.49,75.87) .. controls (392,67.5) and (392.33,67.91) .. (389.54,78.2) -- cycle ; 
 \draw   (389.54,78.2) .. controls (386.75,88.49) and (383.68,98.02) .. (380.52,103.85) .. controls (377.36,109.68) and (375.44,107.73) .. (377.47,100.54) .. controls (379.49,93.34) and (382.98,84.24) .. (387.49,75.87) .. controls (392,67.5) and (392.33,67.91) .. (389.54,78.2) -- cycle ; 

\path  [shading=_ku6oe10ic,_54xptasjb] (389.54,78.2) .. controls (386.75,88.49) and (383.68,98.02) .. (380.52,103.85) .. controls (377.36,109.68) and (375.44,107.73) .. (377.47,100.54) .. controls (379.49,93.34) and (382.98,84.24) .. (387.49,75.87) .. controls (392,67.5) and (392.33,67.91) .. (389.54,78.2) -- cycle ; 
 \draw   (389.54,78.2) .. controls (386.75,88.49) and (383.68,98.02) .. (380.52,103.85) .. controls (377.36,109.68) and (375.44,107.73) .. (377.47,100.54) .. controls (379.49,93.34) and (382.98,84.24) .. (387.49,75.87) .. controls (392,67.5) and (392.33,67.91) .. (389.54,78.2) -- cycle ; 

\path  [shading=_k28y3hq5h,_wcwrmiwnb] (383.08,110.37) .. controls (393.17,113.81) and (402.48,117.48) .. (408.1,121.01) .. controls (413.72,124.54) and (411.65,126.33) .. (404.6,123.85) .. controls (397.55,121.37) and (388.69,117.3) .. (380.62,112.27) .. controls (372.56,107.23) and (372.99,106.93) .. (383.08,110.37) -- cycle ; 
 \draw   (383.08,110.37) .. controls (393.17,113.81) and (402.48,117.48) .. (408.1,121.01) .. controls (413.72,124.54) and (411.65,126.33) .. (404.6,123.85) .. controls (397.55,121.37) and (388.69,117.3) .. (380.62,112.27) .. controls (372.56,107.23) and (372.99,106.93) .. (383.08,110.37) -- cycle ; 

\path  [shading=_o358n8jsf,_ar9gm5bln] (383.08,110.37) .. controls (393.17,113.81) and (402.48,117.48) .. (408.1,121.01) .. controls (413.72,124.54) and (411.65,126.33) .. (404.6,123.85) .. controls (397.55,121.37) and (388.69,117.3) .. (380.62,112.27) .. controls (372.56,107.23) and (372.99,106.93) .. (383.08,110.37) -- cycle ; 
 \draw   (383.08,110.37) .. controls (393.17,113.81) and (402.48,117.48) .. (408.1,121.01) .. controls (413.72,124.54) and (411.65,126.33) .. (404.6,123.85) .. controls (397.55,121.37) and (388.69,117.3) .. (380.62,112.27) .. controls (372.56,107.23) and (372.99,106.93) .. (383.08,110.37) -- cycle ; 

\draw    (291.93,75) -- (308.46,61.28) ;
\draw [shift={(310,60)}, rotate = 140.31] [color={rgb, 255:red, 0; green, 0; blue, 0 }  ][line width=0.75]    (10.93,-3.29) .. controls (6.95,-1.4) and (3.31,-0.3) .. (0,0) .. controls (3.31,0.3) and (6.95,1.4) .. (10.93,3.29)   ;
\draw    (291.93,75) -- (308.46,88.72) ;
\draw [shift={(310,90)}, rotate = 219.69] [color={rgb, 255:red, 0; green, 0; blue, 0 }  ][line width=0.75]    (10.93,-3.29) .. controls (6.95,-1.4) and (3.31,-0.3) .. (0,0) .. controls (3.31,0.3) and (6.95,1.4) .. (10.93,3.29)   ;

\draw    (375,145) -- (356.41,126.41) ;
\draw [shift={(355,125)}, rotate = 45] [color={rgb, 255:red, 0; green, 0; blue, 0 }  ][line width=0.75]    (10.93,-3.29) .. controls (6.95,-1.4) and (3.31,-0.3) .. (0,0) .. controls (3.31,0.3) and (6.95,1.4) .. (10.93,3.29)   ;
\draw    (375,145) -- (346.97,140.33) ;
\draw [shift={(345,140)}, rotate = 9.46] [color={rgb, 255:red, 0; green, 0; blue, 0 }  ][line width=0.75]    (10.93,-3.29) .. controls (6.95,-1.4) and (3.31,-0.3) .. (0,0) .. controls (3.31,0.3) and (6.95,1.4) .. (10.93,3.29)   ;

\draw (236,66) node [anchor=north west][inner sep=0.75pt]   [align=left] {Chiplets};
\draw (376,137) node [anchor=north west][inner sep=0.75pt]   [align=left] {Mini-lenses};

\end{tikzpicture}

%% file: fig10.tex
  
\tikzset {_nj3jgiz6f/.code = {\pgfsetadditionalshadetransform{ \pgftransformshift{\pgfpoint{0 bp } { 0 bp }  }  \pgftransformscale{1 }  }}}
\pgfdeclareradialshading{_ewvczefie}{\pgfpoint{0bp}{0bp}}{rgb(0bp)=(1,1,1);
rgb(0bp)=(1,1,1);
rgb(25bp)=(0.29,0.29,0.29);
rgb(400bp)=(0.29,0.29,0.29)}
\tikzset{every picture/.style={line width=0.75pt}} 

\begin{tikzpicture}[x=0.75pt,y=0.75pt,yscale=-1,xscale=1]

\draw  [fill={rgb, 255:red, 245; green, 166; blue, 35 }  ,fill opacity=1 ] (95.53,120.6) -- (163.49,52.65) -- (290.53,52.65) -- (290.53,57.04) -- (222.58,125) -- (95.53,125) -- cycle ; \draw   (290.53,52.65) -- (222.58,120.6) -- (95.53,120.6) ; \draw   (222.58,120.6) -- (222.58,125) ;
\path  [shading=_ewvczefie,_nj3jgiz6f] (95.53,89.35) -- (163.89,21) -- (290.53,21) -- (290.53,52.65) -- (222.18,121) -- (95.53,121) -- cycle ; 
 \draw  [color={rgb, 255:red, 65; green, 117; blue, 5 }  ,draw opacity=1 ] (95.53,89.35) -- (163.89,21) -- (290.53,21) -- (290.53,52.65) -- (222.18,121) -- (95.53,121) -- cycle ; 
 \draw  [color={rgb, 255:red, 65; green, 117; blue, 5 }  ,draw opacity=1 ] (290.53,21) -- (222.18,89.35) -- (95.53,89.35) ; \draw  [color={rgb, 255:red, 65; green, 117; blue, 5 }  ,draw opacity=1 ] (222.18,89.35) -- (222.18,121) ;
\draw  [draw opacity=0][fill={rgb, 255:red, 245; green, 166; blue, 35 }  ,fill opacity=1 ][line width=1.5]  (164.04,21) -- (290.53,21) -- (221.5,90.04) -- (95,90.04) -- cycle ; \draw  [color={rgb, 255:red, 128; green, 128; blue, 128 }  ,draw opacity=1 ][line width=1.5]  (164.04,21) -- (95,90.04)(169.04,21) -- (100,90.04)(174.04,21) -- (105,90.04)(179.04,21) -- (110,90.04)(184.04,21) -- (115,90.04)(189.04,21) -- (120,90.04)(194.04,21) -- (125,90.04)(199.04,21) -- (130,90.04)(204.04,21) -- (135,90.04)(209.04,21) -- (140,90.04)(214.04,21) -- (145,90.04)(219.04,21) -- (150,90.04)(224.04,21) -- (155,90.04)(229.04,21) -- (160,90.04)(234.04,21) -- (165,90.04)(239.04,21) -- (170,90.04)(244.04,21) -- (175,90.04)(249.04,21) -- (180,90.04)(254.04,21) -- (185,90.04)(259.04,21) -- (190,90.04)(264.04,21) -- (195,90.04)(269.04,21) -- (200,90.04)(274.04,21) -- (205,90.04)(279.04,21) -- (210,90.04)(284.04,21) -- (215,90.04)(289.04,21) -- (220,90.04) ; \draw  [color={rgb, 255:red, 128; green, 128; blue, 128 }  ,draw opacity=1 ][line width=1.5]  (164.04,21) -- (290.53,21)(159.04,26) -- (285.53,26)(154.04,31) -- (280.53,31)(149.04,36) -- (275.53,36)(144.04,41) -- (270.53,41)(139.04,46) -- (265.53,46)(134.04,51) -- (260.53,51)(129.04,56) -- (255.53,56)(124.04,61) -- (250.53,61)(119.04,66) -- (245.53,66)(114.04,71) -- (240.53,71)(109.04,76) -- (235.53,76)(104.04,81) -- (230.53,81)(99.04,86) -- (225.53,86) ; \draw  [color={rgb, 255:red, 128; green, 128; blue, 128 }  ,draw opacity=1 ][line width=1.5]   ;
\draw    (245,123) -- (224.58,123) ;
\draw [shift={(222.58,123)}, rotate = 360] [color={rgb, 255:red, 0; green, 0; blue, 0 }  ][line width=0.75]    (10.93,-3.29) .. controls (6.95,-1.4) and (3.31,-0.3) .. (0,0) .. controls (3.31,0.3) and (6.95,1.4) .. (10.93,3.29)   ;

\draw (137,42) node [anchor=north west][inner sep=0.75pt]  [color={rgb, 255:red, 0; green, 0; blue, 0 }  ,opacity=1 ] [align=left] {\begin{minipage}[lt]{85.09pt}\setlength\topsep{0pt}
\begin{center}
Transmissive RIS \\
(Phase Mask)
\end{center}

\end{minipage}};
\draw (97.53,98) node [anchor=north west][inner sep=0.75pt]   [align=left] {\textcolor[rgb]{1,1,1}{Dielectic Half-lens}};
\draw (246,110) node [anchor=north west][inner sep=0.75pt]   [align=left] {Reflector/Mirror};

\end{tikzpicture}